\documentclass[twocolumn,showpacs,aps,prd,floatfix,superscriptaddress]{revtex4}
\usepackage{graphicx}
\usepackage{dcolumn}
\usepackage{amsmath}
\usepackage{epsfig}

\newcommand{\BaBarYear}       {10}
\newcommand{\BaBarNumber}     {06}
\newcommand{\SLACPubNumber} {14181}

 \newcommand{\BaBarType}      {PUB}  % Journal publication

\def\figurebox#1#2#3{%
    \def\arg{#3}%
    \ifx\arg\empty
    {\hfill\vbox{\hsize#2\hrule\hbox to #2{\vrule\hfill\vbox to #1{\hsize#2\vfill}\vrule}\hrule}\hfill}%
    \else
    {\hfill\epsfbox{#3}\hfill}%
    \fi}

\def\sss{\scriptscriptstyle}
\def\barpd{{\raise.35ex\hbox
{${\sss (}$}}--{\raise.35ex\hbox{${\sss )}$}}}
\def\dbarp{\hbox{$D^{0}$\kern-1.25em\raise1.5ex\hbox{\barpd}}}
\def\dbarpstar{\hbox{$D^{*0}$\kern-1.65em\raise1.5ex\hbox{\barpd}}}
\def\Dzbpar  {\ensuremath{\Dbar^{(*)0}}\xspace}
\def\RDK {\ensuremath{{\cal R}_{DK}}\xspace}
\def\RDKp {\ensuremath{{\cal R}^+_{DK}}\xspace}
\def\RDKm {\ensuremath{{\cal R}^-_{DK}}\xspace}
\def\RDKpth {\ensuremath{{\cal R}^{+(th)}_{DK}}\xspace}
\def\RDKmth {\ensuremath{{\cal R}^{-(th)}_{DK}}\xspace}
\def\RDKpmth {\ensuremath{{\cal R}^{\pm(th)}_{DK}}\xspace}
\def\RDstarKpiz {\ensuremath{{\cal R}^*_{(D \piz)K}}\xspace}
\def\RDstarKgam {\ensuremath{{\cal R}^*_{(D \gamma)K}}\xspace}
\def\RDstarKpizpm {\ensuremath{{\cal R}^{*\pm}_{(D \piz)K}}\xspace}
\def\RDstarKgampm {\ensuremath{{\cal R}^{*\pm}_{(D \gamma)K}}\xspace}
\def\RDstarKpizp {\ensuremath{{\cal R}^{*+}_{(D \piz)K}}\xspace}
\def\RDstarKgamp {\ensuremath{{\cal R}^{*+}_{(D \gamma)K}}\xspace}
\def\RDstarKpizm {\ensuremath{{\cal R}^{*-}_{(D \piz)K}}\xspace}
\def\RDstarKgamm {\ensuremath{{\cal R}^{*-}_{(D \gamma)K}}\xspace}
\def\RDDstarK {\ensuremath{{\cal R}^{(*)}_{DK}}\xspace}
\def\RDDstarKp {\ensuremath{{\cal R}^{(*)+}_{DK}}\xspace}
\def\RDDstarKm {\ensuremath{{\cal R}^{(*)-}_{DK}}\xspace}
\def\RDDstarKpm {\ensuremath{{\cal R}^{(*)\pm}_{DK}}\xspace}
\def\RDpi {\ensuremath{{\cal R}_{D\pi}}\xspace}
\def\RDstarpipiz {\ensuremath{{\cal R}^*_{(D \piz)\pi}}\xspace}
\def\RDstarpigam {\ensuremath{{\cal R}^*_{(D \gamma)\pi}}\xspace}
\def\RDDstarpi {\ensuremath{{\cal R}^{(*)}_{D\pi}}\xspace}
\def\RDDstarpip {\ensuremath{{\cal R}^{(*)+}_{D\pi}}\xspace}
\def\RDDstarpim {\ensuremath{{\cal R}^{(*)-}_{D\pi}}\xspace}
\def\RDDstarpipm {\ensuremath{{\cal R}^{(*)\pm}_{D\pi}}\xspace}
\def\ADK {\ensuremath{{\cal  A}_{DK}}\xspace}
\def\ADstarKpiz {\ensuremath{{\cal  A}^*_{(D \piz)K}}\xspace}
\def\ADstarKgam {\ensuremath{{\cal  A}^*_{(D \gamma)K}}\xspace}
\def\ADDstarK {\ensuremath{{\cal  A}^{(*)}_{DK}}\xspace}
\def\ADpi {\ensuremath{{\cal  A}_{D\pi}}\xspace}
\def\ADstarpipiz {\ensuremath{{\cal  A}^*_{(D \piz)\pi}}\xspace}
\def\ADstarpigam {\ensuremath{{\cal  A}^*_{(D \gamma)\pi}}\xspace}
\def\ADDstarpi {\ensuremath{{\cal  A}^{(*)}_{D\pi}}\xspace}
\input pubboard/babarsym

\long\def\inst#1{\par\nobreak\kern 4pt\nobreak
    {\it #1}\par\vskip 10pt plus 3pt minus 3pt}

\begin{document}

%\linenumbers

\preprint{\babar-\BaBarType-\BaBarYear/\BaBarNumber }
\preprint{SLAC-PUB-\SLACPubNumber}

\begin{flushleft}
%BAD {\badnumber} Version {\badversion} \\
\babar-PUB-{\BaBarYear/\BaBarNumber} \\
SLAC-PUB-{\SLACPubNumber}
\end{flushleft}

% Title of the paper
\title{
\boldmath Search for $ b \to u$ Transitions in $\Bm \to D \Km$ and
$D^* \Km$ Decays}

% Input author list file
%% author list as of 04-Mar-2010 (445 authors)
%
\author{P.~del~Amo~Sanchez}
\author{J.~P.~Lees}
\author{V.~Poireau}
\author{E.~Prencipe}
\author{V.~Tisserand}
\affiliation{Laboratoire d'Annecy-le-Vieux de Physique des Particules (LAPP), Universit\'e de Savoie, CNRS/IN2P3,  F-74941 Annecy-Le-Vieux, France}
\author{J.~Garra~Tico}
\author{E.~Grauges}
\affiliation{Universitat de Barcelona, Facultat de Fisica, Departament ECM, E-08028 Barcelona, Spain }
\author{M.~Martinelli$^{ab}$}
\author{A.~Palano$^{ab}$ }
\author{M.~Pappagallo$^{ab}$ }
\affiliation{INFN Sezione di Bari$^{a}$; Dipartimento di Fisica, Universit\`a di Bari$^{b}$, I-70126 Bari, Italy }
% ...
\collaboration{The \babar\ Collaboration}
\noaffiliation

\date{\today}% It is always \today, today, but you may specify any date with \date.

\begin{abstract}
We report results from an updated study of the suppressed decays
$\Bm \to D \Km$ and $\Bm \to D^* \Km$ followed by $D\to \Kp \pim$,
where $D^{(*)}$ indicates a $D^{(*)0}$ or a $\bar D^{(*)0}$ meson,
and $D^* \to D \piz$ or $D^* \to D \gamma$. These decays are
sensitive to the CKM unitarity triangle angle $\gamma$ due to
interference between the $b\rightarrow c$ transition $\Bm \to
D^{(*)0}K^-$ followed by the doubly Cabibbo-suppressed decay
$\Dz\to\Kp\pim$, and the $b\rightarrow u$ transition $\Bm \to \bar
D^{(*)0}K^-$ followed by the Cabibbo-favored decay
$\Dzb\to\Kp\pim$. We also report an analysis of the decay $\Bm \to
D^{(*)} \pim$ with the $D$ decaying into the doubly
Cabibbo-suppressed mode $D \to \Kp\pim$. Our results are based on
467 million $\FourS \to B\Bbar$ decays collected with the \babar\
detector at SLAC. We measure the ratios ${\cal R}^{(*)}$ of the
suppressed ($[K^+\pi^-]_D K^-/\pi^-$) to favored ($[K^-\pi^+]_D
K^-/\pi^-$) branching fractions as well as the \CP asymmetries
${\cal A}^{(*)}$ of those modes. We see indications of signals for
the $\Bm\to D\Km$ and $\Bm\to D^*_{D\piz}\Km$ suppressed modes,
with statistical significances of 2.1 and 2.2$\sigma$,
respectively, and we measure:
$$ \RDK = (1.1\pm 0.6 \pm 0.2)\times 10^{-2},\ \ADK = -0.86 \pm 0.47 \ ^{+0.12}_{-0.16},$$
$$ \RDstarKpiz = (1.8\pm 0.9 \pm 0.4)\times 10^{-2},\ \ADstarKpiz = +0.77 \pm 0.35\pm 0.12,$$
$$ \RDstarKgam = (1.3\pm 1.4\pm 0.8 )\times 10^{-2},\ \ADstarKgam= +0.36 \pm 0.94\ ^{+0.25}_{-0.41},$$
where the first uncertainty is statistical and the second is
systematic. We use a frequentist approach to obtain  the magnitude
of the ratio $r_B \equiv |A(\Bm \to \Dzb \Km) / A(\Bm \to \Dz
\Km)|= (9.5^{+5.1}_{-4.1})\%$, with $r_B<16.7\%$ at 90\%
confidence level. In the case of $\Bm \to D^* \Km$ we find
$r^*_B\equiv |A(\Bm \to \Dstarzb \Km) / A(\Bm \to \Dstarz \Km)|=
(9.6^{+3.5}_{-5.1})\%$, with $r^*_B<15.0\%$ at 90\% confidence
level.

\end{abstract}

% PACS, the Physics and Astronomy Classification Scheme.

\maketitle

% reset footnote counter
\setcounter{footnote}{0}

% The body of the paper starts here
\section{Introduction}
\label{sec:Introduction}

\begin{figure}[hb]
    \begin{center}
    \epsfig{file=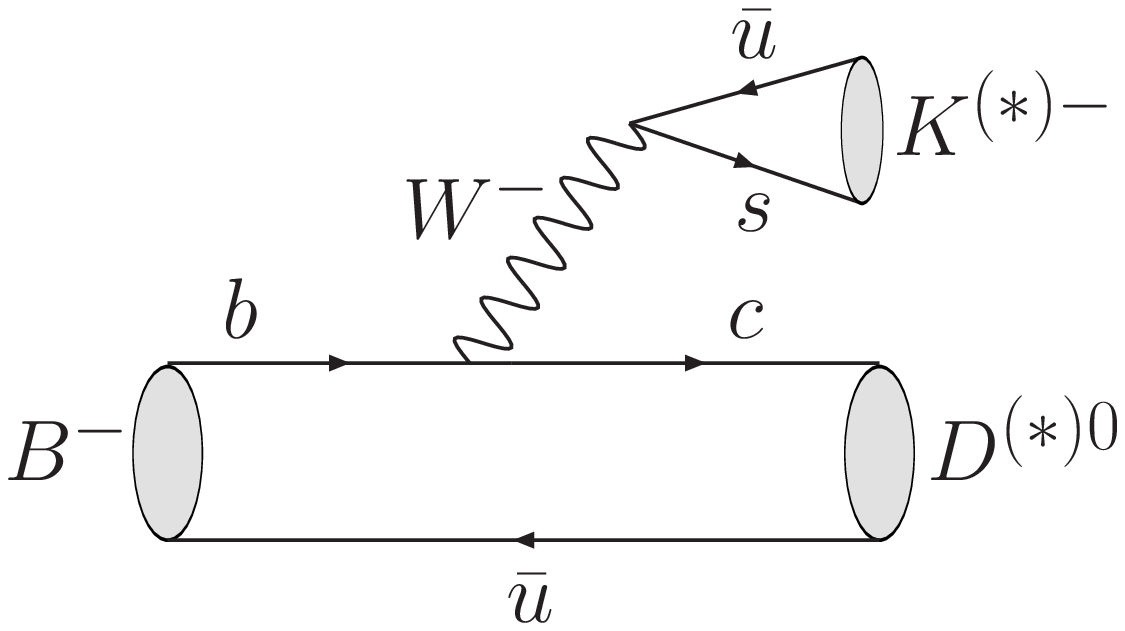,width=0.4\linewidth}
    \hspace{1cm}
    \epsfig{file=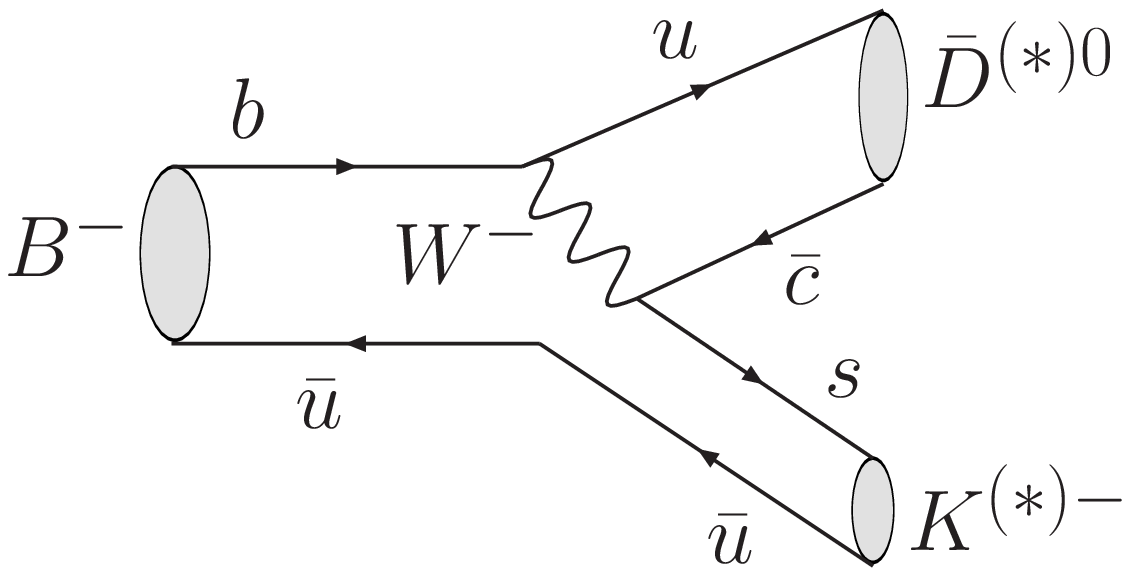,width=0.4\linewidth}
    \caption{ Feynman diagrams for $B^- \to D^{(*)0} K^{(*)-}$ and $\Dzbpar K^{(*)-}$.
   The latter is CKM and color-suppressed with respect to the former.
  }
    \label{fig:feynmandk}
    \end{center}
  \end{figure}

The Standard Model accommodates \CP violation through a single
phase in the Cabibbo-Kobayashi-Maskawa (CKM) quark mixing matrix
$V$~\cite{ckm}. In the Wolfenstein parameterization~\cite{wolf},
the angle $\gamma = \arg{(-V_{ud}V_{ub}^*/V_{cd}V_{cb}^*)}$ of the
unitarity triangle is related to the complex phase of the CKM
matrix element $V_{ub}$ through $V_{ub}=|V_{ub}|e^{-i\gamma}$. A
theoretically clean source of information on the angle $\gamma$ is
provided by $B^- \rightarrow D^{(*)} \  K^-$ decays, where
$D^{(*)}$ represents an admixture of $D^{(*)0}$ and \Dzbpar
states. These decays exploit the interference between $B^- \to
D^{(*)0}K^-$ and $B^- \to \Dzbpar K^-$ (Fig.~\ref{fig:feynmandk})
that occurs when the $D^{(*)0}$ and the \Dzbpar decay to common
final states.

In the Atwood-Dunietz-Soni (ADS) method \cite{ads}, the \Dz from
the favored $\b\to c$ amplitude is reconstructed in the doubly
Cabibbo-suppressed decay $K^+\pi^-$ , while the \Dzb from the
$\b\to u$ suppressed amplitude is reconstructed in the favored
decay $K^+\pi^-$. The product branching fractions for these final
states, which we denote as $[K^+\pi^-]_D K^-$ ($\Bm\to D \Km$) and
$[K^+\pi^-]_{D^*} K^-$ ($\Bm \to D^* \Km$), are small ($\sim
10^{-7}$), but the two interfering amplitudes are of the same
order of magnitude, and large \CP asymmetries are therefore
possible. The favored decay mode $\Bm \to [\Km \pip]_{D^{(*)}}
\Km$ is used to normalize the measurement and cancel many
systematic uncertainties. Thus, ignoring possible small effects
due to $D$ mixing and assuming no \CP violation in the
normalization modes, we define the charge-specific ratios for \Bp
and \Bm decay rates to the ADS final states as
     \begin{eqnarray}
   \label{eqn:rpm}
\RDK^{\pm} & \equiv &
   \frac{\Gamma([K^\mp \pi^\pm]_D K^\pm)}{\Gamma([K^\pm \pi^\mp]_D K^\pm)}\nonumber
   \\
   & = & r_B^2 + r_D^2 + 2 \: r_B r_D \cos(\pm \gamma + \delta),
    \end{eqnarray}
where $r_B = |A(\Bm \to \Dzb \Km) / A(\Bm \to \Dz \Km)|\approx
10\%$~\cite{ADS-BABAR,BaBarDalitz, BelleADS, BelleDalitz}  and
$r_D = |A(\Dz \to \Kp \pim) / A(\Dz \to \Km \pip)| = (5.78\pm
0.08)\%$~\cite{HFAG} are the suppressed to favored $B$ and $D$
amplitude ratios.  The rates in Eq.~(\ref{eqn:rpm}) depend on the
relative weak phase $\gamma$ and the relative strong phase $\delta
\equiv \delta_B + \delta_D$ between the interfering amplitudes,
where $\delta_B$ and $\delta_D$ are the strong phase differences
between the two $B$ and $D$ decay amplitudes, respectively. The
value of $\delta_D$ has been measured to be
$\delta_D=(201.9^{+11.3}_{-12.4})^\circ$~\cite{HFAG}, where we
have accounted for a phase shift of $180^\circ$ in the definition
of $\delta_D$ between Ref.~\cite{HFAG} and this analysis.

The main experimental observables are the charge-averaged decay
rate and the direct \CP asymmetry, which can be written as

  \begin{eqnarray}
  \label{eqn:rkpi}
   \RDK & \equiv & \frac{1}{2}\left(\RDK^+ + \RDK^-\right)\nonumber \\
     & = & r_B^2 + r_D^2 + 2 \: r_B r_D \cos \gamma \cos \delta
  \end{eqnarray}

   \begin{eqnarray}
  \label{eqn:akpi}
   \ADK  & \equiv & \frac{\RDK^- - \RDK^+}{\RDK^- + \RDK^+}\nonumber \\
      & = & 2 \: r_B r_D \sin \gamma \sin \delta / \RDK.
  \end{eqnarray}
The treatment for the $D^{*} K$ mode is identical to the $D K$
one, but the parameters $r_B^*$ and $\delta_B^*$ are not expected
to be numerically the same as those of the $D K$ mode. Taking into
account the effective strong phase difference of $\pi$ between the
\Dstar decays to $D \gamma$ and $D \piz$\cite{Bondar},
   we define the charge-specific ratios for $D^*$ as:
   \begin{eqnarray}
   \label{eqn:rpmdst1}
   \RDstarKpizpm & \equiv &
   \frac{\Gamma([K^\mp \pi^\pm]_{D^*\to D\piz} K^\pm)}{\Gamma([K^\pm \pi^\mp]_{D^*\to D\piz}
   K^\pm)}\nonumber \\
   & = & r_B^{*2} + r_D^2 + 2 \: r_B^* r_D \cos(\pm \gamma + \delta^*),
   \end{eqnarray}

\begin{eqnarray}
   \label{eqn:rpmdst2}
   \RDstarKgampm & \equiv &
   \frac{\Gamma([K^\mp \pi^\pm]_{D^*\to D\gamma} K^\pm)}{\Gamma([K^\pm \pi^\mp]_{D^*\to D\gamma}
   K^\pm)}\nonumber \\
   & = & r_B^{*2} + r_D^2 - 2 \: r_B^* r_D \cos(\pm \gamma + \delta^*),
   \end{eqnarray}
with $r_B^* = |A(\Bm \to \bar D^{*0} \Km) / A(\Bm \to D^{*0}
\Km)|$ and $\delta^*\equiv \delta_B^* + \delta_D$, where
$\delta_B^*$ is the strong phase difference between the two $B$
decay amplitudes. The charge averaged ratios for $D^{*}\to D\piz$
and $D^{*}\to D\gamma$ are then:

  \begin{eqnarray}
  \label{eqn:rkpi2}
   \RDstarKpiz & \equiv & \frac{1}{2}\left(\RDstarKpizp + \RDstarKpizm \right)\nonumber \\
     & = & r_B^{*2} + r_D^2 + 2 \: r_B^* r_D \cos \gamma \cos
     \delta^*,
  \end{eqnarray}
 \begin{eqnarray}
  \label{eqn:rkpi3}
   \RDstarKgam & \equiv & \frac{1}{2}\left(\RDstarKgamp + \RDstarKgamm \right)\nonumber \\
      & = & r_B^{*2} + r_D^2 - 2 \: r_B^* r_D \cos \gamma
      \cos \delta^*.
  \end{eqnarray}
Definitions of the direct \CP asymmetries \ADstarKpiz and
\ADstarKgam follow Eq.~(\ref{eqn:akpi}).

 This paper is an update of our previous ADS analysis
   in Ref.~\cite{ADS-BABAR}, which used $232 \times 10^6 B\Bbar$
   pairs and set 90\% C.L. upper limits $\RDK<0.029$, $\RDstarKpiz<0.023$ and $\RDstarKgam<0.045$.
   In addition to an increased data sample, new features in the analysis include a multi-dimensional fit involving the
   neural network output used to discriminate the signal from the
   continuum background, rather than a simple cut on this variable as was done in the
   previous analysis. We also include measurements of
   the ratios of the doubly Cabibbo-suppressed to Cabibbo-favored $D^{(*)}\pi$ decay rates,
   \begin{equation}
   \label{eqn:rdpi}
   \RDDstarpipm \equiv
   \frac{\Gamma(B^\pm \to [K^\mp \pi^\pm]_{D^{(*)}} \pi^\pm)}{\Gamma(B^\pm \to [K^\pm \pi^\mp]_{D^{(*)}}
   \pi^\pm)},
   \end{equation}
   and of the corresponding asymmetries. These measurements are
   used as a check for the $\Bm
\to [\Kp\pim]_{D^{(*)}} \Km$ ADS analysis. In the $D^{(*)}\pi$
case, we expect that the ratio $r_B^{(*)(D\pi)}$ of the $V_{ub}$
to $V_{cb}$ amplitudes is suppressed by a factor $|V_{cd}
V_{us}/V_{ud}V_{cs}|$ compared to the $D^{(*)}K$ case, if we
assume the same color suppression factor for both decays. One
expects therefore $r_B^{(*)(D\pi)}\approx r^{(*)}_B \times \tan^2
\theta_c \approx 5\times 10^{-3}\ll r_D$, where $\theta_c$ is the
Cabibbo angle and where we have assumed $r^{(*)}_B=10\%$.
Neglecting higher order terms, $\RDDstarpi \simeq r_D^2$ and
$\ADDstarpi \simeq 2 r_B^{(*)} \tan^2 \theta_c \sin \gamma \sin
\delta^{(*)}/r_D $. Hence, the maximum asymmetry possible for
$D^{(*)}\pi$ ADS decays is $2 r^{(*)}_B \tan^2 \theta_c
/r_D\approx 18\%$.

\section{\boldmath The \babar\ detector and dataset}
\label{sec:babar}
   The results presented in this paper are based on $467 \times 10^6$ $\FourS\to B\Bbar$ decays,
    corresponding to an integrated luminosity of 426 fb$^{-1}$ (on-peak data). The data were collected between
    1999 and 2007 with the \babar\ detector~\cite{babar} at the \pep2\ \epem collider at SLAC.
   In addition, a 44~fb$^{-1}$ data sample, with center-of-mass (CM) energy 40~\mev
   below the \FourS resonance (off-peak data), is used to study backgrounds from continuum events, $e^+ e^- \to q \bar{q}$
   ($q=u,d,s,$ or $c$).

   The \babar\  detector response to various
   physics processes as well as to varying beam and environmental
   conditions is modeled with simulation software based on the
   {\tt Geant4}~\cite{geant4} tool kit. We use {\tt EVTGEN}~\cite{evtgen} to model the
   kinematics of $B$ meson decays and {\tt JETSET}~\cite{jetset} to model continuum
   processes $e^+ e^- \to q \bar{q}$.

%-------------------------------------------------------------------------------------
\section{\boldmath ANALYSIS METHOD}
\label{sec:Analysis}

   \subsection{Basic Requirements}
   \label{sec:basic}
   We reconstruct $\Bm \to D^{(*)}K^-$ and $\Bm \to
   D^{(*)}\pi^-$ with the $D$ decaying to $\Km\pip$ (right-sign (RS) decays) and $\Kp\pim$
(wrong-sign (WS) decays). For decays involving a
   $D^{*}$, both $D^{*}\to D\piz$ and $D^{*}\to D\gamma$ modes are
   reconstructed. Charged kaon and pion candidates must satisfy identification
   criteria that are typically 85\% efficient, depending on
   momentum and polar angle. The misidentification rates  are at
   the few percent level.  We select
   $D$ candidates with an invariant mass within 20\mevcc (about 3 standard deviations) of the known
   \Dz mass~\cite{PDG}. All $D$ candidates are mass and vertex constrained.
   For modes with $D^{*}\to D\piz$ or $D^*\to D\gamma$, the mass difference $\Delta m$ between the $D^*$
   and the $D$ must be within 4\mevcc ($\simeq 4\sigma$) or 15\mevcc
   ($\simeq 2\sigma$), respectively, of the nominal mass difference~\cite{PDG}.

   For the WS decays $B^\pm \to [K^\mp \pi^\pm]_D K^\pm$,
   two important sources of background arise: the first from $B^\pm \to [\pi^\mp K^\pm]_D
   K^\pm$ (in which the $K$ and $\pi$ in the $D$ decay are
   misidentified as $\pi$ and $K$) and the second from $B^\pm \to [K^\mp
   K^\pm]_D\pi^\pm$ (when the $K^\mp$ $\pi^\pm$ pair has an invariant mass within 20\mevcc of the nominal \Dz
   mass). To eliminate the first background, we recompute the invariant mass
($M_{{\rm switch}}$) of the $h^+h^{\prime -}$ pair in $\Dz \to
h^+h^{\prime -}$ switching the
   mass assumptions on the $h^+$ and the $h^{\prime -}$.  We veto
    candidates with $M_{{\rm switch}}$ within 20\mevcc of the \Dz  mass~\cite{PDG}.
   To eliminate the second background, we also veto any candidate where
   the $KK$ invariant mass is within 20\mevcc of the \Dz
   mass. To ensure the same selection efficiencies, these criteria are
   applied both to $B^\pm \to [K^\mp \pi^\pm]_{D^{(*)}} K^\pm$ and to $B^\pm \to [K^\pm \pi^\mp]_{D^{(*)}}
   K^\pm$ candidates.  These veto cuts
   are 88\% efficient on signal decays.

   We identify $B$ candidates using two nearly independent
   kinematic variables that are customarily used when reconstructing
   $B$-meson decays at the $\FourS$.  These variables are
   the energy-substituted mass,
   $\mes \equiv \sqrt{(\frac{s}{2}  + \vec{p}_0\cdot \vec{p}_B)^2/E_0^2 - p_B^2}$
   and energy difference $\Delta E \equiv E_B^*-\frac{1}{2}\sqrt{s}$,
   where $E$ and $p$ are energy and momentum, the asterisk
   denotes the CM frame, the subscripts $0$ and $B$ refer to the
   \FourS and $B$ candidate, respectively, and $s$ is the square
   of the CM energy.  For signal events $\mes = m_{B^+}$~\cite{PDG} and
   $\Delta E = 0$ within the
   resolutions of about 2.6 \mevcc and 17 MeV, respectively. We require that all candidates have $|\Delta E|<40\mev$ and
   we use \mes in the fit to extract the number of signal events.

The average number of $B\to D^{(*)}K$ candidates reconstructed per
selected event is about 1.4 in $B\to DK$ signal Monte Carlo (MC)
events and about 2 for $B\to D^{*}K$ signal MC events. This is
mostly due to the cross-feed between the $D K$ and the $D^* K$
final states. For all events with multiple $B\to D^{(*)}K$
candidates, we retain only one candidate per event, based on the
smallest value of $|\Delta E|$. This method does not bias the
sample since $\Delta E$ is not used to extract the number of
signal events. After this arbitration, less than 0.4\% (0.5\%) of
the $B\to DK$ ($B\to D^{*}K$) signal MC events selected are
reconstructed as $B\to D^{*}K$ ($B\to DK$). About 10\% of the
$B\to D^{*}_{D\piz}K$ events selected are reconstructed as $B\to
D^{*}_{D\gamma}K$ and about 2\% of the $B\to D^{*}_{D\gamma}K$
events selected are reconstructed as $B\to D^{*}_{D\piz}K$.

The $B\to D^{(*)}\pi$ analysis is performed independently of the
$B\to D^{(*)}K$ analysis, but uses the same multiple candidate
selection algorithm. A summary of the selection efficiencies for
the WS modes $[K^\pm \pi^\mp]_{D^{(*)}} h^\mp$ ($h$=$K,\pi$) and
the RS modes $[\pi^\pm K^\mp]_{D^{(*)}} h^\mp$ is given in
Table~\ref{tab:summaryeff}.

\begin{table}
\caption{Selection efficiencies, after correction for known
data/MC differences, for $B^\mp \to[K^\pm \pi^\mp]_{D^{(*)}}
h^\mp$ ($\epsilon_{WS}$) and $B^\mp \to[K^\mp \pi^\pm]_{D^{(*)}}
h^\mp$ ($\epsilon_{RS}$), and efficiency ratio
$\epsilon_{WS}/\epsilon_{RS}$.}

\begin{center}
\begin{tabular}{lccc}
  \hline\hline
 Channel  &  $\epsilon_{WS}$ (\%)  &  $\epsilon_{RS}$ (\%)  & $\epsilon_{WS}/\epsilon_{RS}$ ($10^{-2}$) \\ \hline
  $D K$                     & 26.5$\pm$0.1 & 26.6$\pm$0.1 &  99.6$\pm$0.5\\
  $D^{*}_{D\piz}K$         & 13.3$\pm$0.1 & 13.2$\pm$0.1 & 100.6$\pm$1.1\\
$D^{*}_{D\gamma}K$         & 17.4$\pm$0.1 & 17.5$\pm$0.1 &  99.8$\pm$0.8\\
\hline
  $D \pi$                   & 26.0$\pm$0.1 & 26.5$\pm$0.1 &  97.9$\pm$0.5\\
  $D^{*}_{D\piz}\pi$       & 14.3$\pm$0.1 & 14.8$\pm$0.1 &  96.4$\pm$0.9\\
$D^{*}_{D\gamma}\pi$       & 18.8$\pm$0.1 & 19.5$\pm$0.1 &  96.3$\pm$0.7\\
\hline\hline
\end{tabular}
\label{tab:summaryeff}
\end{center}
\end{table}
%-------------------------------------------------------------------------------------

\subsection{Neural Network}
   \label{sec:nn}

\begin{figure*}
\begin{center}
\epsfig{file=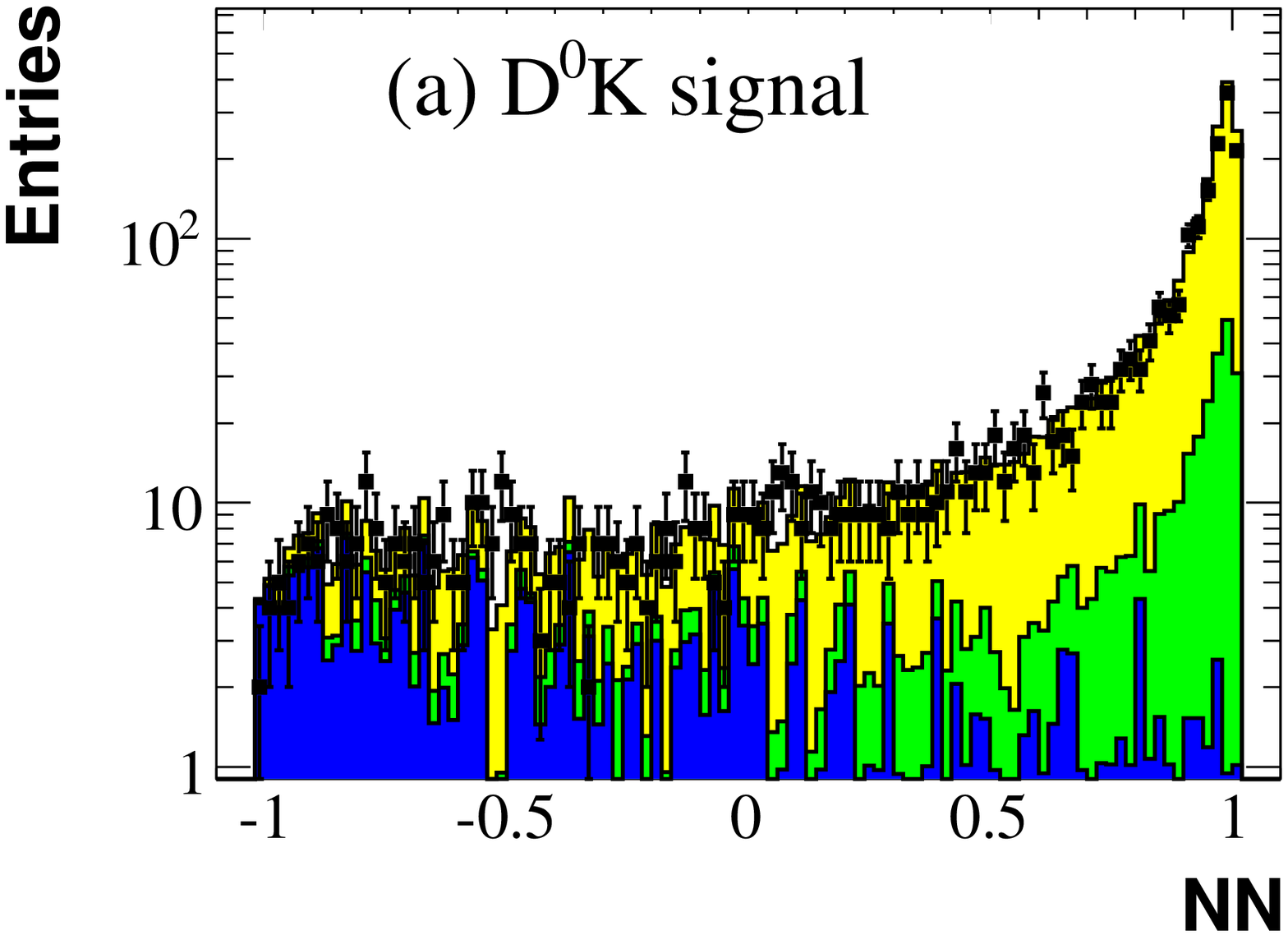,width=0.319\linewidth}
\epsfig{file=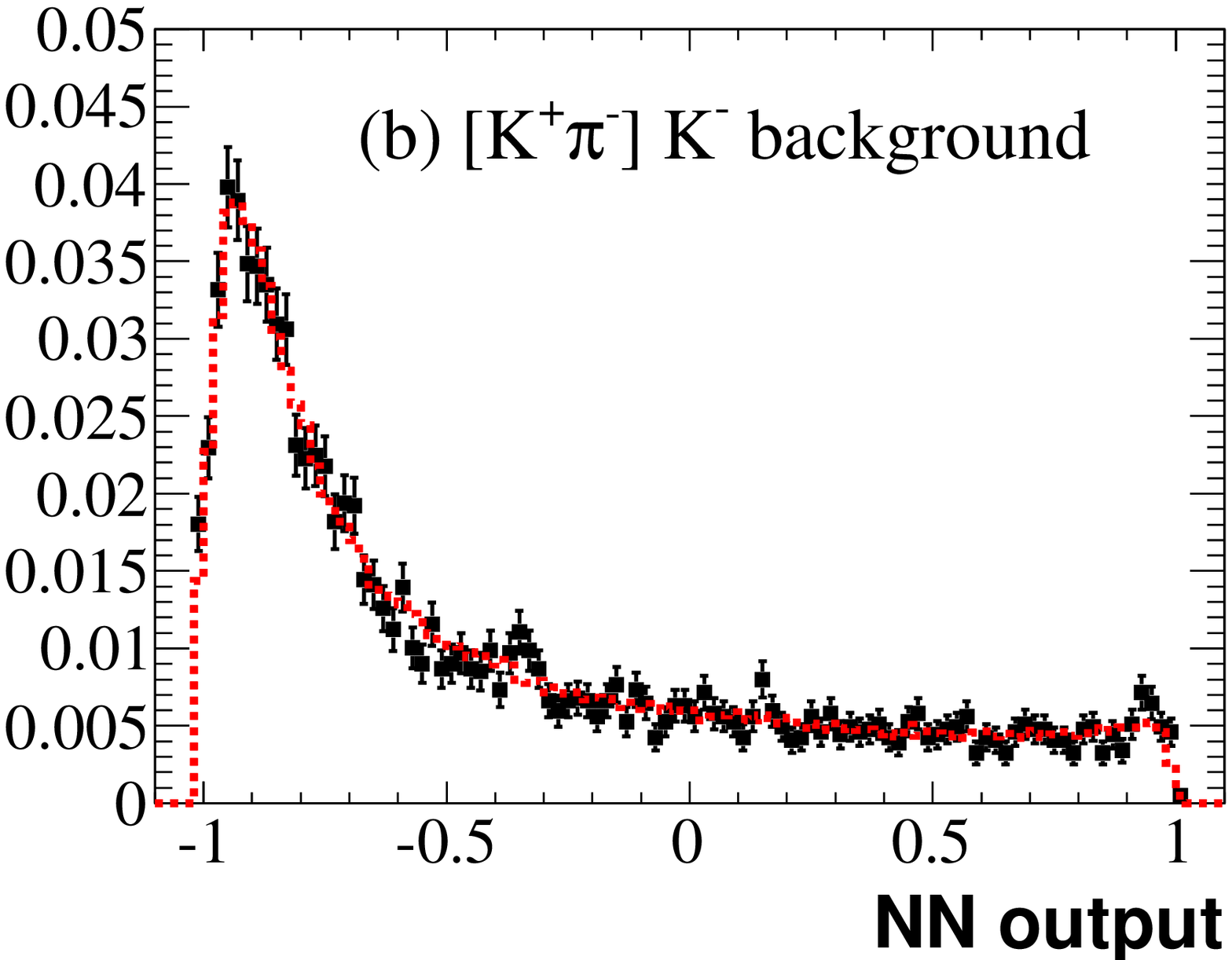,width=0.319\linewidth}
\epsfig{file=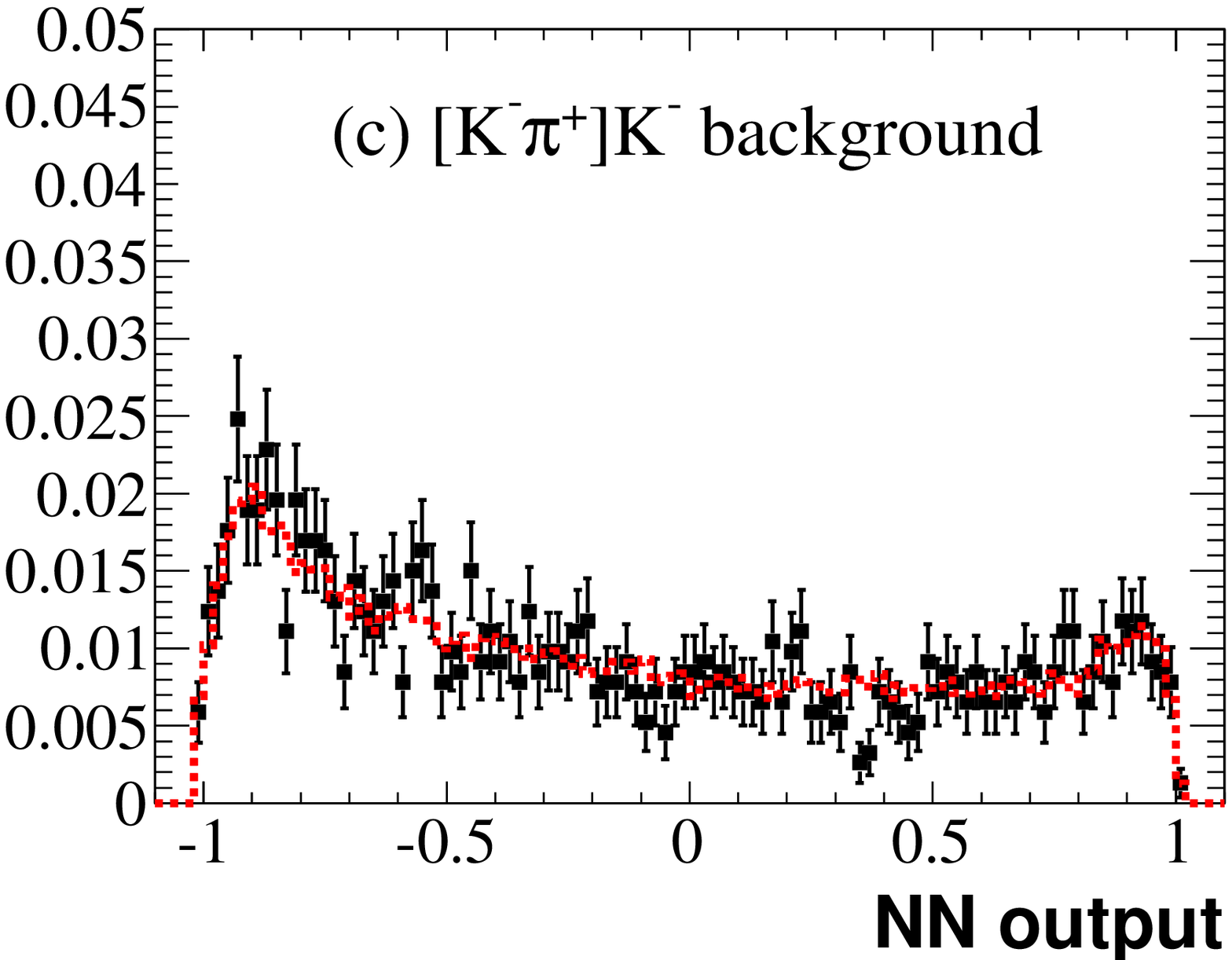,width=0.319\linewidth}
\epsfig{file=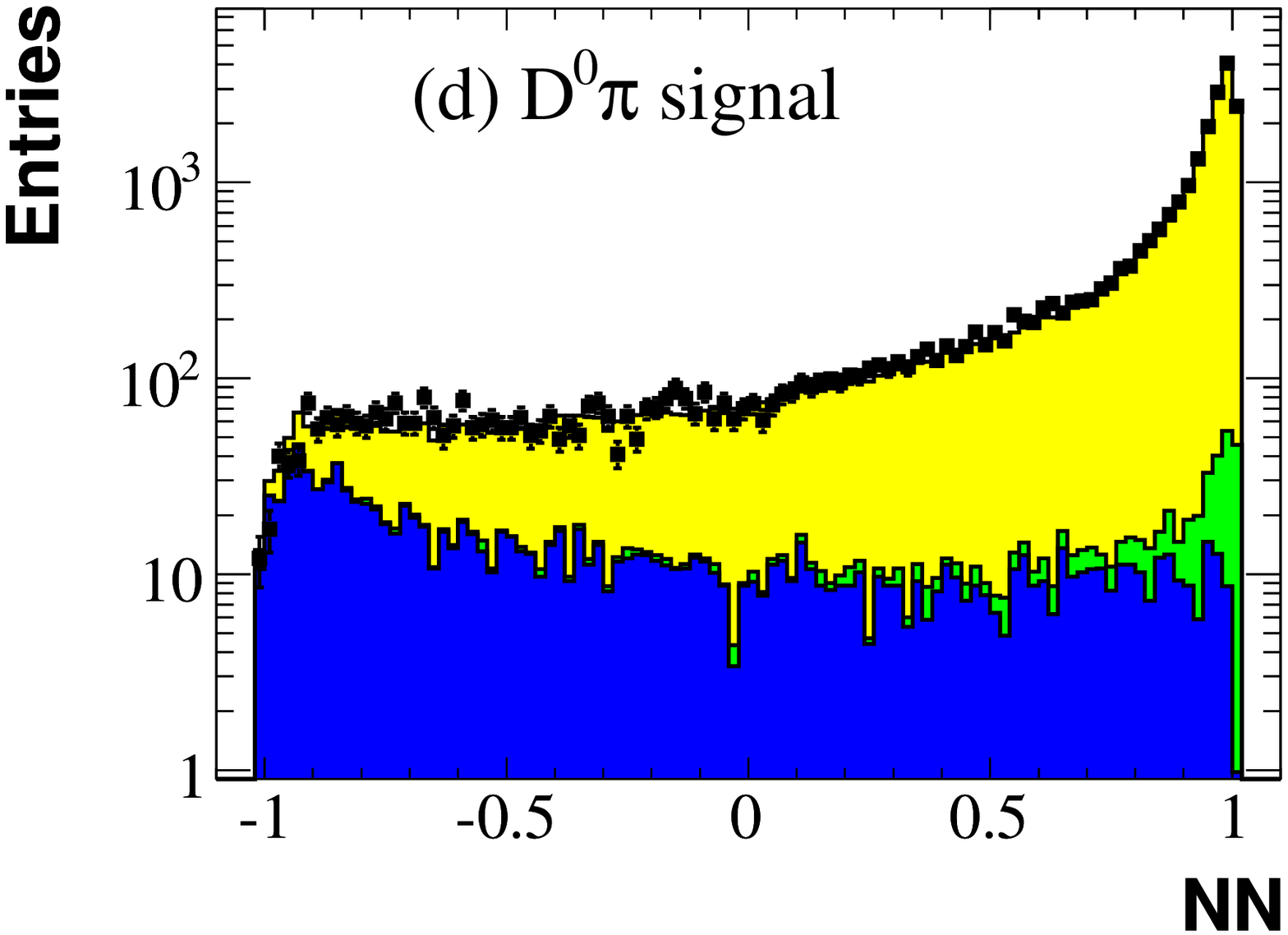,width=0.319\linewidth}
\epsfig{file=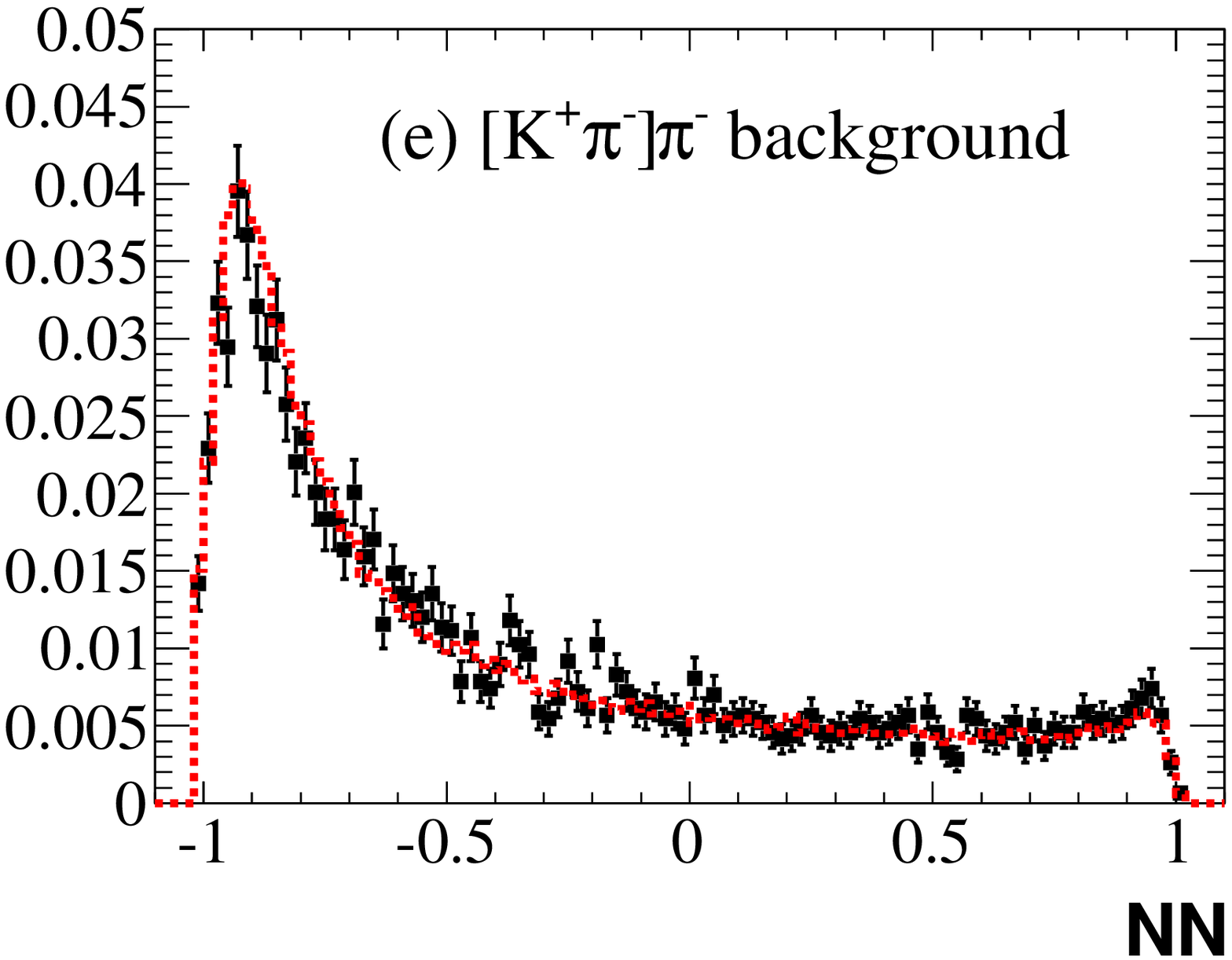,width=0.319\linewidth}
\epsfig{file=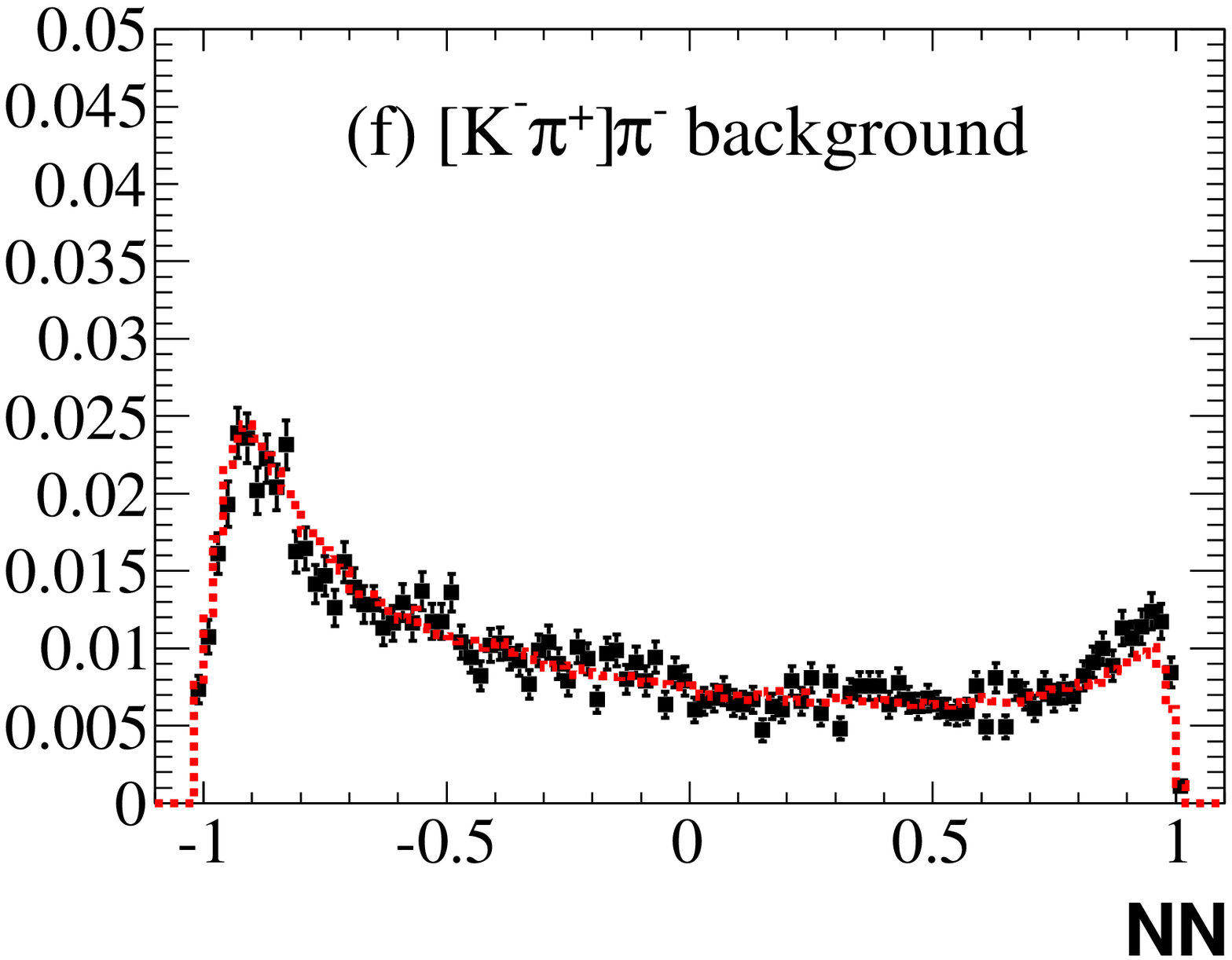,width=0.319\linewidth} \caption{(color
online). Signal and background distributions of the neural network
output, and results of the $NN$ verifications for $DK$ (a),
$D^{(*)}K$ (b, c), $D\pi$ (d) and $D^{(*)}\pi$ (e, f) candidates.
(a,d): $D h^\pm$ right-sign candidates, signal-enriched by a cut
on the $\Delta E$, \mes signal region. Shaded plain histograms are
MC expectations for $q\bar q$ background (dark gray/blue), $b\bar
b$ background (middle gray/green) and $B^\pm\to D h^\pm$ signal
events (light gray/yellow). Points with error bars are on-peak
data. (b,e): $D^{(*)}h^\pm$ wrong-sign background. (c,f):
$D^{(*)}h^\pm$ right-sign background. Plots b, c, e, and f are
normalized to unity. The dotted line histograms show the
distribution of simulated continuum events. The off-peak data used
to check the $NN$ are overlaid as data points. To increase the
statistics, the $\mes$ and $\Delta E$ requirements on the off-peak
and continuum MC events have been relaxed, and $D h^\pm$ and $D^*
h^\pm$ contributions have been summed.} \label{fig:nnplots}
\end{center}
\end{figure*}

After these initial requirements, backgrounds dominantly arise
   from continuum events, especially $e^+ e^- \to c \bar{c}$, with
   $\bar{c} \to \Dzb X$, $\Dzb \to K^+ \pi^-$ and $c \to \Dz X$,
   $\Dz \to K^- +$anything. The continuum background is reduced by using neural network
   techniques. To select the discriminating variables used in the
   neural network, we rely on a study performed for the
previous version of this analysis~\cite{ADS-BABAR}, and we
consider the seven quantities listed below:
%, based both on the topology of the events and on $B$-tagging like quantities:

  \begin{enumerate}
\item Two event shape moments $L_0 = \sum_i{p_i}$, and $L_2 =
\sum_i{p_i \cos^2\theta_i}$, calculated in the CM frame. Here,
$p_i$ is the momentum and $\theta_i$ is the angle with respect to
the thrust axis of the $B$ candidate; the index $i$ runs over all
tracks and clusters not used to reconstruct the $B$ meson (rest of
the event). These variables are sensitive to the shape of the
event, separating jet-like continuum events from more spherical
$B\Bbar$ events.
   \item The absolute value of the cosine of the angle in
   the CM frame between the thrust axes of the $B$ candidate and
   the detected remainder of the event, $|\cos \theta_T|$.  The distribution of
   $|\cos \theta_T|$ is approximately uniform for signal and strongly
   peaked at one for continuum background.
   \item The absolute value of the cosine of the CM angle between the $B$ candidate momentum and the beam axis, $|\cos \theta_B|$.
   In this variable, the signal follows a $1-\cos^2 \theta_B$ distribution, while the background is approximately uniform.
   \item The charge difference $\Delta Q$ between the sum
   of the charges of tracks in the $D^{(*)}$ hemisphere
   and the sum of the charges of the tracks in the opposite
   hemisphere, excluding the tracks used in the reconstructed $B$,
   and where the partitioning of the event into two hemispheres is done in the CM frame.
   This variable exploits the correlation occurring in $c\bar{c}$ events between the charge of the $c$ (or
   $\bar c$) in a given hemisphere and the sum of the charges of
   all particles in that hemisphere. For signal events, the average charge difference is $\langle \Delta Q \rangle = 0$, whereas
   for the $c\bar{c}$ background $\langle \Delta Q \rangle \approx \frac{7}{3}\times Q_B$,
   where $Q_B$ is the charge of the $B$ candidate.
   \item The product $Q_B \cdot Q_K$, where $Q_K$ is the sum of the charges of all
   kaons in the rest of the event.
   In many signal events, there is a charged kaon among the decay
   products of the other $B$ in the event.  The charge
   of this kaon tends to be highly correlated with the charge of the
   $B$.  Thus, signal events tend to have $Q_B \cdot Q_K \leq -1$.  On
   the other hand, most continuum events have no kaons outside of the
   reconstructed $B$, and therefore $Q_K = 0$.
\item A quantity ${\cal M}_{K\ell}$, defined to be zero if there
are no leptons ($e$ or $\mu$) in the event, and, if a lepton is
found, taken to be equal to the invariant mass of this lepton and
the kaon from $B$ (bachelor $K$). This quantity differentiates
between continuum background and signal because continuum events
have fewer leptons than \BB events. Furthermore, a large fraction
of leptons in $c\bar{c}$ background events are from $D \to K \ell
\nu$, where the kaon becomes the bachelor kaon candidate, so that
the average ${\cal M}_{K\ell}$ in $c\bar{c}$ events is lower than
in $B$ signal events.
\item The absolute value of the measured proper time interval
between the two $B$ decays, $|\Delta t|$. This is calculated from
the measured separation, $\Delta z$, between the decay points of
the reconstructed $B$ and the other $B$ along the beam direction,
and the known Lorentz boost of the initial $e^+e^-$ state. For
continuum background, $|\Delta t|$ is peaked at 0, with most
events having $|\Delta t|<2\ps$, while it is less peaked and can
extend beyond $5\ps$ for $B^\pm\to D^{(*)} h^\pm$ signal events.
\end{enumerate}

The neural network is trained with simulated continuum
   and signal $[K^\pm \pi^\mp]_{D^{(*)}} K^\mp$ events. Only wrong-sign $D^{(*)}K$ candidates are used in the training,
   but the neural network is used in the analysis of all the $D^{(*)}h^\mp$ channels.
   The distributions of the neural network output ($NN$) for signal-enriched right-sign control samples
   are compared with expectations from the MC simulation
   in Fig.~\ref{fig:nnplots}(a) ($D K$) and Fig.~\ref{fig:nnplots}(d) ($D\pi$). The agreement is satisfactory. In
   the same figure, the $NN$ spectra of background control samples
   (off-peak data) are compared with expectations from continuum
   $q\bar q$ MC. Since we do not expect
   these distributions to be exactly the same for the right-sign
   and wrong-sign background samples, they are shown separately for the $[K^\pm \pi^\mp]_{D^{(*)}}
   K^\mp$(Fig.~\ref{fig:nnplots}(b)), $[K^\mp \pi^\pm]_{D^{(*)}} K^\mp$ (Fig.~\ref{fig:nnplots}(c)), $[K^\pm \pi^\mp]_{D^{(*)}}
   \pi^\mp$(Fig.~\ref{fig:nnplots}(e))
   and $[K^\mp \pi^\pm]_{D^{(*)}} \pi^\mp$ (Fig.~\ref{fig:nnplots}(f)) channels. To increase the statistics, the $\mes$ and $\Delta E$
requirements on the off-resonance and continuum MC events have
been relaxed, and the $Dh^\pm$ and $D^{*}h^\pm$ contributions have
been summed, after checking that they are in agreement with each
other. Good agreement
   between data and the simulation is observed in all
   channels. Good agreement between the $D^{(*)}K$ and the
   $D^{(*)}\pi$ background $NN$ distributions is also
   visible in Fig.~\ref{fig:nnplots}, while on the contrary the
   background $NN$ distribution of wrong-sign decays is clearly different from
   the background $NN$ distribution of right-sign decays.
 We have examined the distributions of all variables used in
the neural network, and found good agreement between the
simulation and the data control samples. Finally, we examined the
$NN$ distributions in the signal MC for the different $B$ signal
channels, right-sign and wrong-sign separately ($D\pi$, $D^*\pi$,
$DK$, $D^*K$) and did not observe any significant difference
between these channels.

   \subsection{Fitting for event yields and
    ${\cal R}^{(*)}$}
   \label{sec:fit}
The ratios ${\cal R}^{(*)}$ are extracted by performing extended
unbinned maximum likelihood fits to the set of variables \mes,
$NN$, and $I_{\mathrm {sign}}$, where $I_{\mathrm {sign}}$ is a
discrete variable equal to 0 for
 WS events and to 1 for RS events. We write the
extended likelihood $\cal L$ as

$${\cal L} = \frac{e^{-N'}}{N!} N'^{N} \prod_{j=1}^N
f(\textbf{x}_j|\theta),$$

\noindent where the vector $\textbf{x}$ indicates the variables
($\mes, NN$, and $I_{\mathrm {sign}}$) and $\theta$ indicates the
set of parameters which are fitted from the data. $N$ is the total
number of signal and background events, and $N'=\sum_i N_i$ is the
expectation value for the total number of events. The sum runs
over the different signal and background categories $i$ which will
be detailed below. The probability density function (PDF)
$f(\textbf{x}_j|\theta)$ is written as the sum over the different
signal and  background categories
$$ f(\textbf{x}_j|\theta,N') = \frac{\sum_i N_i f_i(\textbf{x}_j|\theta)}{N'},$$
where $f_i(\textbf{x}|\theta)$ is the product $F(\mes)\times
G(NN)\times H(I_{\mathrm {sign}})$ of an \mes component $F(\mes)$,
a $NN$ component $G(NN)$ and a two-bin histogram $H(I_{\mathrm
{sign}})$ set to (1,0) for the WS category and (0,1) for the RS
category. The $NN$ distributions are all modeled by histograms
with 102 bins between $-$1.02  and 1.02.

 The fits are performed separately to each of the
$D\pi$, $D^{*}_{D\piz}\pi$, $D^{*}_{D\gamma}\pi$, $DK$,
$D^{*}_{D\piz}K$ and $D^{*}_{D\gamma}K$ samples. They are
configured in such a way that ${\cal R}^{(*)}$ is an explicit fit
parameter: for the $B$ signal, we fit for the number of right-sign
decays $N_{RS}$ and the ratio ${\cal R^{(*)}}$ = $N_{WS}/(c\times
N_{RS})$, where $N_{WS}$ is the number of wrong-sign signal events
and $c$ is the ratio of the wrong-sign to right-sign selection
efficiencies. For $B\to D^{(*)}K$, the factor $c$ is consistent
with unity within the statistical precision of the simulation
(Table~\ref{tab:summaryeff}) and is set to this value in the fits.
For $B\to D^{(*)}\pi$, $c$ differs slightly from unity due to
different particle identification cuts applied at an early stage
of the event selection and we use therefore the values of
Table~\ref{tab:summaryeff} in the fits.

The following signal and background categories are used to
describe each sample in the fits:

\begin{enumerate}
      \item The right-sign signal $B^-\to
      [K^-\pi^+]_{D^{(*)}}K^-/\pi^-$: its \mes spectrum is modeled by a Gaussian function ${\cal G}_{sig}(\mes)$ whose mean and width are
    determined from the fit to data. The $NN$ PDF ${\cal NN}_{sig}$ is constructed from the  $NN$ spectrum of the
    $\Bm\to D h^-$  signal MC.

    \item The wrong-sign signal $B^-\to [K^+\pi^-]_{D^{(*)}} K^-/\pi^-$: its \mes and $NN$ spectra have the same
    parametrizations ${\cal G}_{sig}(\mes)$ and ${\cal NN}_{sig}$ as the right-sign signal.

    \item The right-sign combinatorial background from $q\bar q$ ($q=u,d,s, c$) events into $[K^-\pi^+]K^-$ ($DK$)
    or $[K^-\pi^+]\pi^-$ ($D\pi$): its \mes component is modeled with the
    ARGUS function~\cite{ARGUS} ${\cal A}_{q\bar q}(\mes)$ whose shape
 and endpoint parameters, $\zeta_{q\bar q}$ and $m_0$,  are allowed to
vary in the fit. The $NN$ PDF ${\cal NN}_{q\bar q}^{(RS)}$ is
constructed from the $NN$ spectrum of $[K^-\pi^+]K^-$ ($DK$)
    or $[K^-\pi^+]\pi^-$ ($D\pi$) candidates in the $q\bar q$ continuum MC (Figs.~\ref{fig:nnplots}c and ~\ref{fig:nnplots}f),
    where the $\Delta E$ requirement has been extended
 to $|\Delta E|<200\mev$ and the $D K$ and $D^*K$ (or $D
\pi$ and $D^*\pi$) samples have been summed to increase  the
statistics.

\item The wrong-sign combinatorial background from
    $q\bar q$ events into $[K^+\pi^-]K^-$ ($DK$) or
    $[K^+\pi^-]\pi^-$ ($D\pi$): its \mes component is parameterized  by the same ARGUS function ${\cal A}_{q\bar q}(\mes)$
used for the right-sign component. The $NN$ PDF ${\cal NN}_{q\bar
q}^{(WS)}$ is constructed from the $NN$ spectrum of
$[K^+\pi^-]K^-$ ($DK$) or $[K^+\pi^-]\pi^-$ ($D\pi$) candidates in
the $q\bar q$ continuum MC (Figs.~\ref{fig:nnplots}b and
~\ref{fig:nnplots}e).

    \item The right-sign combinatorial background from \BB  events into $[K^-\pi^+]K^-$ ($DK$) or
    $[K^-\pi^+]\pi^-$ ($D\pi$), excluding the peaking background which is considered in category 7: its \mes component  is described by
    an ARGUS function~\cite{ARGUS} ${\cal A}_{B}^{(RS)}(\mes)$ with shape
parameter $\zeta_B^{(RS)}$ fixed to its value determined from \BB
     MC, after removal of the  $B\to D^{(*)}K/\pi$ signal events.
    The $NN$ PDF used to describe this background is the PDF ${\cal NN}_{sig}$ describing the  $NN$ spectrum of the
    $\Bm \to D^{(*)} h^-$  signal MC. The number of \BB right-sign combinatorial background events is allowed to vary in the $D h^-$ fits
    but is fixed to the MC prediction in the $D^* h^-$ fits (see below).

   \item The wrong-sign combinatorial background from \BB  events into $[K^+\pi^-]K^-$ ($DK$) or
    $[K^+\pi^-]\pi^-$ ($D\pi$), excluding the peaking background which is considered in category 8: its \mes component  is described by
    an ARGUS function~\cite{ARGUS} ${\cal A}_{B}^{(WS)}(\mes)$ with shape parameter
    $\zeta_B^{(WS)}$ fixed to its value determined from the \BB MC, after removal of the  $B\to D^{(*)}K/\pi$ signal events.
    The $NN$ PDF used to describe this background is the PDF ${\cal NN}_{sig}$ describing the  $NN$ spectrum of the
    $\Bm \to D^{(*)} h^-$  signal MC. The number of \BB  wrong-sign combinatorial background events  is allowed to vary  in the $D h^-$ fits
    but is fixed in the $D^* h^-$ fits (see below).

    \item The background from \BB  events in the right-sign component peaking in \mes inside the signal region (peaking
    background): this background is discussed in more detail in
Section~\ref{sec:Peaking}. For the $D K^\pm$, $D \pi^{\pm}$ and
$D^*_{D\piz}K^\pm$
    categories, the peaking part of the \BB background \mes spectrum is described by the same
Gaussian function ${\cal G}_{sig}(\mes)$ as the signal. This
component is therefore indistinguishable from the signal and its
rate has to be fixed to the MC predictions. For the
$D^*_{D\piz}\pi^\pm$, $D^*_{D\gamma}\pi^\pm$ and the
$D^*_{D\gamma}K^\pm$ categories, the \mes component is described
by an asymmetric Gaussian whose shape parameters and amplitude for
each category are determined from a fit to the \mes spectrum of
\BB MC events, after vetoing the $B^\pm\to D^{(*)}h^\pm$ signal
component. For all categories, the $NN$ PDF used to describe this
background is the PDF ${\cal NN}_{sig}$ describing the  $NN$
spectra of the $B\to D^{(*)} h^\pm$  signal MC.

 \item The peaking background from \BB  events in the
 wrong-sign component: the treatment is similar to the previous
 component but ${\cal G}_{sig}(\mes)$ is used to describe the \mes spectrum of the $DK^\pm$,
 $D\pi^{\pm}$, $D^{*}_{D\piz}K^\pm$ and $D^{*}_{D\gamma}K^\pm$ categories,
while an asymmetric Gaussian is used to describe the \mes spectrum
of the $D^{*}_{D\piz}\pi^\pm$ and $D^{*}_{D\gamma}\pi^\pm$
categories.

\end{enumerate}

To summarize, we fit for the number of right-sign signal events
$N_{RS}$, the ratio ${\cal R}$ = $N_{WS}/(c\times N_{RS})$ of
wrong-sign to right-sign events, the number of wrong-sign and
right-sign $q\bar q$ combinatorial background events, $N^{(q\bar
q)}_{WS}$ and $N^{(q\bar q)}_{RS}$, and for $Dh^\pm$ the number of
wrong-sign and right-sign \BB  combinatorial background events,
$N^{(B\bar B)}_{WS}$ and $N^{(B\bar B)}_{RS}$. We fix to their MC
expectations the numbers of wrong-sign and right-sign \BB peaking
background, $N^{(B\bar B,pk)}_{WS}$ and $N^{(B\bar B,pk)}_{RS}$,
as well as the number of \BB combinatorial background events for
$D^{*}h^\pm$. The other parameters fitted are the reconstructed
\mes peak and resolution, $m_B$ and $\sigma_{m_B}$, and the $q\bar
q$ continuum background shape parameter and endpoint,
$\zeta_{q\bar q}$ and $m_0$.

\section{STUDY OF \BB  BACKGROUNDS}
\label{sec:Peaking}

\begin{table}
     \caption{ Charmless background channels and branching fractions, $Dh^\pm $ channels affected by this
     background and background yields expected in our data sample. }
      \label{tab:charmless}
\begin{center}
\begin{tabular}{llcc}
  \hline\hline
       & Affected    &                                 &  Estimated \\
  Mode & channel     & ${\cal B}$($10^{-6}$) &  Yield     \\
  \hline
  %\pim\pip\pim & none        & $16.2\pm 1.5$ &   - \\
  \Km\pip\pim & $D\pi$ RS & $55\pm 7$ \cite{PDG}    &  67.1$\pm$9.7 \\
  \Kp\pim\pim & $D\pi$ WS & $<0.9$ \cite{KpKppimbabar}       &  $<$1.1 \\
  \Km\pip\Km  & $DK$ RS   & $<0.2$ \cite{KpKppimbabar}       &  $<$0.2 \\
  \Kp\pim\Km  & $DK$ WS   & $ 5.0\pm 0.7$ \cite{KpKmpipbabar}&  6.0$\pm$0.8 \\
  \hline\hline
\end{tabular}

\end{center}
\end{table}

\begin{table*}[htb]
   \caption{Expected numbers of signal and \BB  background events, peaking background parametrization
    and  dominant sources of peaking backgrounds for $B\to D^{(*)}\pi$ and $B\to D^{(*)}K$.
   $N_{\BB}^{(comb)}$ is the combinatorial part of the background, parametrized by an ARGUS function,
    and $N_{\BB}^{(peak)}$ is the component peaking in \mes, parametrized by either a Gaussian function or a
    bifurcated Gaussian function. The average event yield expected for the WS signal is computed assuming $r^{(*)}_B=10\%$ and no
    interference term ($\cos \gamma \times \cos \delta = 0$).}
   \begin{center}
   \begin{tabular}{lrcccll} \hline \hline
Mode & Signal yield  &  $N_{\BB}^{(comb)}$ && $N_{\BB}^{(peak)}$ & Peaking bkgd. parametrization &  Peaking bkgd. sources \\
\hline
 $D\pim$ WS                   & 86    &   93.7$\pm$6.0 && $10.6\pm 3.0$ & Gaussian & $D_0^{*-} e^+ \nu_e$ \\
 $D^*_{D\piz}\pim$ WS      & 31    &   24.7$\pm$8.3 && $29.0\pm 8.7$ & Bifurcated Gaussian & $D_0^{*-} e^+ \nu_e$, $D_1^{'-} e^+ \nu_e$ \\
 $D^*_{D\gamma}\pim$ WS    & 25    &    111$\pm$9 && $47\pm 7$       & Bifurcated Gaussian & $D_0^{*-} e^+ \nu_e$, $D_1^{'-} e^+ \nu_e$,
 and $D^{(*)0}\rho^0$ \\ \hline
 $D\pim$ RS                   & 24240 & 307.3$\pm$11.7 && 222.0$\pm$10.3   & Gaussian & $\Km\pip\pim$, $(c\bar c)\Km$ \\
 $D^*_{D\piz}\pim$ RS      & 8931  & 620.7$\pm$33.7 && $507.3\pm 33.3$  & Bifurcated Gaussian & $D^{*}\rho^-$, $D^{*+}\pim$ \\
 $D^*_{D\gamma}\pim$ RS    & 7242  & 1225$\pm$64 && $2432\pm 67$ & Bifurcated Gaussian & $D^{*}\rho^-$, $D^{*+}\pim$,
 and $D^{*}_{D\piz}\pim$ \\
\hline
 $D\Km$ WS                 & 26.3 &107.0$\pm$6.3  && $12.6\pm 3.1$   & Gaussian & $D h^-$, $\Km \Kp \pim$ \\
 $D^*_{D\piz}\Km$ WS    &  8.5 & 17.3$\pm$2.7  && $ 2.7\pm 1.6$   & Gaussian & ---\\
 $D^*_{D\gamma}\Km$ WS  &  6.8 & 68.3$\pm$5.3  && $ 6.0\pm 2.4$   & Gaussian & ---\\ \hline
 $D\Km$ RS                 & 1944 & 50.7$\pm$5.3  && 299.3$\pm$10.7  & Gaussian & $D \pi^-$ \\
 $D^*_{D\piz}\Km$ RS    &  618 & 56.0$\pm$6.7  && $127.0\pm 8.3$  & Gaussian & $D^{*}_{D\piz} \pi^-$ \\
 $D^*_{D\gamma}\Km$ RS  &  503 & 66.0$\pm$14.7 && $326.7\pm 17.3$ & Bifurcated Gaussian & $D^{*}_{D\gamma} \pi^-$, $D^{*}_{D\piz}K^-$ \\
\hline \hline
   \end{tabular}
   \end{center}
   \label{tab:peaking}
   \end{table*}

We study the \BB  background for each signal category ($D\pi$,
$D^*\pi$ $DK$, $D^*K$) and charge combination (right-sign and
wrong-sign) using a sample of $e^+e^-\to \FourS \to B \bar B$ MC
events corresponding to about 3 times the data luminosity. In
addition, dedicated Monte Carlo signal samples are used to
estimate the background from $B^- \to D h^-$ events and the
background from the charmless decay $B^- \to K^+ \pi^- K^-$.
 We identify three main classes of background events which can peak in \mes inside the signal region
and mimic the $D^{(*)}\pi$ and $D^{(*)}K$ signal:

   \begin{enumerate}
   \item Charmless $B$ decays $B^- \to h^+ h^- h^-$ ($h=\pi, K$):
    we list in Table~\ref{tab:charmless} the 3-body charmless decays
   affecting our analysis, their branching fractions~\cite{PDG} and the
   numbers of reconstructed events expected in the affected modes after the selection.
   Due to the particle identification criteria used in the analysis only
   decays with the same final state particles as our signal modes contribute significantly to the background. These events are
   indistinguishable from the $D h^\pm$ signal if the $K^-\pi^+$
   invariant mass is consistent with the $D$ mass. The two
   decays affected by a significant charmless background are right-sign $B^-
   \to [K^-\pi^+]_D \pi^-$ and wrong-sign $B^-\to [K^+\pi^-]_D K^-$.
    Using $B^- \to K^- \pi^+ \pi^-$ events selected in the \BB  Monte
   Carlo sample, we estimate the efficiency of $B^- \to K^- \pi^+ \pi^-$ events to be reconstructed as
   a $[K^-\pi^+]_D \pi^-$ candidate as $(0.26\pm 0.02)\%$. The
   corresponding background is estimated to be 67.1$\pm$9.7
   events, where the error is dominated by the statistical uncertainty on the $B^- \to
K^- \pi^+ \pi^-$ branching fraction.
 The efficiency  of $B^- \to K^+ \pi^- K^-$ events to be reconstructed as
    $[K^+\pi^-]_D K^-$ WS candidates is determined from a high statistics dedicated $B^- \to K^+ \pi^- K^-$ signal Monte Carlo
    sample, and is found to be $(0.27\pm 0.01)\%$. The corresponding peaking
    background from $B^- \to K^+ \pi^- K^-$ events mimicking $B^- \to
[K^+\pi-]_D K^-$ WS decays is estimated to be 6.0$\pm$0.8 events,
where the error is dominated by the statistical uncertainty on the
$B^- \to K^+ \pi^- K^-$ branching fraction. From a fit to data
selected in the $D$ mass sidebands, we cross-check this prediction
and find $6.5\pm 4.0$ peaking events, in good agreement with the
MC prediction. We also check that, because of the tight $\Delta m$
cut applied to the $D^*$ decay products, the $B^- \to D^*h^-$
channels are not affected by charmless peaking backgrounds.

   \item Events of the type $B^- \to D h^-$: this background is estimated by
running the analysis on a sample of $B^- \to D h^-$ signal MC
events properly renormalised to the data sample, and fitting the
\mes spectra of the selected events to the sum of a Gaussian
signal and a combinatorial background. We find that a peaking
background of $2.6\pm 0.4$ events is predicted in the $B^- \to
[K^+\pi-]_D K^-$ WS channel. This component is dominated (2 events
out of 2.6) by decays $B^- \to [K^-K^+]_D \pi^-$ failing the $D$
mass veto and by WS decays $B^- \to [K^+\pi^-]_D \pi^-$ where the
\pim is misidentified as a \Km. For the $D^*K$ channels, the $B^-
\to [K^-K^+]_D \pi^-$ contribution is suppressed by the $\Delta m$
cut on the $D^*$-$D$ mass difference, and the WS $D^*\pi$
contribution is 0.5$\pm$0.1 events for $D^*\to D\piz$ and
0.6$\pm$0.2 events for $D^*\to D\gamma$. Another background of the
same type occurs in the right-sign $DK$ decays. It consists of
events $B^- \to [K^-\pi^+]_{D^{(*)}} \pi^-$ where the bachelor
$\pi^-$ is misidentified as a $K^-$, which fake the RS signal $B^-
\to [K^-\pi^+]_{D^{(*)}} K^-$. This contribution is predicted by
the simulation and has been verified in the data by fitting the
$\Delta E$ spectrum of $D^{(*)}K$ candidates in the \mes signal
region, which shows a second peak due to $D^{(*)}\pi$ candidates,
 shifted by ~50\mev with respect to the signal.
 \item Other decays: this component is estimated by fitting
the \mes spectra of \BB MC events, after removing the charmless
and $\Bm \to D h^-$ components. For $\Bm \to [\Kp\pim]_D \Km$ WS
decays, the peaking component is estimated to be $4\pm 3$ events,
where the uncertainty is dominated by the statistical error on the
simulated data. The main sources of peaking background which could
be identified are listed in Table~\ref{tab:peaking}. They include
$\Bzb \to D^{*+}h^-$ reconstructed as $\Bm \to D^{*0}h^-$,
semi-leptonic decays $\Bz\to D^{**-} e^+ \nueb$ ($D^{**-}\to \bar
D^{(*)0} \pi^-$, $\bar D^0\to K^+\pi^-$) where the $e^+$ is
missed, faking the WS signal $B^- \to [K^+\pi^-]_{D^{(*)}}\pi^-$,
and decays $\Bm\to D^{(*)}\rho^-$ faking the RS signal $B^- \to
[K^-\pi^+]_{D^{(*)}}\pi^-$.
   \end{enumerate}

A summary of the \BB  background studies is given in
Table~\ref{tab:peaking}, for $B\to D^{(*)}\pi$ and $B\to
D^{(*)}K$. For each channel, the \mes spectra of events selected
in the \BB  MC simulation (after removing the corresponding
signal) were fitted by the sum of a combinatorial background
component and a peaking component, using the same parametrization
described in Sec.~\ref{sec:fit}. The average number of \BB
combinatorial and peaking background events predicted by the
simulation are given in Table~\ref{tab:peaking}, together with the
main sources of peaking events and the functional shapes chosen to
describe the peaking background. The numbers of signal events
expected are also given for comparison. For the $B\to D^*K$ WS
channels, we could not identify a specific source of peaking
background due to the lack of statistics in the simulation. For
all channels, we use the values of the peaking components
summarized in Table~\ref{tab:peaking} in the maximum likelihood
fit. Statistical uncertainties in the expected yields are
incorporated in the corresponding systematic uncertainties.

\section{RESULTS}
   \label{sec:results}

   \subsection{Results for $B\to D^{(*)}\pi$}
   \label{sec:resultsdpi}

\begin{figure*}[htb]
\begin{center}
\epsfig{file=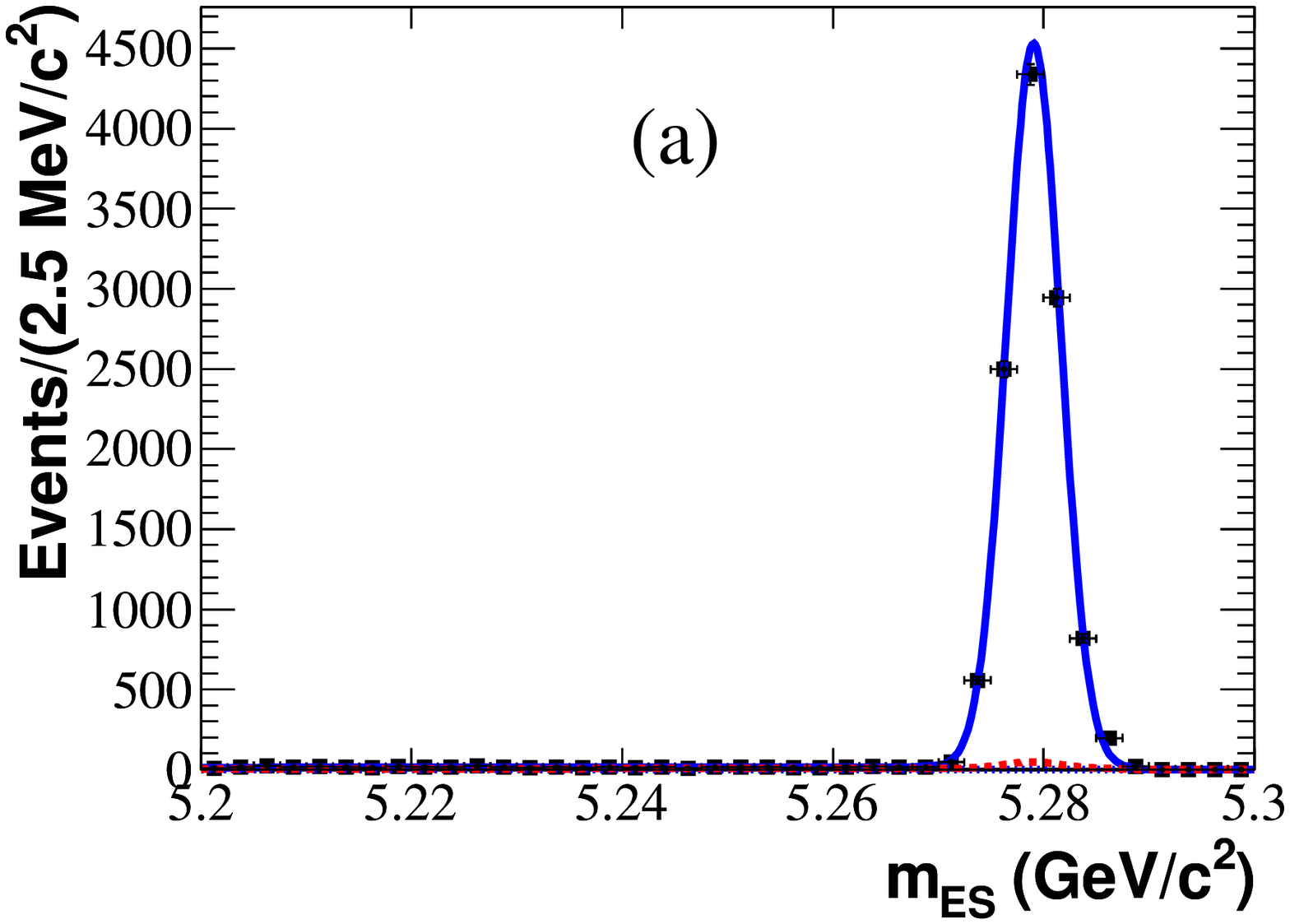,width=0.319\linewidth}
\epsfig{file=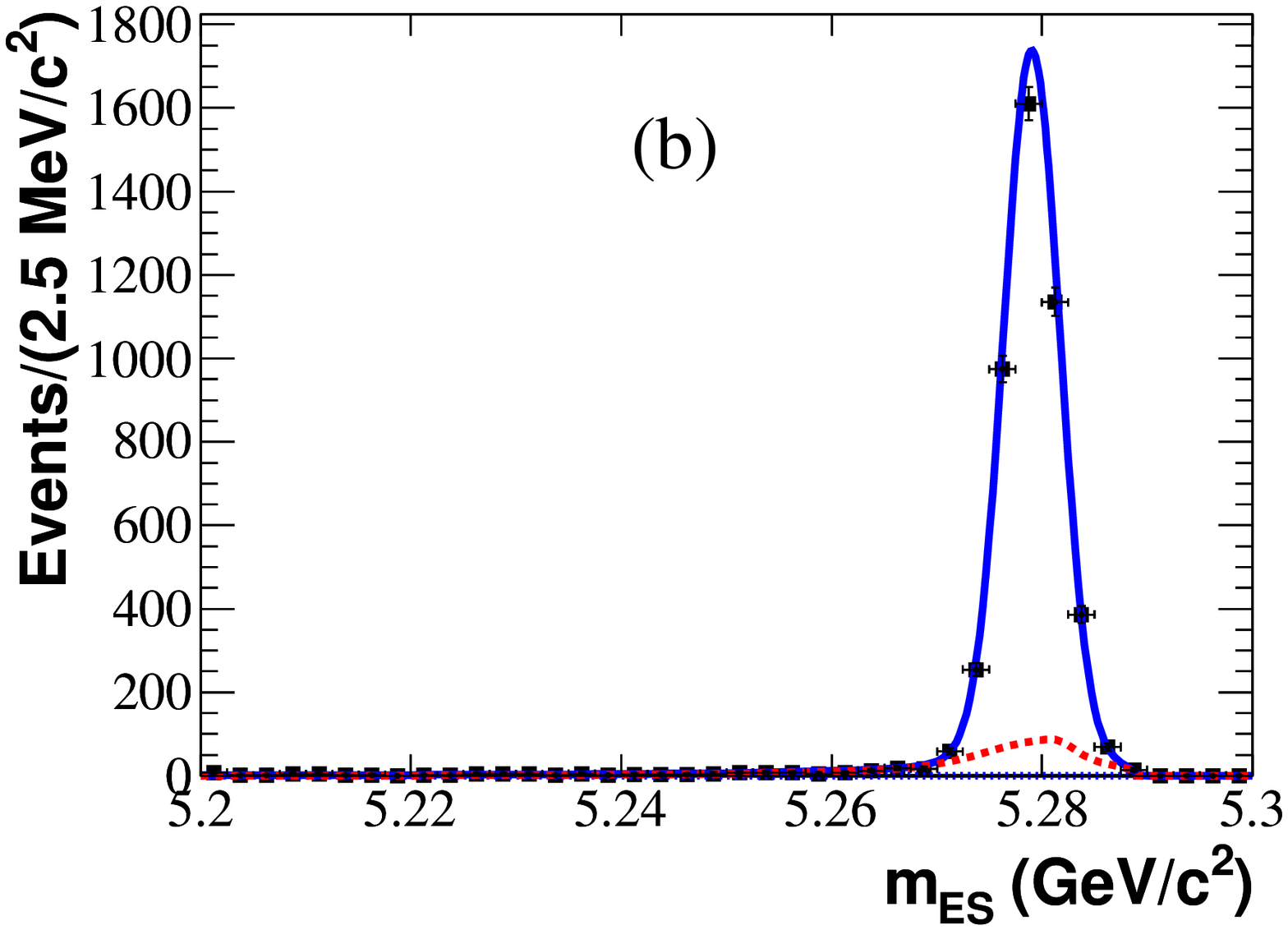,width=0.319\linewidth}
\epsfig{file=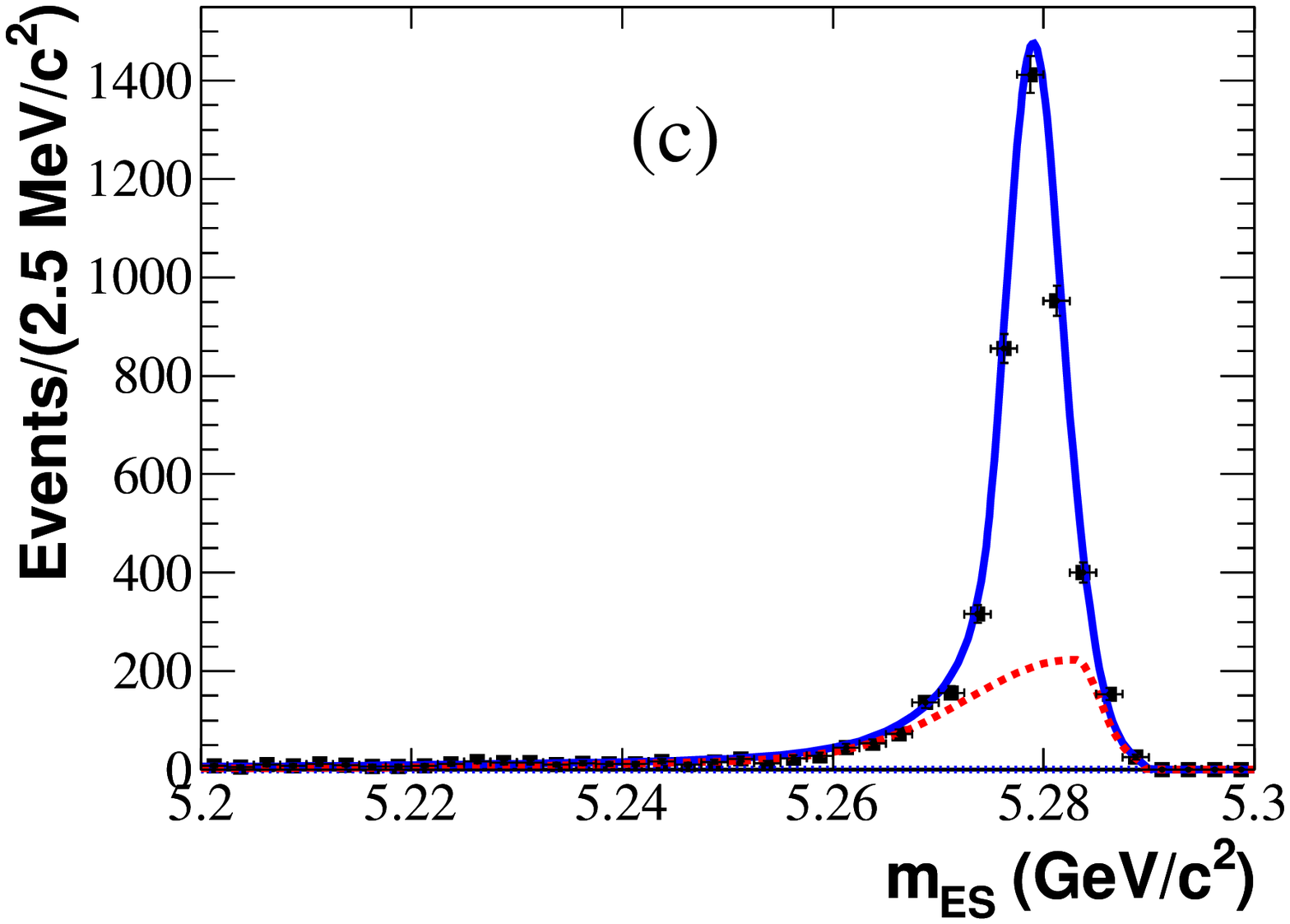,width=0.319\linewidth}
\epsfig{file=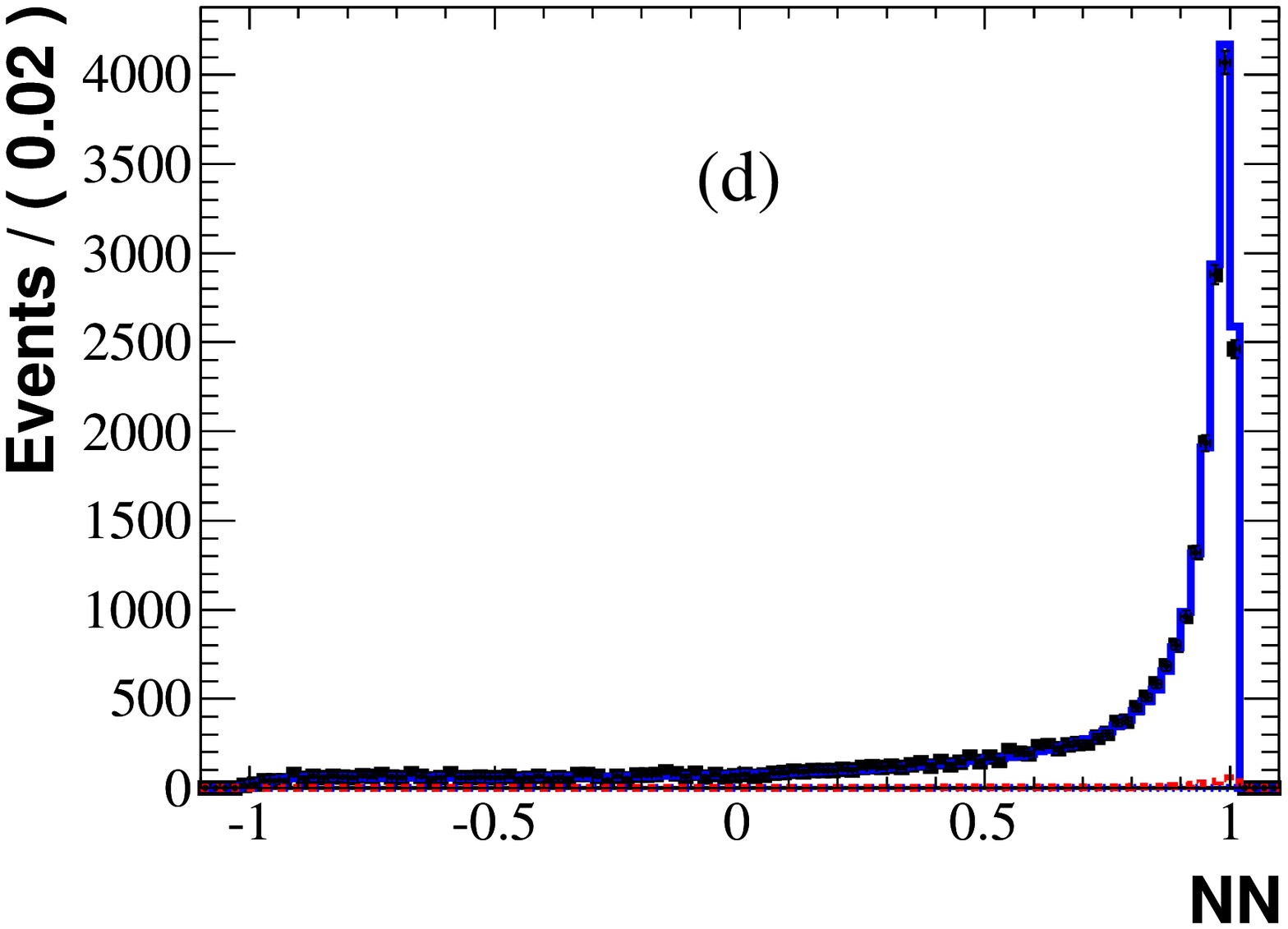,width=0.319\linewidth}
\epsfig{file=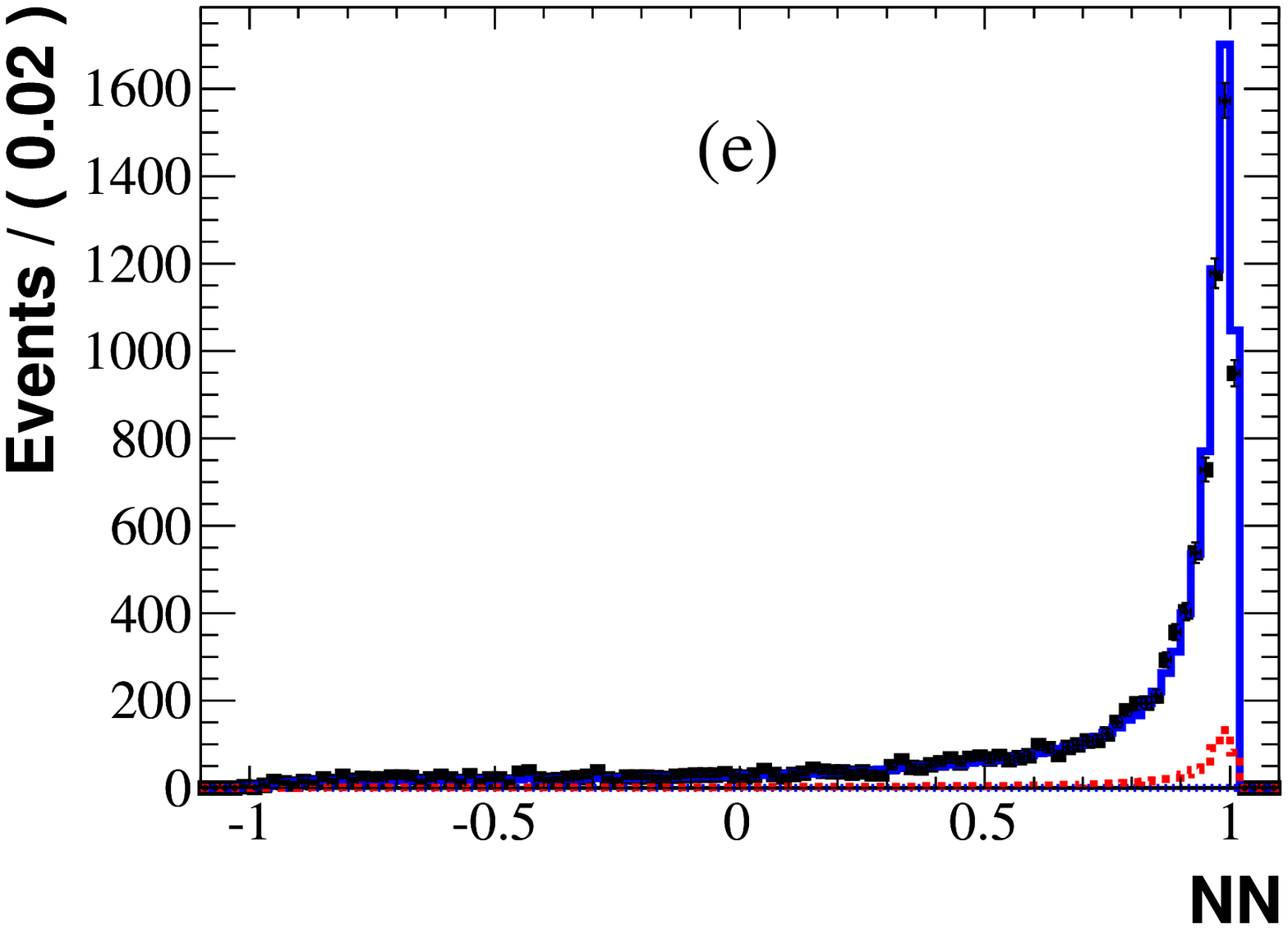,width=0.319\linewidth}
\epsfig{file=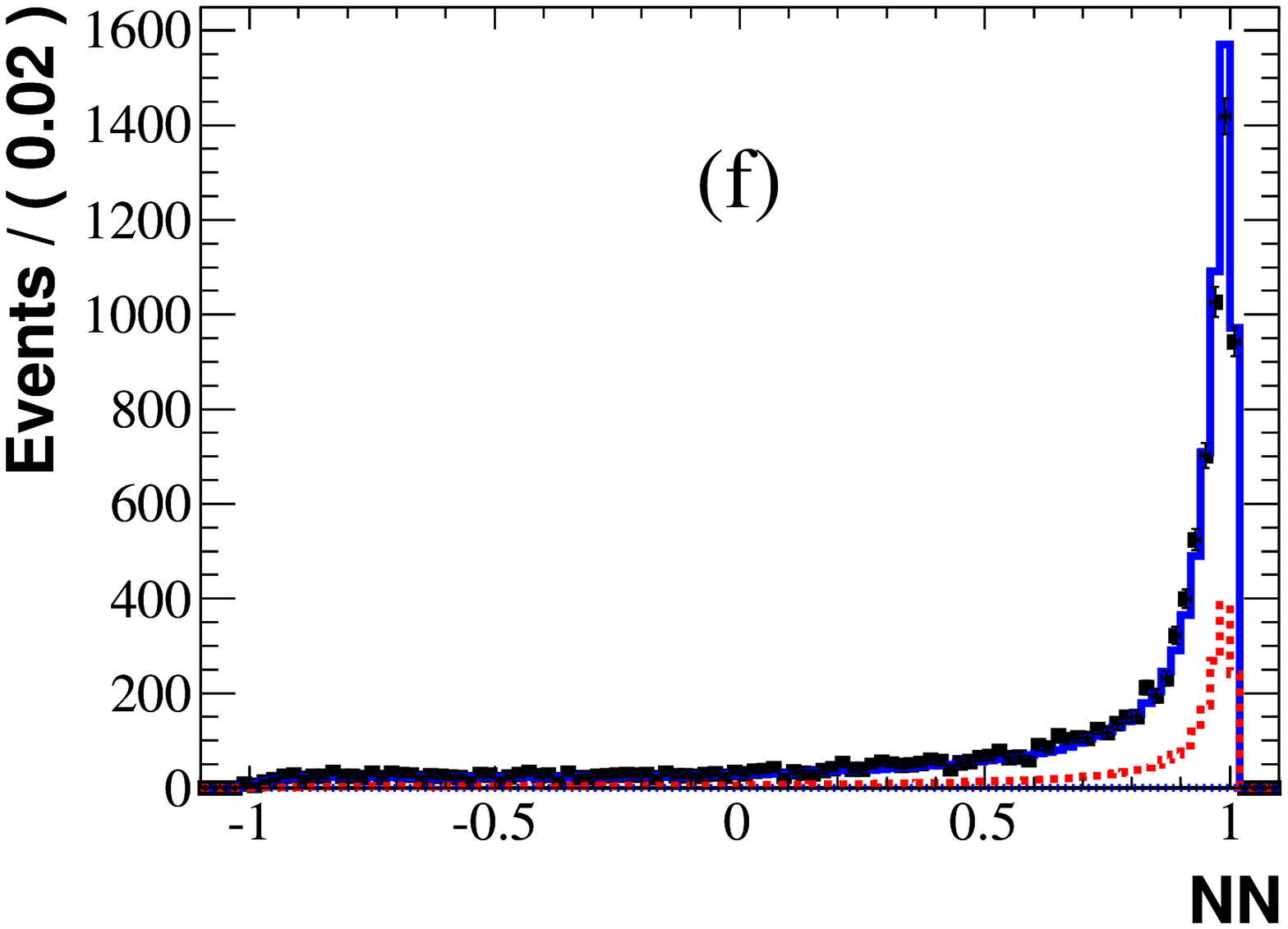,width=0.319\linewidth} \caption{(color
online). Projections on \mes (top) and $NN$ (bottom) of the fit
results for $D\pi$ (a,d), $D^*_{D \piz}\pi$ (b,e) and $D^*_{D
\gamma}\pi$ (c,f) RS decays, for samples enriched in signal with
the requirements $NN>0.94$ (\mes projections) or
$5.2725<\mes<5.2875\gevcc$ ($NN$ projections). The points with
error bars are data. The curves represent the fit projections for
signal plus background (solid) and background (dashed).}
\label{fig:dpiresultscab}
\end{center}
\end{figure*}

\begin{figure*}[htb]
\begin{center}
\epsfig{file=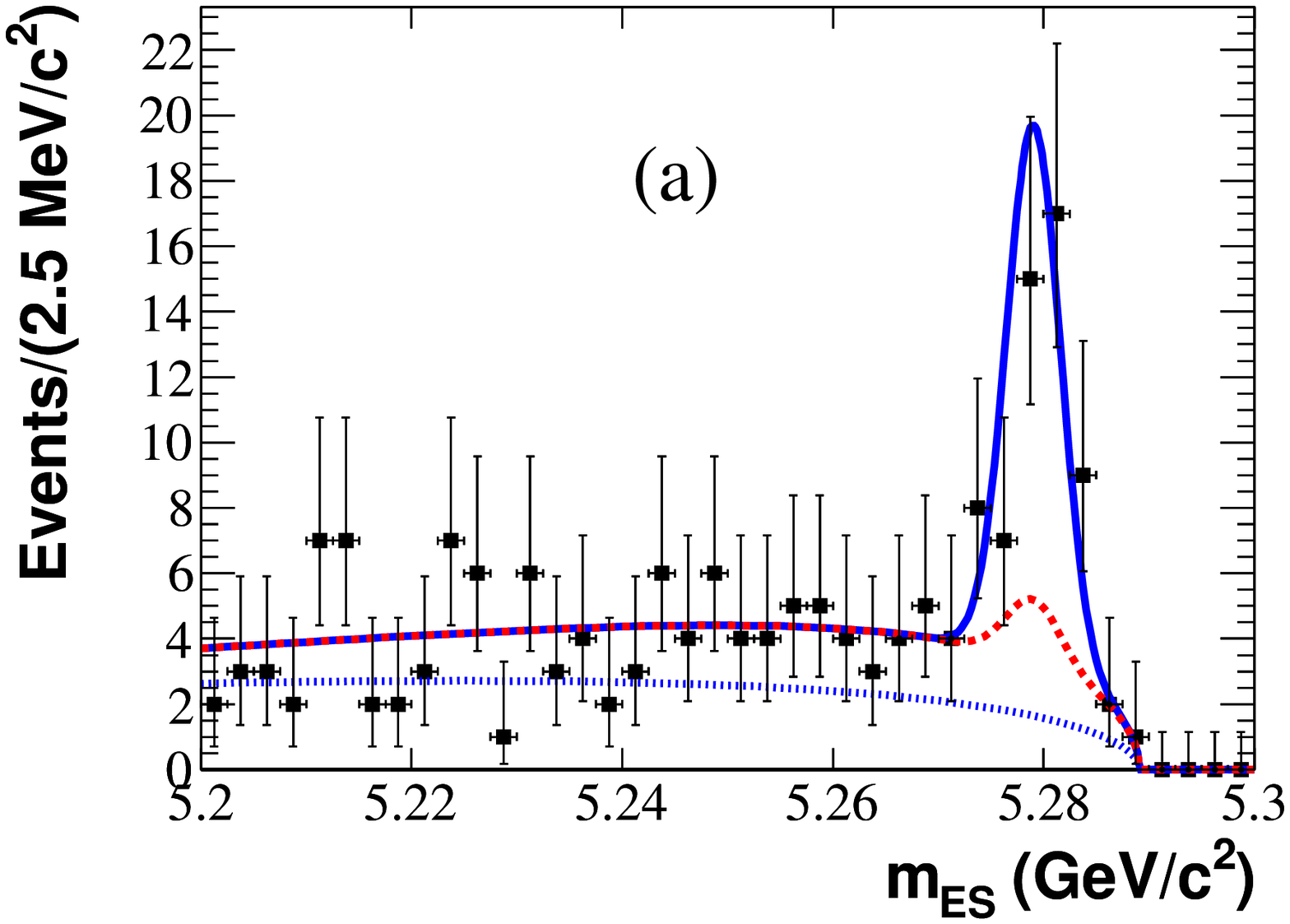,width=0.319\linewidth}
\epsfig{file=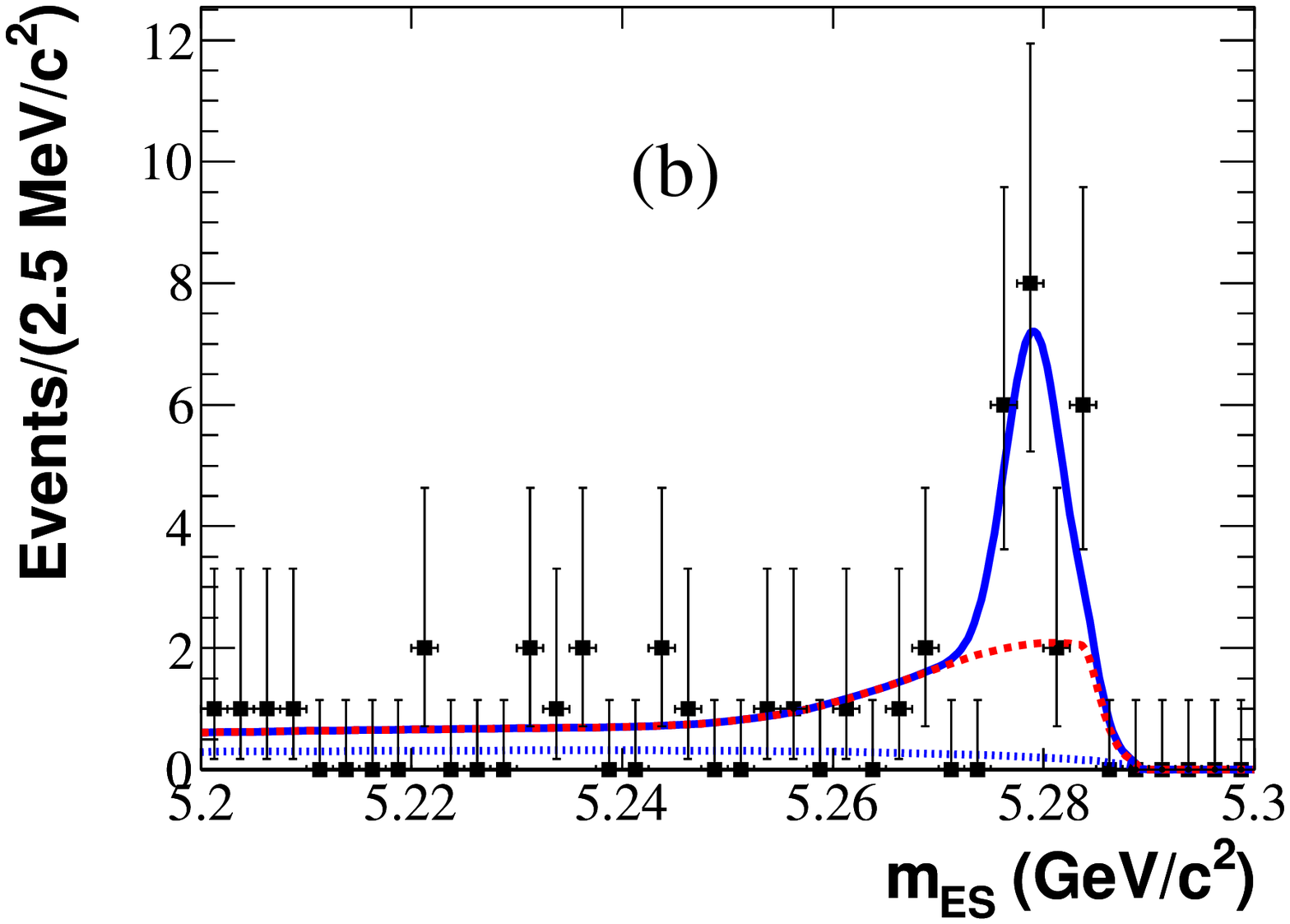,width=0.319\linewidth}
\epsfig{file=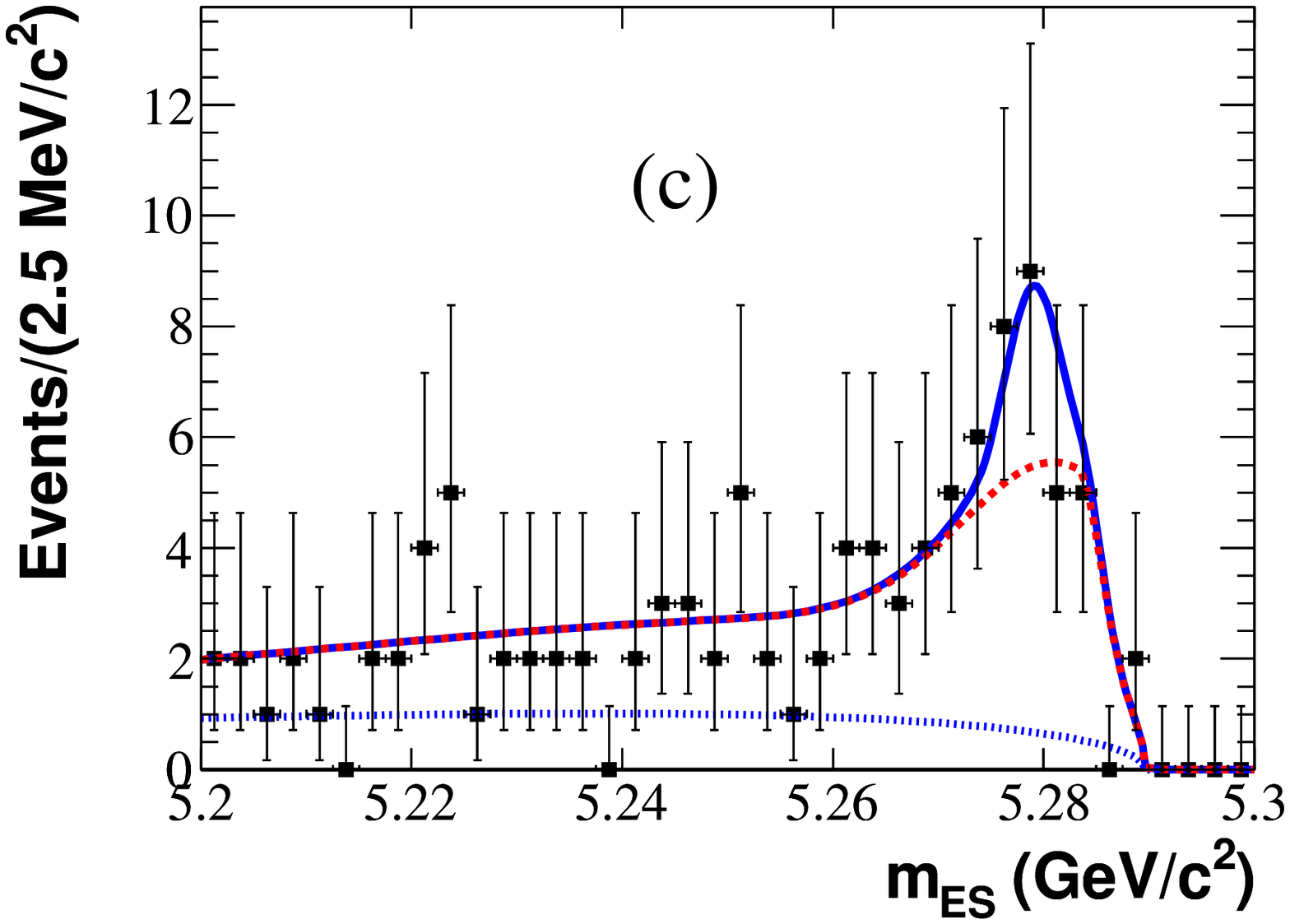,width=0.319\linewidth}
\epsfig{file=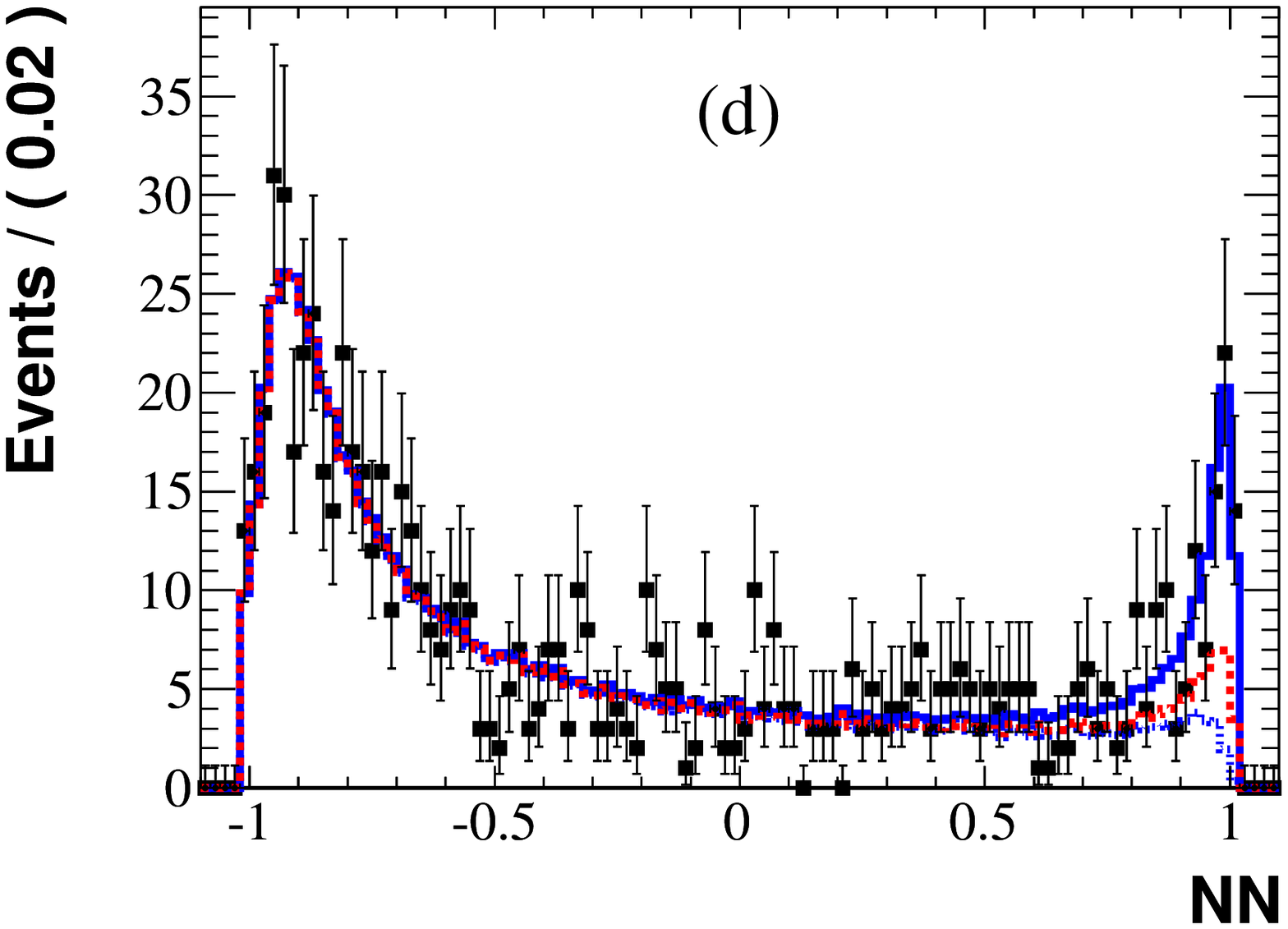,width=0.319\linewidth}
\epsfig{file=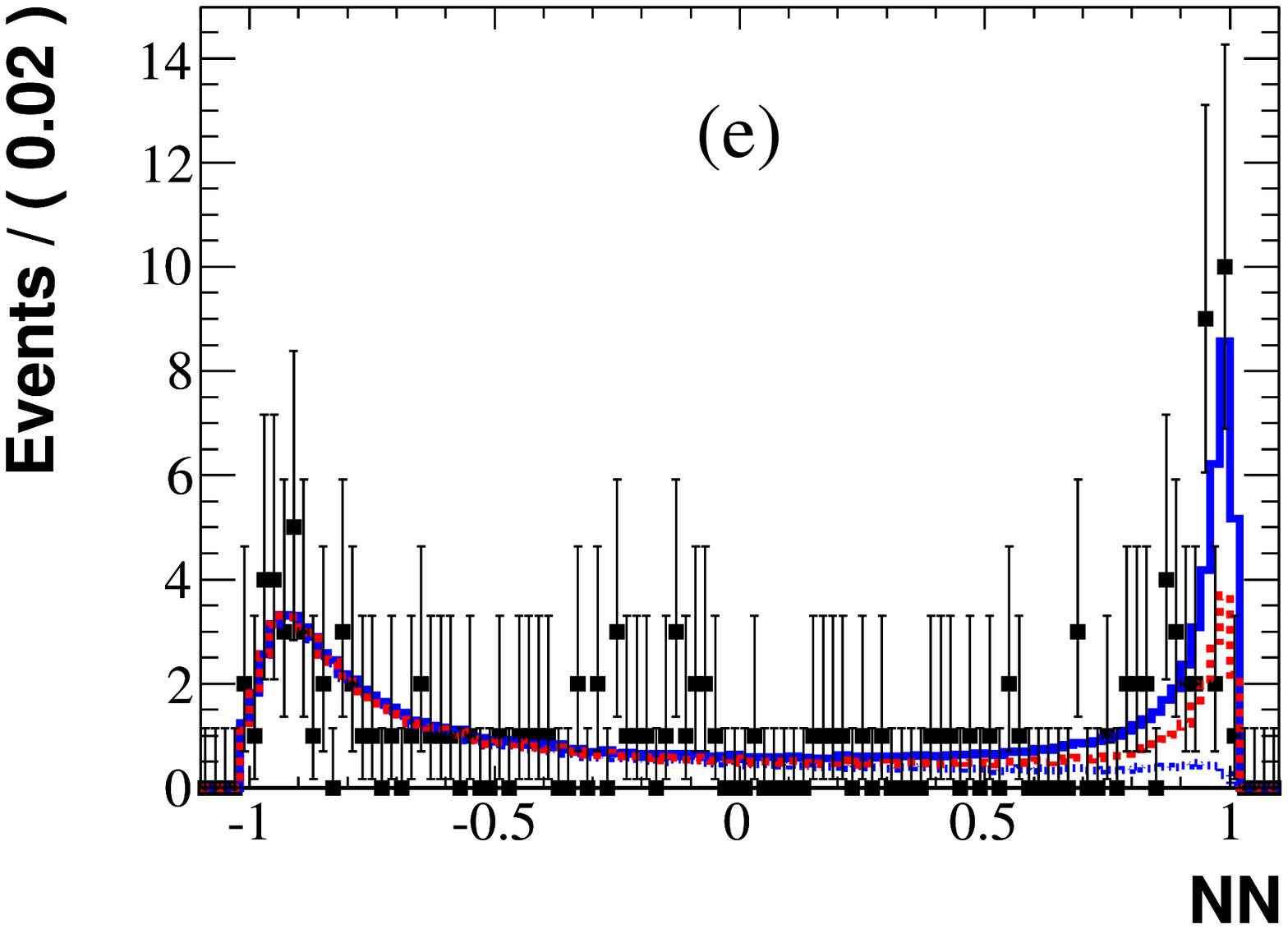,width=0.319\linewidth}
\epsfig{file=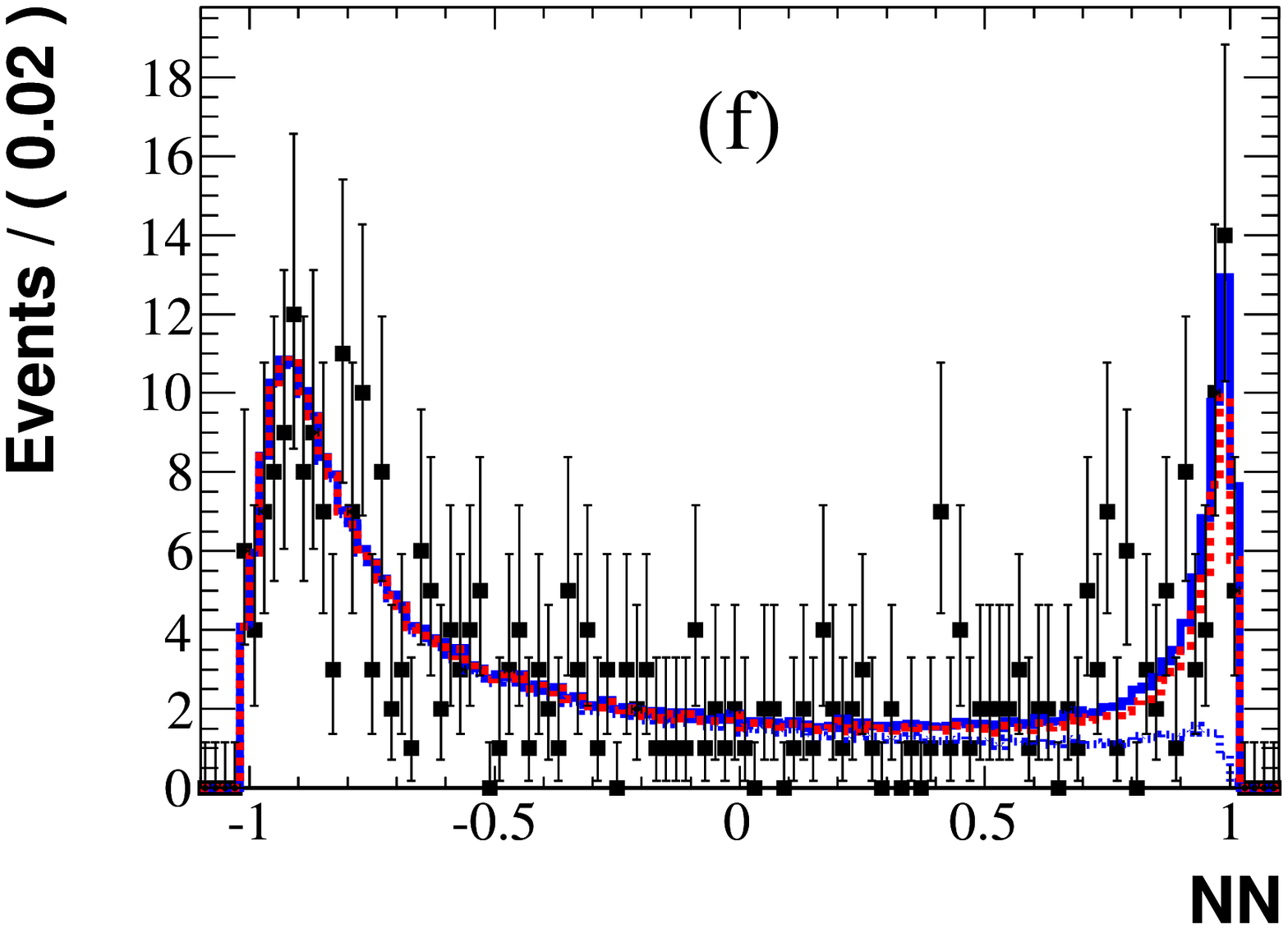,width=0.319\linewidth} \caption{(color
online). Projections on \mes (top) and $NN$ (bottom) of the fit
results for $D\pi$ (a,d), $D^{*}_{D \piz}\pi$ (b,e) and $D^{*}_{D
\gamma}\pi$ (c,f) WS decays, for samples enriched in signal with
the requirements $NN>0.94$ (\mes projections) or
$5.2725<\mes<5.2875\gevcc$ ($NN$ projections). The curves
represent the fit projections for signal plus background (solid),
the sum of all background components(dashed), and $q\bar q$
background only (dotted).} \label{fig:dpiresultsads}
\end{center}
\end{figure*}

\begin{table*}[htb]
   \caption{Summary of fit results for $D^{(*)}\pi$.}
   \begin{center}
   \begin{tabular}{lrclrclrcl} \hline\hline
    Mode & \multicolumn{3}{c}{$D\pi$}  &  \multicolumn{3}{c}{$D^*_{D\piz}\pi$}  & \multicolumn{3}{c}{$D^*_{D\gamma}\pi$} \\ \hline
   Ratio of rates, \RDDstarpi  $(10^{-3})$&
   $ 3.3$ & $\pm$ & $ 0.6$ &
   $ 3.2$ & $\pm$ & $ 0.9$ &
   $ 2.7$ & $\pm$ & $ 1.4$ \\
   Number of signal events $N_{WS}$ & $79.8$ & $\pm$ & $ 13.8$ & $28.3$ & $\pm$ & $ 7.7$ &  $18.7$ & $\pm$ & $ 9.7$ \\
   Number of normalization events $N_{RS}$& $  24662$ & $\pm$ & $ 160$ & $9296 $ & $\pm$ & $ 102$ & $7214 $ & $\pm$ & $ 105$ \\
   $B^+$ ratio of rates, \RDDstarpip $(10^{-3})$&
   $ 3.2$ & $\pm$ & $ 0.8$ &
   $ 3.5$ & $\pm$ & $ 1.2$ &
   $ 4.6$ & $\pm$ & $ 2.2$ \\
   $B^-$ ratio of rates, \RDDstarpim $(10^{-3})$&
   $ 3.4$ & $\pm$ & $ 0.8$ &
   $ 2.9$ & $\pm$ & $ 1.2$ &
   $ 1.0$ & $\pm$ & $ 1.8$ \\
   Asymmetry \ADDstarpi & $0.03$ & $\pm$ & $ 0.17$ & $-0.09$ & $\pm$ & $ 0.27$ & $-0.65$ & $\pm$ & $ 0.55$\\
   \hline\hline
   \end{tabular}
   \end{center}
   \label{tab:fit-results-dpi}
   \end{table*}

The results for $B\to D^{(*)}\pi$ are displayed in
Fig.~\ref{fig:dpiresultscab} (right-sign modes) and
Fig.~\ref{fig:dpiresultsads} (wrong-sign modes). They are
summarized in Table~\ref{tab:fit-results-dpi}. Clear signals are
observed in the $B\to D\pi$ and in the $B\to D^{*}_{D\piz}\pi$ WS
modes, with statistical significances of $7\sigma$ and
$4.8\sigma$, respectively. The significance is defined as
$\sqrt{-2\ln({\cal L}_0/{\cal L}_{\mathrm{max}})}$, where ${\cal
L}_{\mathrm{max}}$ and ${\cal L}_0$ are the likelihood values with
the nominal and with zero WS signal yield, respectively. For $B\to
D^{*}_{D\gamma}\pi$ WS decays, the significance is only $2\sigma$,
due to the large peaking background. Below we discuss the sources
of systematic uncertainties that contribute to our \RDDstarpi
measurements:

\begin{enumerate}
    \item Signal $NN$ shape: in the nominal fit, we use the $NN$ PDF from the $B$ signal MC. To estimate the related systematics, we
     refit the data using a signal $NN$ PDF extracted from the high purity and high statistics $B\to D\pi$
     RS data, after subtracting the residual continuum background contamination predicted by the
     simulation. We set the systematic uncertainty to the difference with the nominal fit
     result.
    \item \B background $NN$ shape: from a study of generic
    \BB  MC, it appears that the $NN$ spectra of $B$
    background events in the \mes-\DeltaE signal box
    are similar to the signal (but suffer from very low statistics), while
    the $NN$ spectra of background events in an enlarged \mes-\DeltaE region
    differ significantly from the signal and show less peaking
    close to 1. In the nominal fit we assumed that both
    the peaking and the non-peaking \BB  background components could be described by the $B\to D\pi$
    signal $NN$ PDF.
    To estimate the related systematic error, we used \BB  generic background events selected in a
    $\Delta E$-\mes enlarged window $|\Delta E|<200\mev$ and
    $\mes>5.20\gevcc$ to build the $NN$ PDF of the non-peaking part of the \BB
    background (keeping the signal $NN$ PDF to describe the peaking
    part of this background) and repeated the fits, taking the difference of the results as
    the associated systematic uncertainty.
    \item Continuum background $NN$ shape: to account for possible differences between the simulation
     and the data, we  used the $NN$ spectrum from off-peak data instead of $q\bar
     q$ MC ($q=u,d,s,c$) to model this component. We set the associated systematic
     uncertainty to the difference of the two results, but the
     error is dominated by the large statistical uncertainty on the
     off-peak data sample.
    \item The shape parameters $\zeta_B^{(WS)}$ and $\zeta_B^{(RS)}$ of the ARGUS functions describing the suppressed and favored
    \BB combinatorial background: in the nominal fits, these parameters are fixed to their values as determined from \BB simulated events.
    To account for possible disagreement between data and simulation, we repeated the fits varying these
    parameters in a conservative range.%, $\zeta=-20$ and $\zeta=-80$.
    \item Peaking component in the \B background: we varied the yield of the peaking
   component by $\pm 1\sigma$, where $\sigma$ is either the statistical error from
   a fit to generic \BB  MC or the uncertainty on
   the branching fraction for known sources of peaking background.
   \item Uncertainty on the number of \BB  combinatorial
   background events: in the $D^*\pi$ (and $D^*K$) fits where this component
   has been fixed, we vary it by $\pm 25\%$ (the level of agreement between data and simulation observed in the $D\pi$
   and $DK$ fits) and we take the difference with the nominal fit result as a systematic uncertainty.
   \end{enumerate}
   The resulting systematic uncertainties are listed in
   Table~\ref{tab:systematics-dpi}. We add them in quadrature and
   quote the results:

    \begin{eqnarray}
     \RDpi &=& (3.3 \pm 0.6 \pm 0.4)\times 10^{-3},\nonumber \\
\RDstarpipiz &=& (3.2 \pm 0.9 \pm 0.8)\times 10^{-3},\nonumber \\
\RDstarpigam &=& (2.7 \pm 1.4 \pm 2.2)\times 10^{-3},\nonumber
    \end{eqnarray}

where the first uncertainty is statistical and the second is
systematic. The values of \RDDstarpi are in good agreement with
the world average $R_D=r_D^2={\cal B}(\Dz\to \Kp \pim)/{\cal
B}(\Dz \to \Km \pip)$, $R_D=(3.36\pm 0.08)\times
10^{-3}$~\cite{HFAG}.

A separate fit to \Bp and \Bm candidates provides a
   measurement of the corresponding asymmetries. We obtain the
   following results:

   \begin{eqnarray}
 \ADpi        &=& 0.03 \pm 0.17 \pm 0.04,\nonumber \\
\ADstarpipiz &=&-0.09 \pm 0.27\pm 0.05, \nonumber \\
 \ADstarpigam &=&-0.65 \pm 0.55\pm 0.22,\nonumber
    \end{eqnarray}

\noindent where the uncertainties are dominated by the statistical
error. No significant asymmetry is observed for the $D^{(*)}\pi$
WS decays. The largest source of systematic uncertainty on the
$D^{(*)}\pi$ asymmetries is from the uncertainty on the \B
background peaking component.

   \begin{table}[htb]
   \caption{Summary of systematic uncertainties on ${\cal R}$ for $D^{(*)}\pi$, in units of $10^{-3}$.}
   \begin{center}
\begin{tabular}{lccc}
     \hline\hline
     Source & $\Delta {\cal R}(10^{-3})$ & $\Delta {\cal R}(10^{-3})$ & $\Delta {\cal R}(10^{-3})$ \\
      & $D \pi$ & $D^{*}_{D\piz}\pi$ & $D^{*}_{D\gamma}\pi$ \\
    \hline
     Signal $NN$                  & $\pm 0.1$  & $\pm 0.1$ & $\pm 0.1$ \\
     \BB  background $NN$         & $\pm 0.1$  & $\pm 0.1$ & $\pm 0.9$ \\
     $udsc$ background $NN$       & $\pm 0.1$  & $\pm 0.1$ & $\pm 0.3$ \\
     \BB  comb. bkg shape (\mes)  & $\pm 0.2$  & $\pm 0.1$ & $\pm 0.2$ \\
      Peaking background WS       & $\pm 0.2$  & $\pm 0.8$ & $\pm 2.0$ \\
     Peaking background RS        & $\pm 0.0$  & $\pm 0.1$ & $\pm 0.1$ \\
     \BB  comb. bkg               & -          & $\pm 0.0$ & $\pm 0.4$ \\ \hline
     Combined                     & $\pm 0.4$  & $\pm 0.8$ & $\pm 2.2$ \\
     \hline\hline
   \end{tabular}
\end{center}
   \label{tab:systematics-dpi}
   \end{table}

\subsection{Results for $B\to D^{(*)}K$}
   \label{sec:resultsdk}

\begin{figure*}[htb]
\begin{center}
\epsfig{file=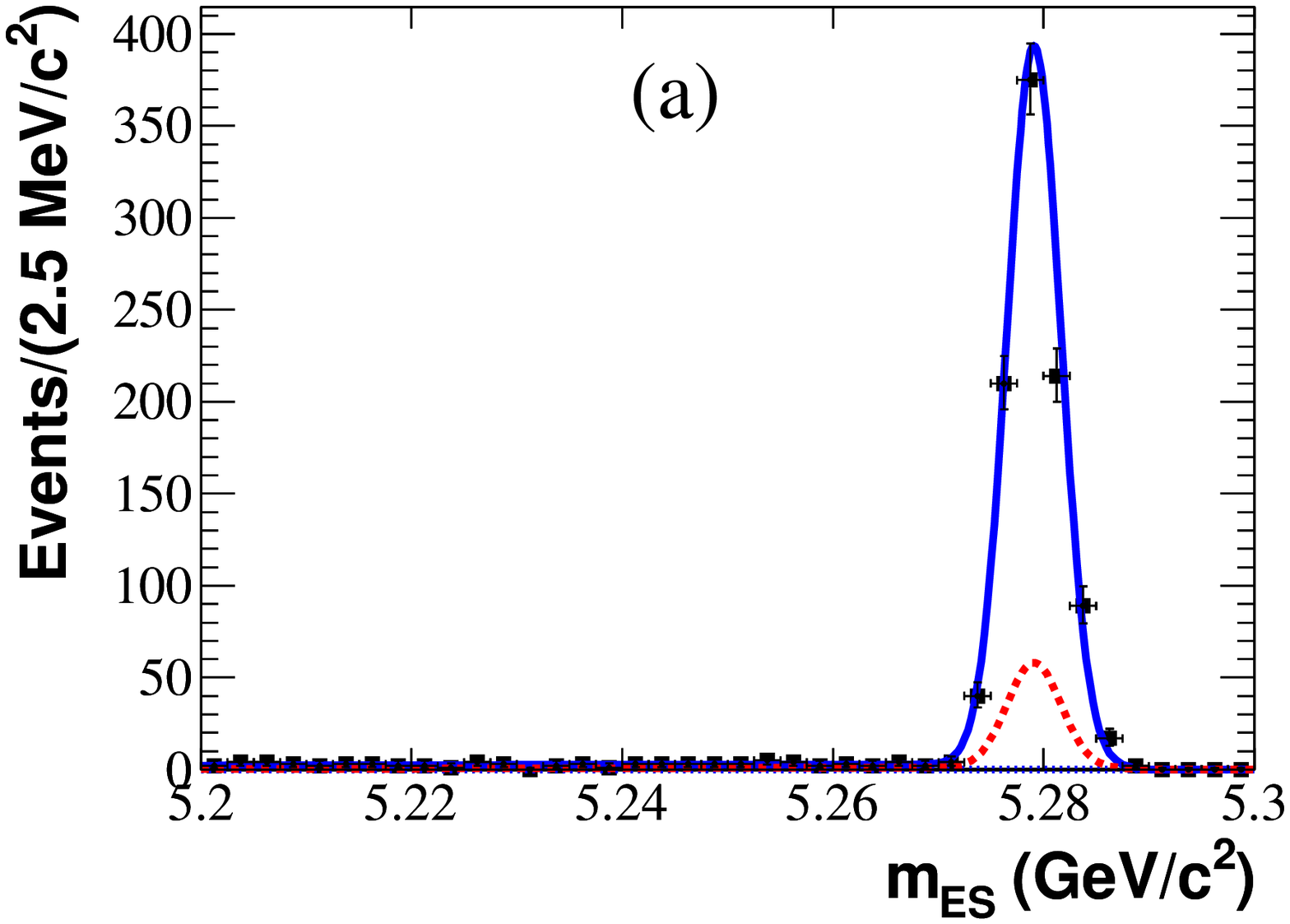,width=0.319\linewidth}
\epsfig{file=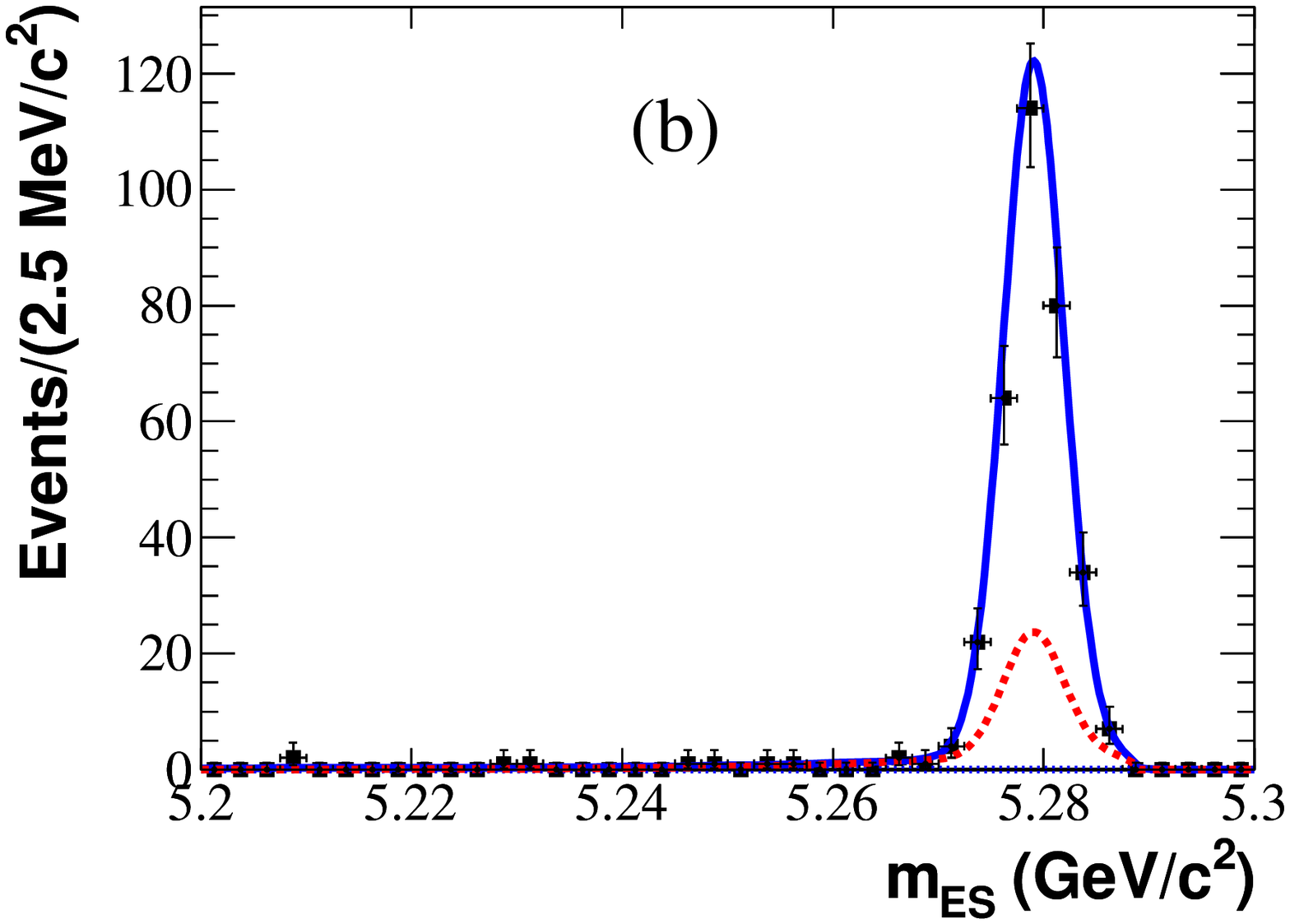,width=0.319\linewidth}
\epsfig{file=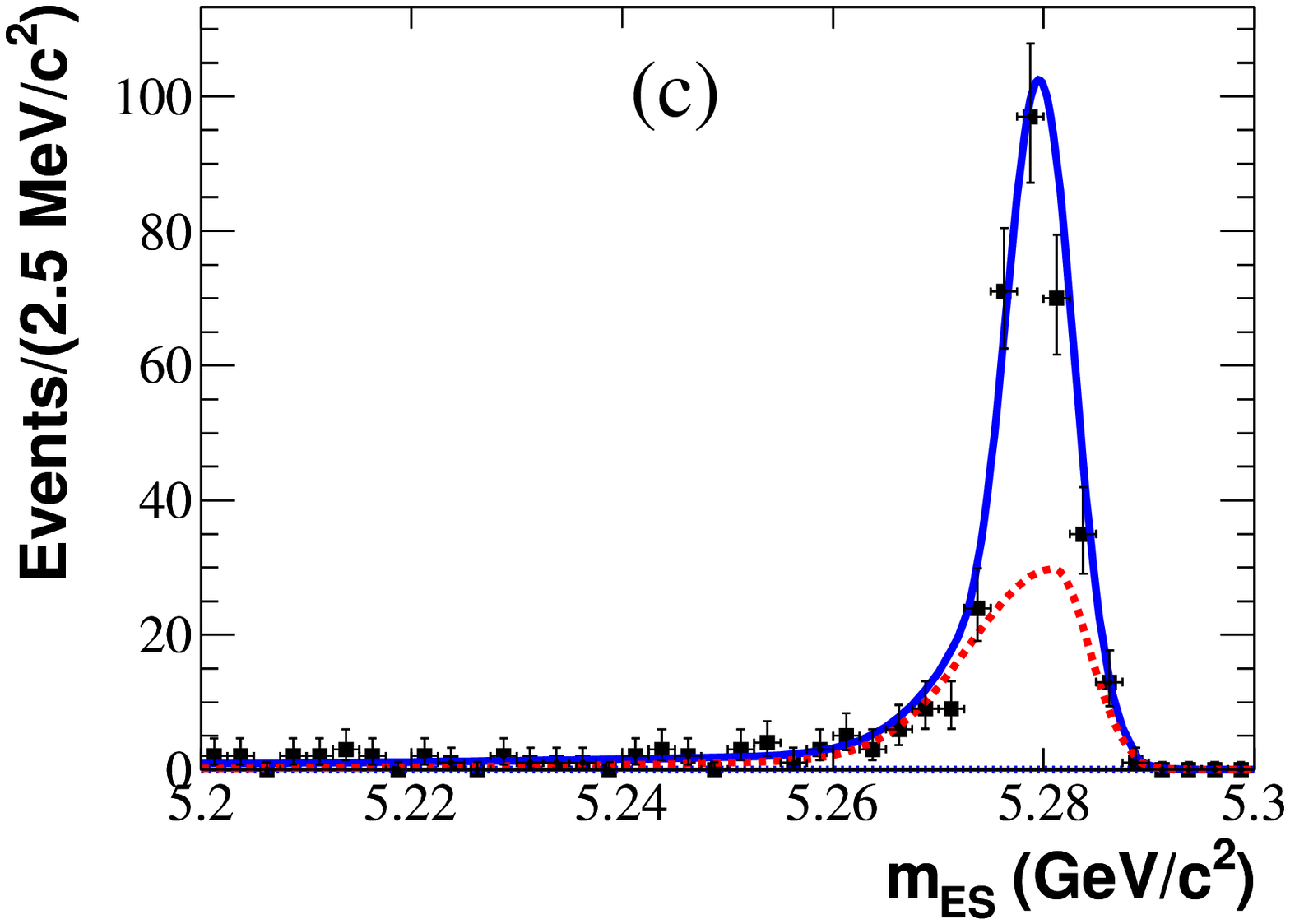,width=0.319\linewidth}
\epsfig{file=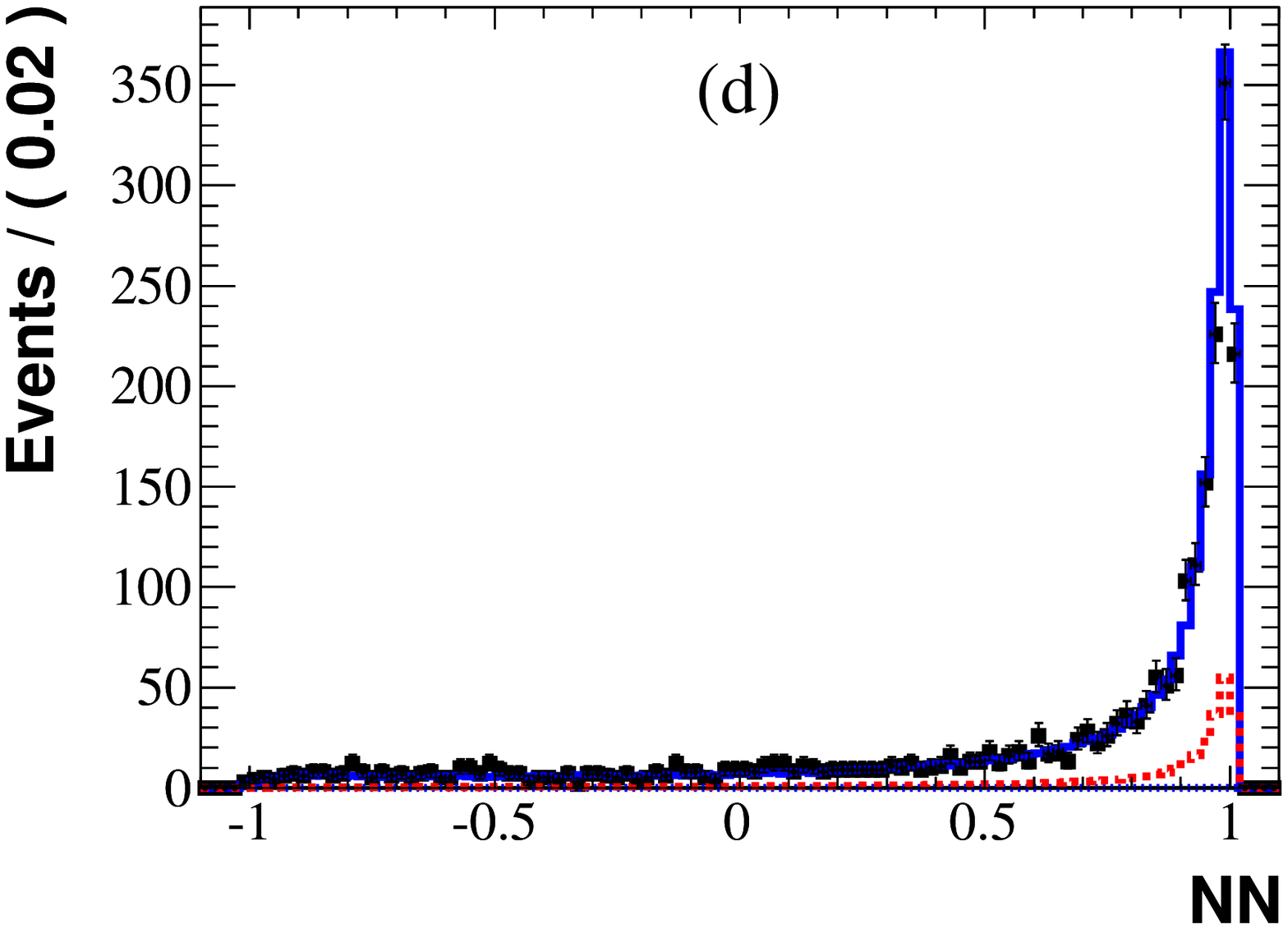,width=0.319\linewidth}
\epsfig{file=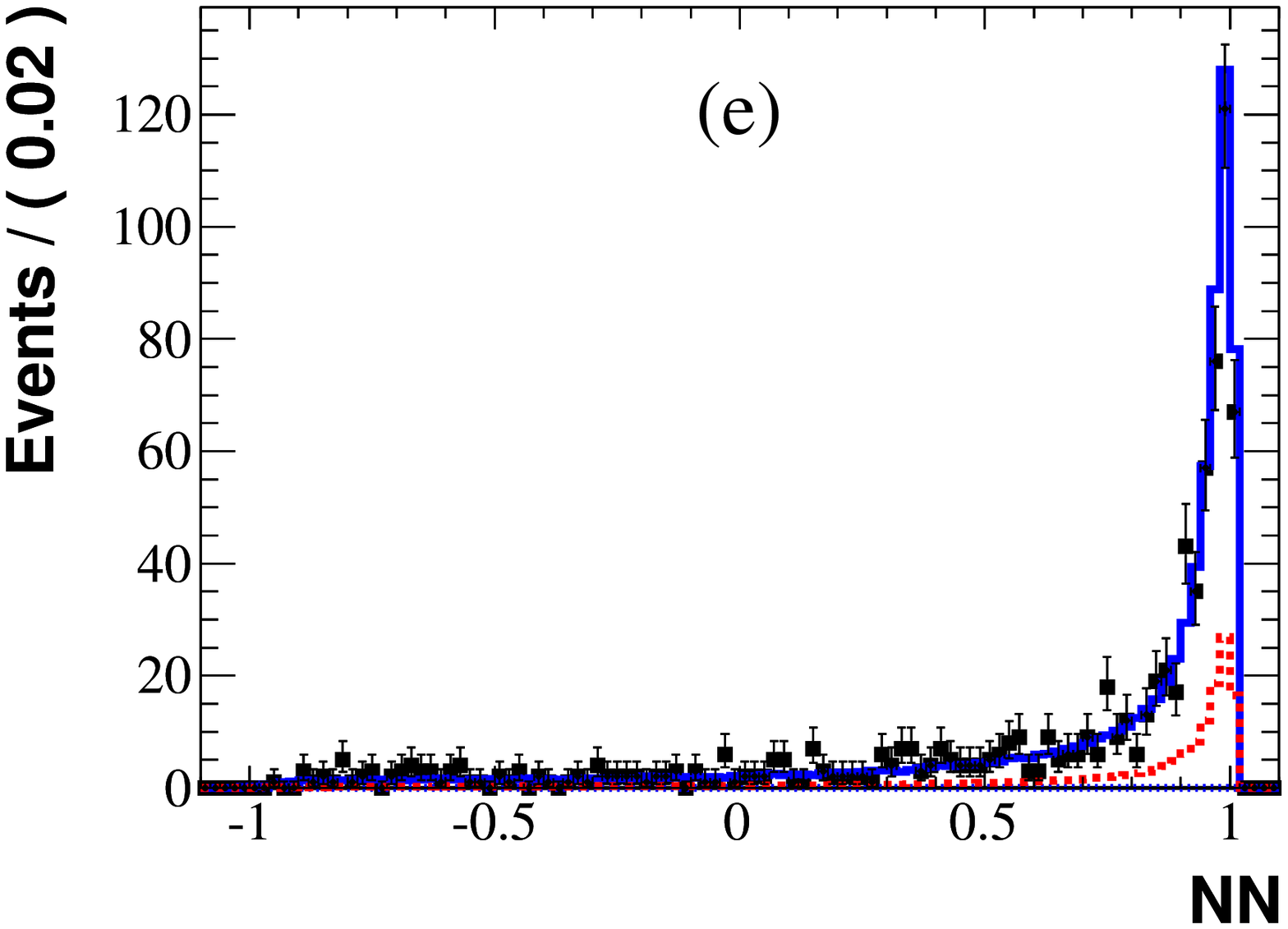,width=0.319\linewidth}
\epsfig{file=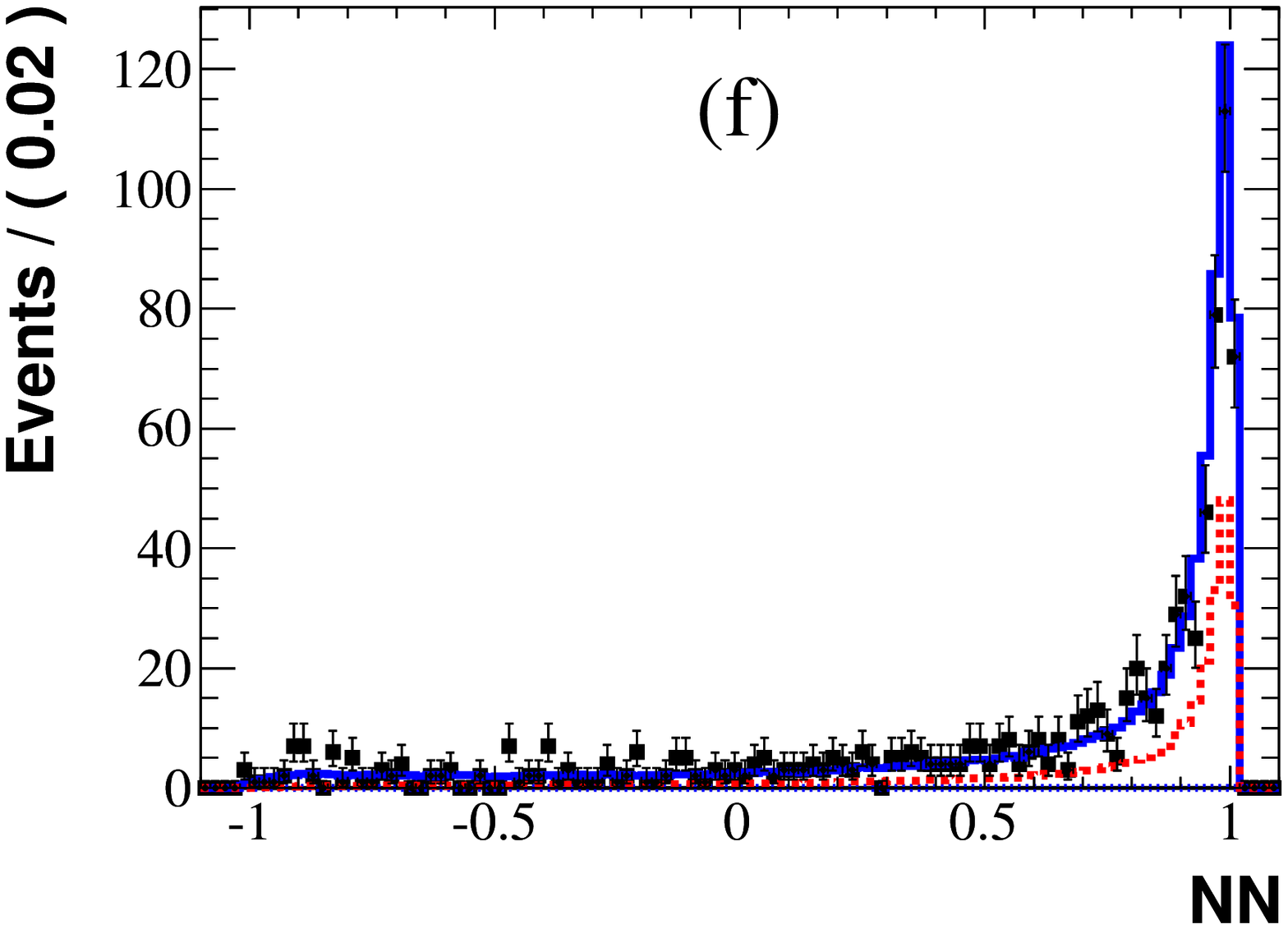,width=0.319\linewidth} \caption{(color
online). Projections on \mes (a, b, c) and $NN$ (d, e, f) of the
fit results for $DK$ (a, d), $D^{*}_{D \piz}K$ (b, e) and
$D^{*}_{D \gamma}K$ (c, f) RS decays, for samples enriched in
signal with the requirements $NN>0.94$ (\mes projections) or
$5.2725<\mes<5.2875\gevcc$ ($NN$ projections). The points with
error bars are data. The curves represent the fit projections for
signal plus background (solid) and background (dashed).}
\label{fig:dKresultscab}
\end{center}
\end{figure*}

\begin{figure*}[htb]
\begin{center}
\epsfig{file=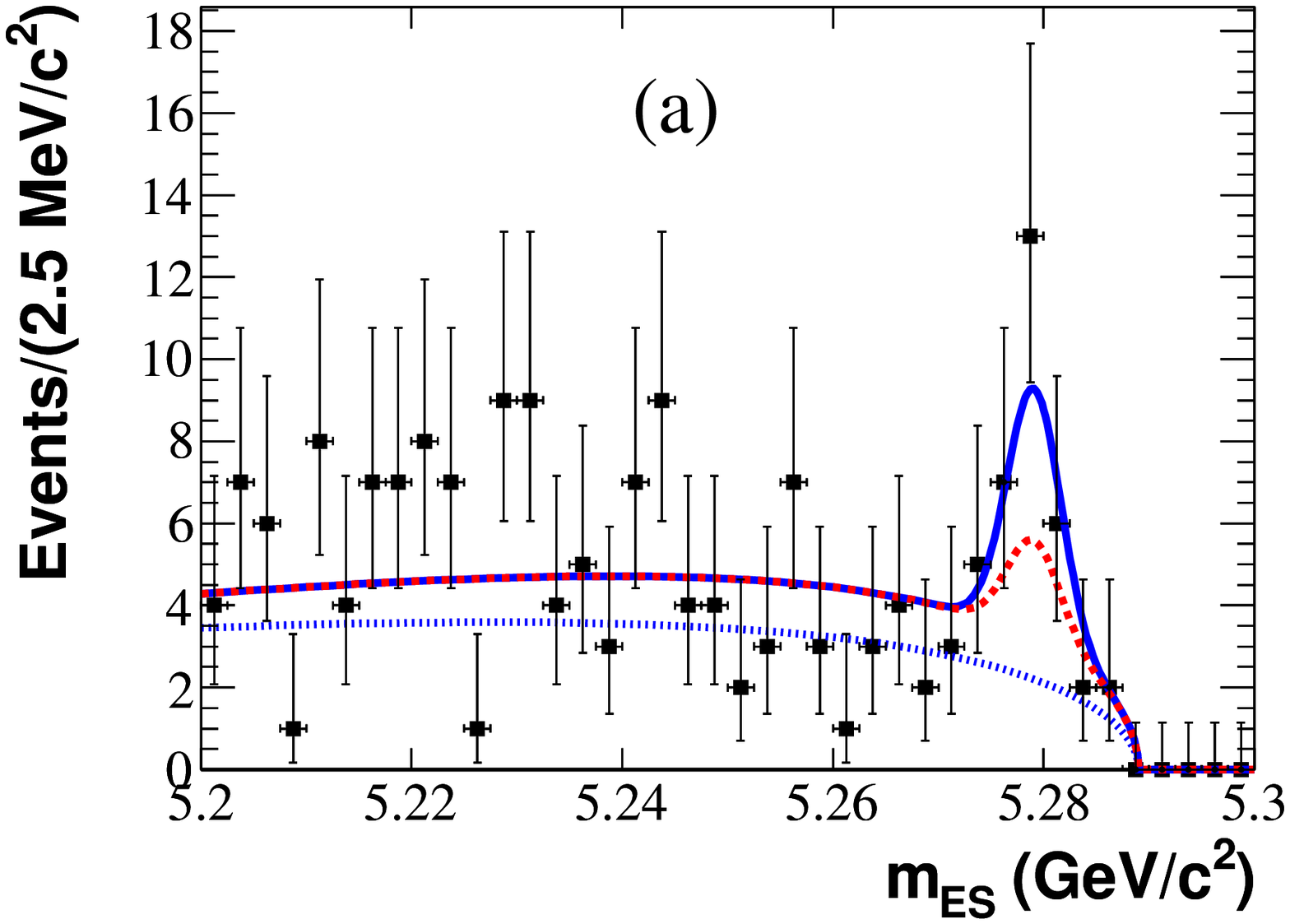,width=0.319\linewidth}
\epsfig{file=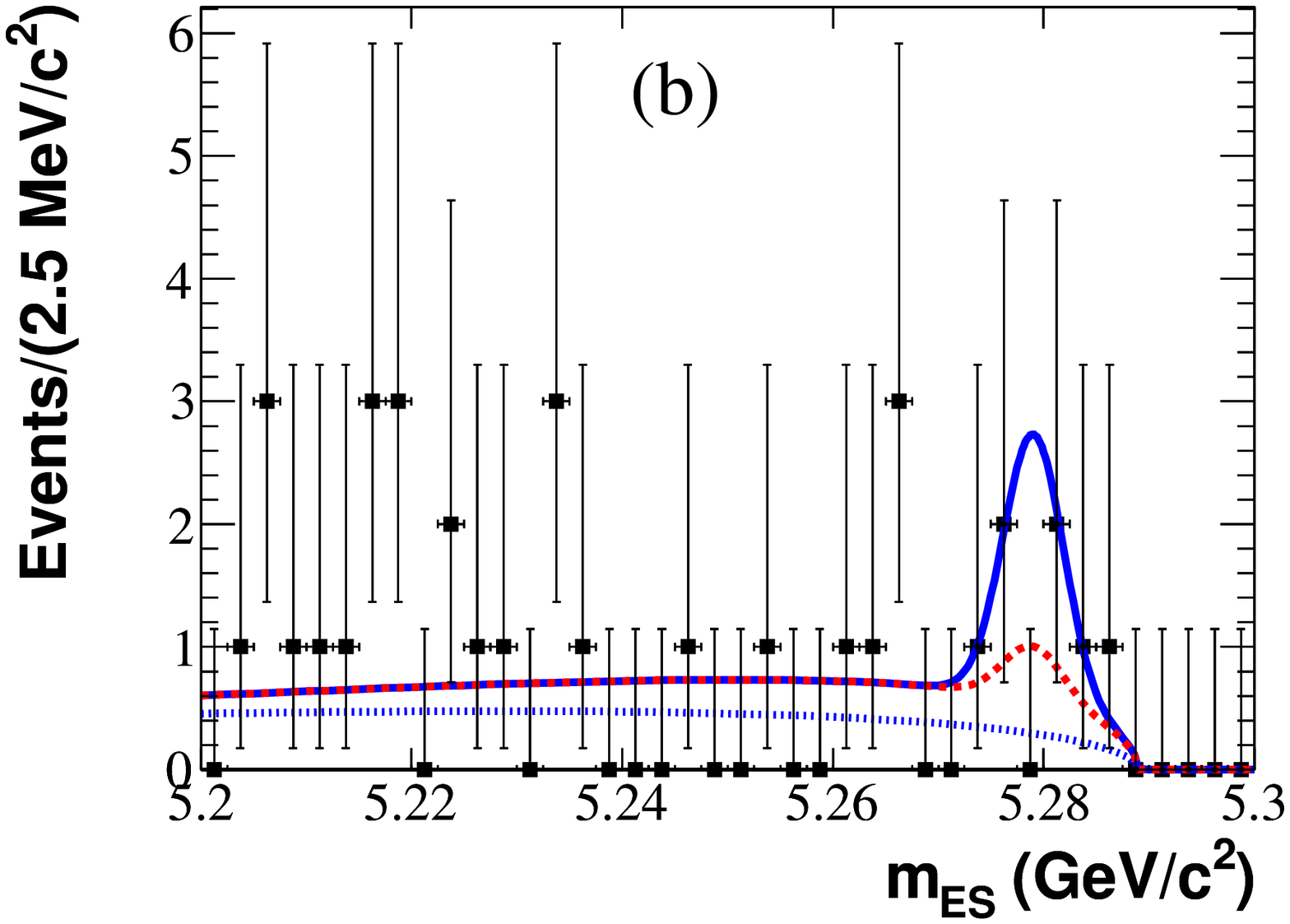,width=0.319\linewidth}
\epsfig{file=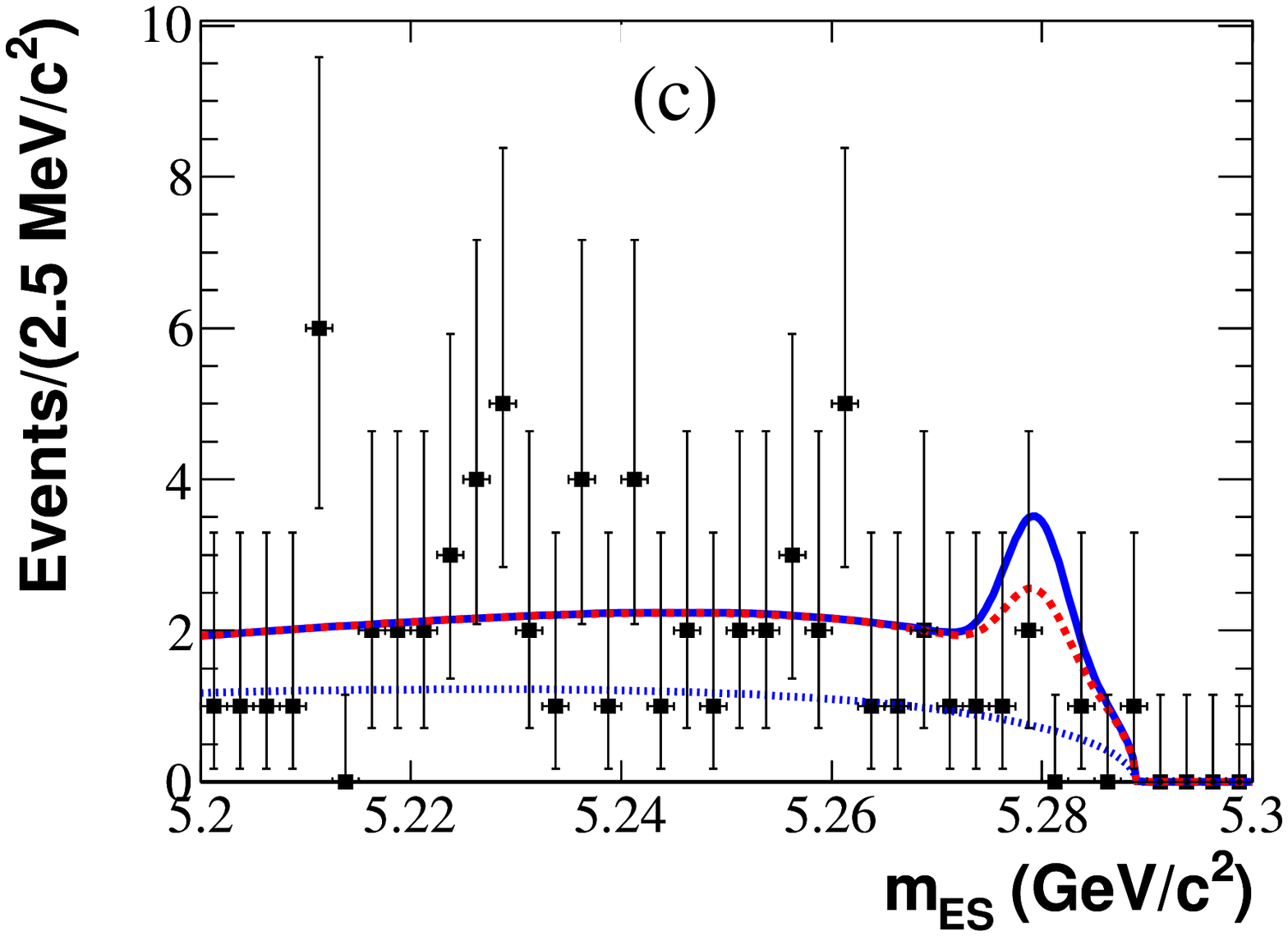,width=0.319\linewidth}
\epsfig{file=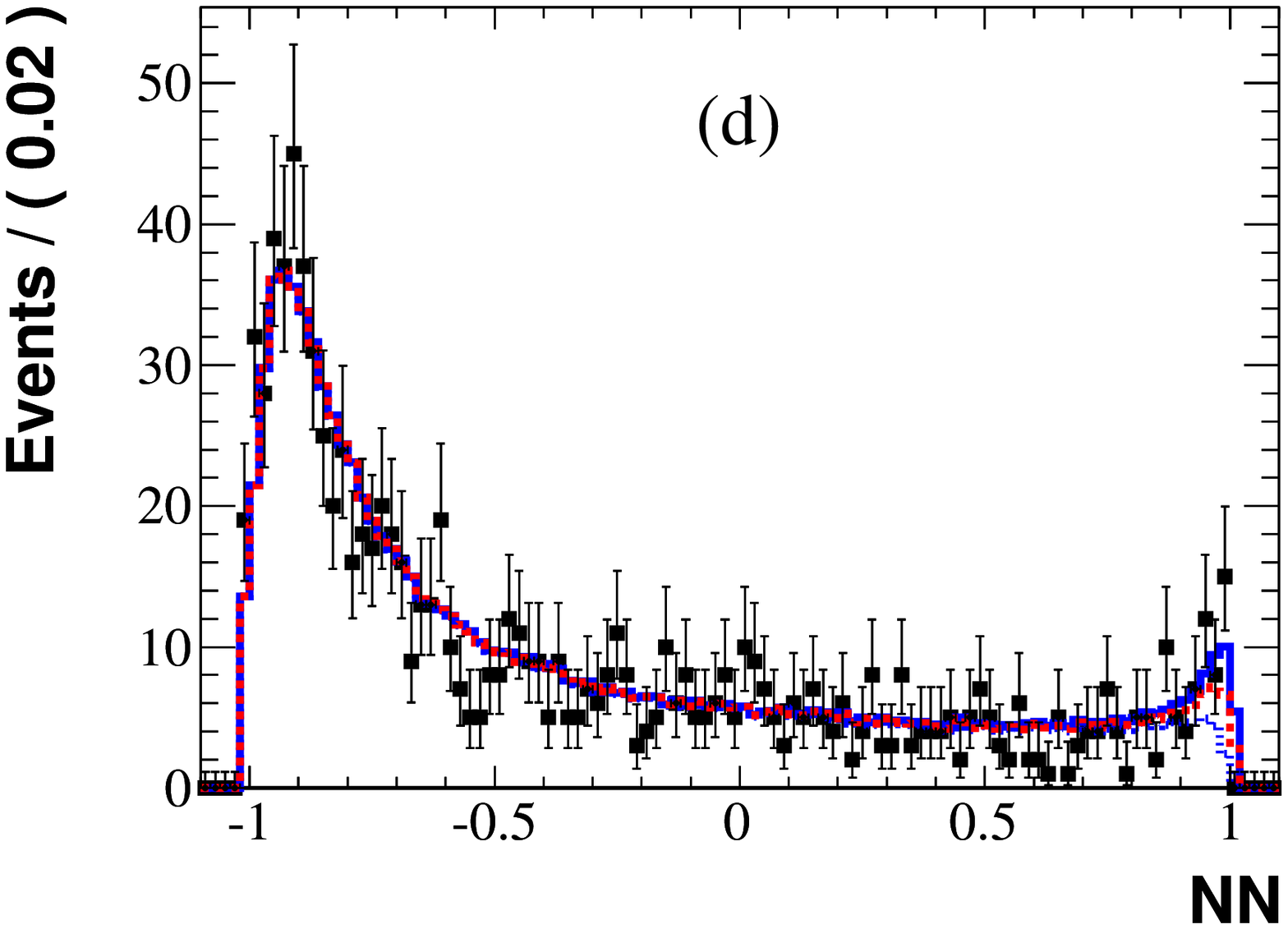,width=0.319\linewidth}
\epsfig{file=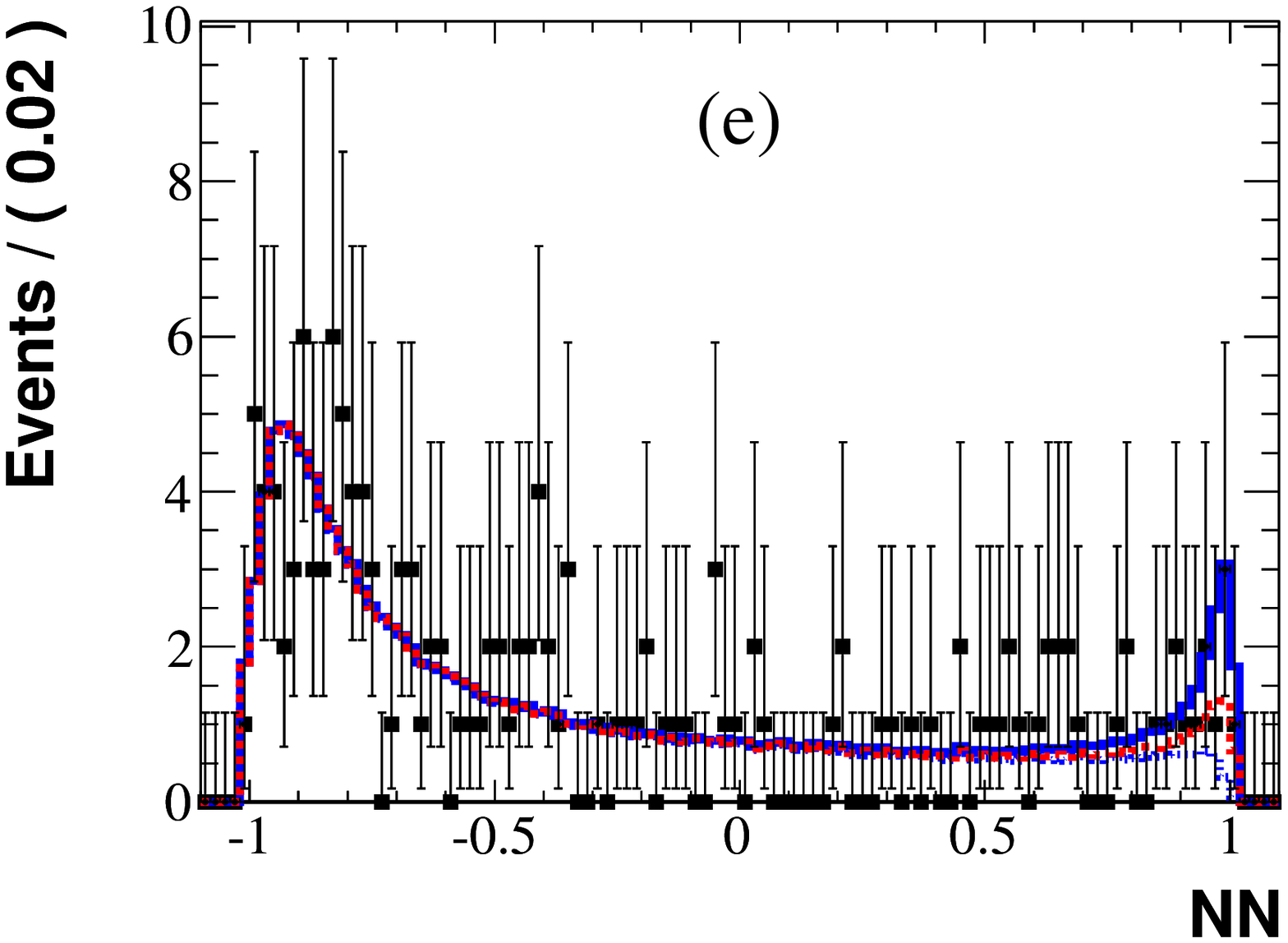,width=0.319\linewidth}
\epsfig{file=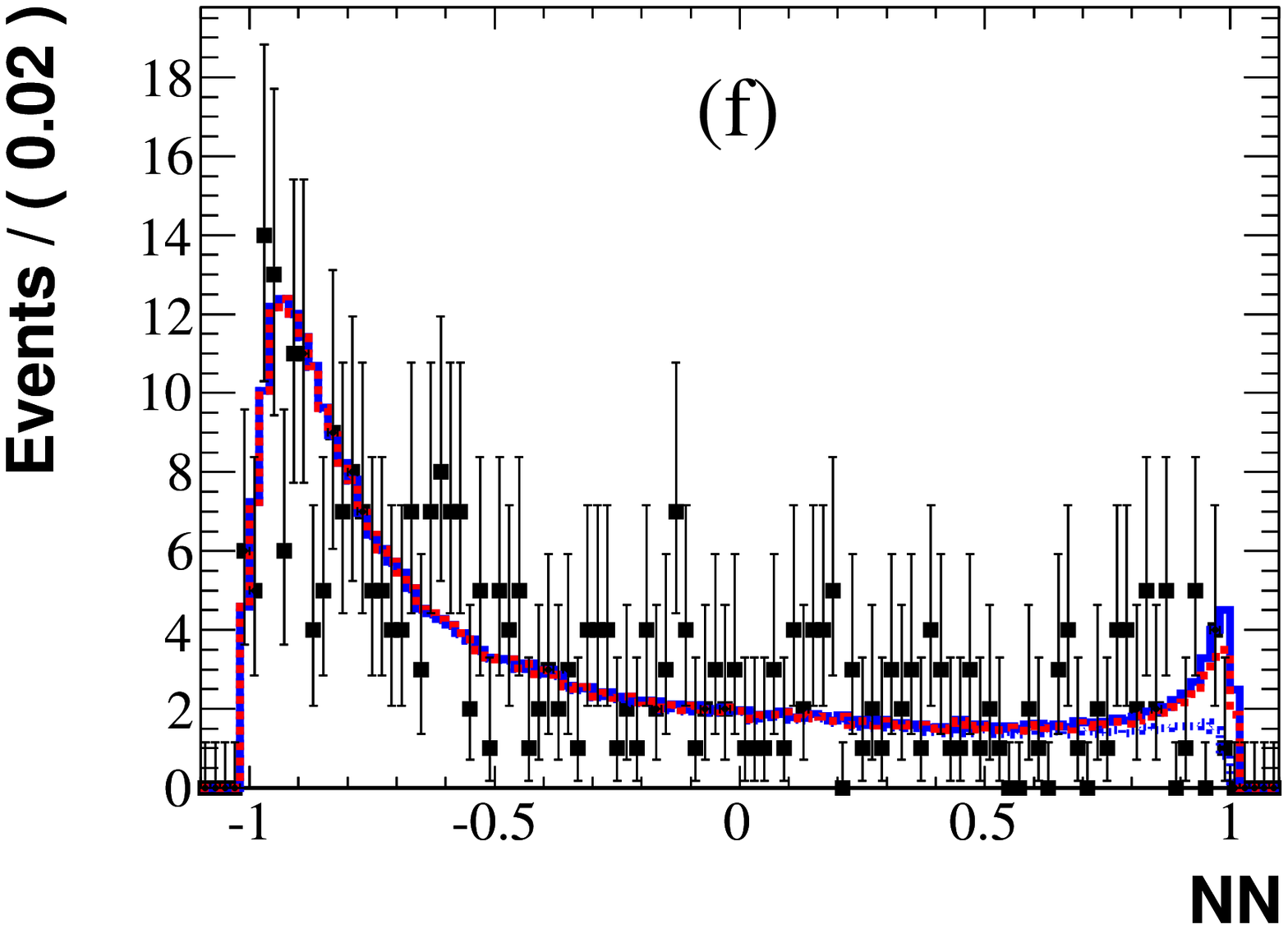,width=0.319\linewidth} \caption{(color
online). Projections on \mes (a, b, c) and $NN$ (d, e, f) of the
fit results for $DK$ (a, d), $D^{*}_{D \piz}K$ (d, e) and
$D^{*}_{D \gamma}K$ (c, f) WS decays, for samples enriched in
signal with the requirements $NN>0.94$ (\mes projections) or
$5.2725<\mes<5.2875\gevcc$ ($NN$ projections). The points with
error bars are data. The curves represent the fit projections for
signal plus background (solid), the sum of all background
components (dashed), and $q\bar q$ background only (dotted).}
\label{fig:dKresultsads}
\end{center}
\end{figure*}

\begin{table*}[htb]
   \caption{Summary of fit results for $D^{(*)}K$.}
   \begin{center}
   \begin{tabular}{lrclrclrcl} \hline\hline
    Mode & \multicolumn{3}{c}{$DK$}  &  \multicolumn{3}{c}{$D^*_{D\piz}K$}  & \multicolumn{3}{c}{$D^*_{D\gamma}K$} \\ \hline
   Ratio of rates, \RDDstarK $(10^{-3})$&
    $11.1$ & $\pm$ & $ 5.5$ &
    $17.6$ & $\pm$ & $ 9.3$ &
    $13$ & $\pm$ & $ 14$ \\
   No. of signal events $N_{WS}$ & $19.4$ & $\pm$ & $ 9.6$ & $10.3$ & $\pm$ & $ 5.5$ &  $5.9$ & $\pm$ & $ 6.4$ \\
   No. of normalization events $N_{RS}$& $  1755$ & $\pm$ & $ 48$ & $587 $ & $\pm$ & $ 28$ & $455 $ & $\pm$ & $ 29$ \\
   $B^+$ Ratio of rates, \RDDstarKp $(10^{-3})$&
   $21.9$ & $\pm$ & $ 9.0$ &
   $4.9$ & $\pm$ & $ 7.9$ &
   $9$ & $\pm$ & $ 16$ \\
$B^-$ Ratio of rates, \RDDstarKm $(10^{-3})$&
   $1.7$ & $\pm$ & $ 5.9$ &
   $37$ & $\pm$ & $ 18$ &
   $19$ & $\pm$ & $ 23$ \\
Asymmetry \ADDstarK & $-0.86$ & $\pm$ & $ 0.47$ & $0.77$ & $\pm$ & $ 0.35$ & $0.36$ & $\pm$ & $ 0.94$\\
   \hline\hline
   \end{tabular}
   \end{center}
   \label{tab:fit-results-dk}
   \end{table*}

\begin{figure*}
\begin{center}
\epsfig{file=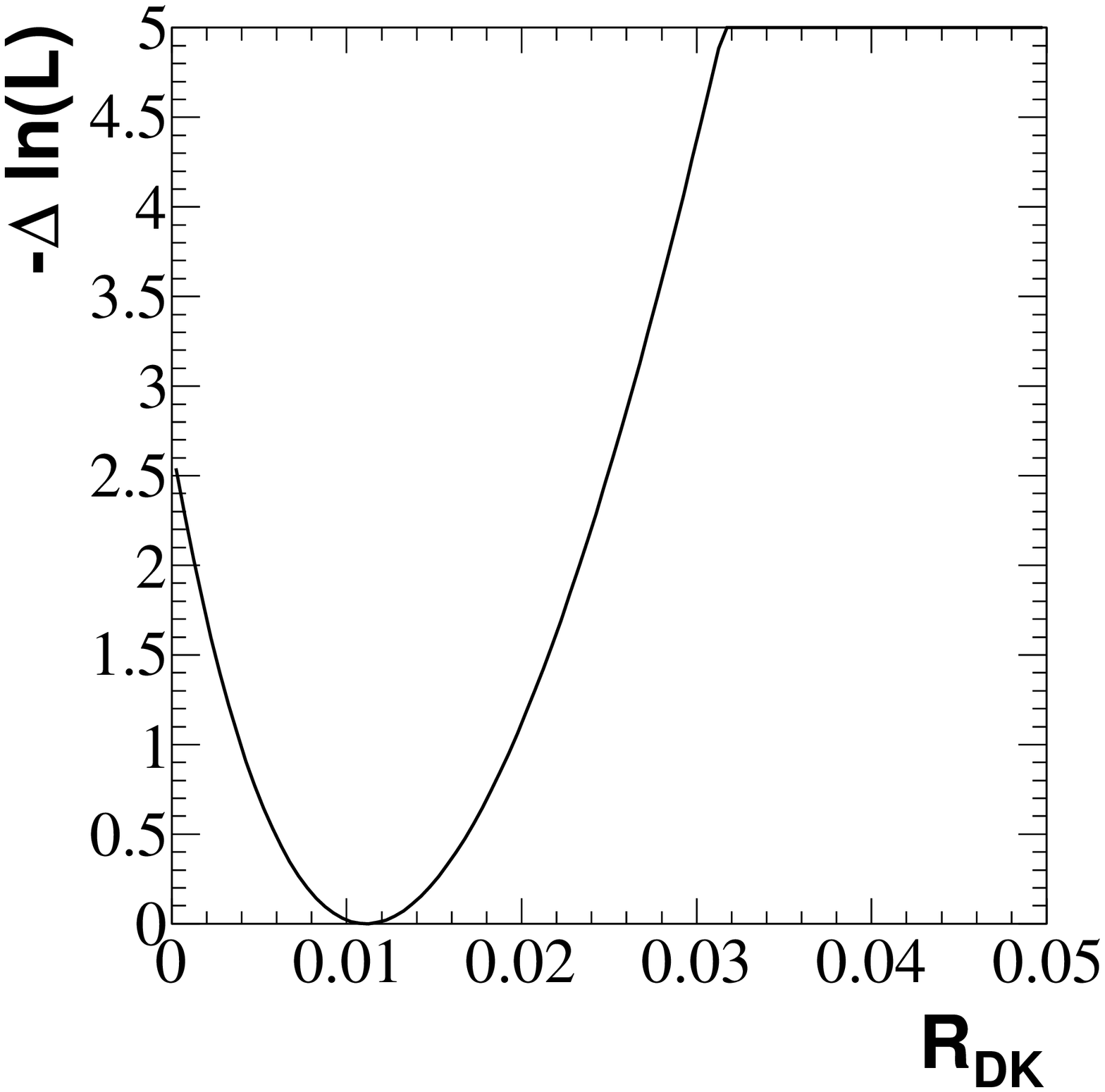,width=0.32\linewidth}
\epsfig{file=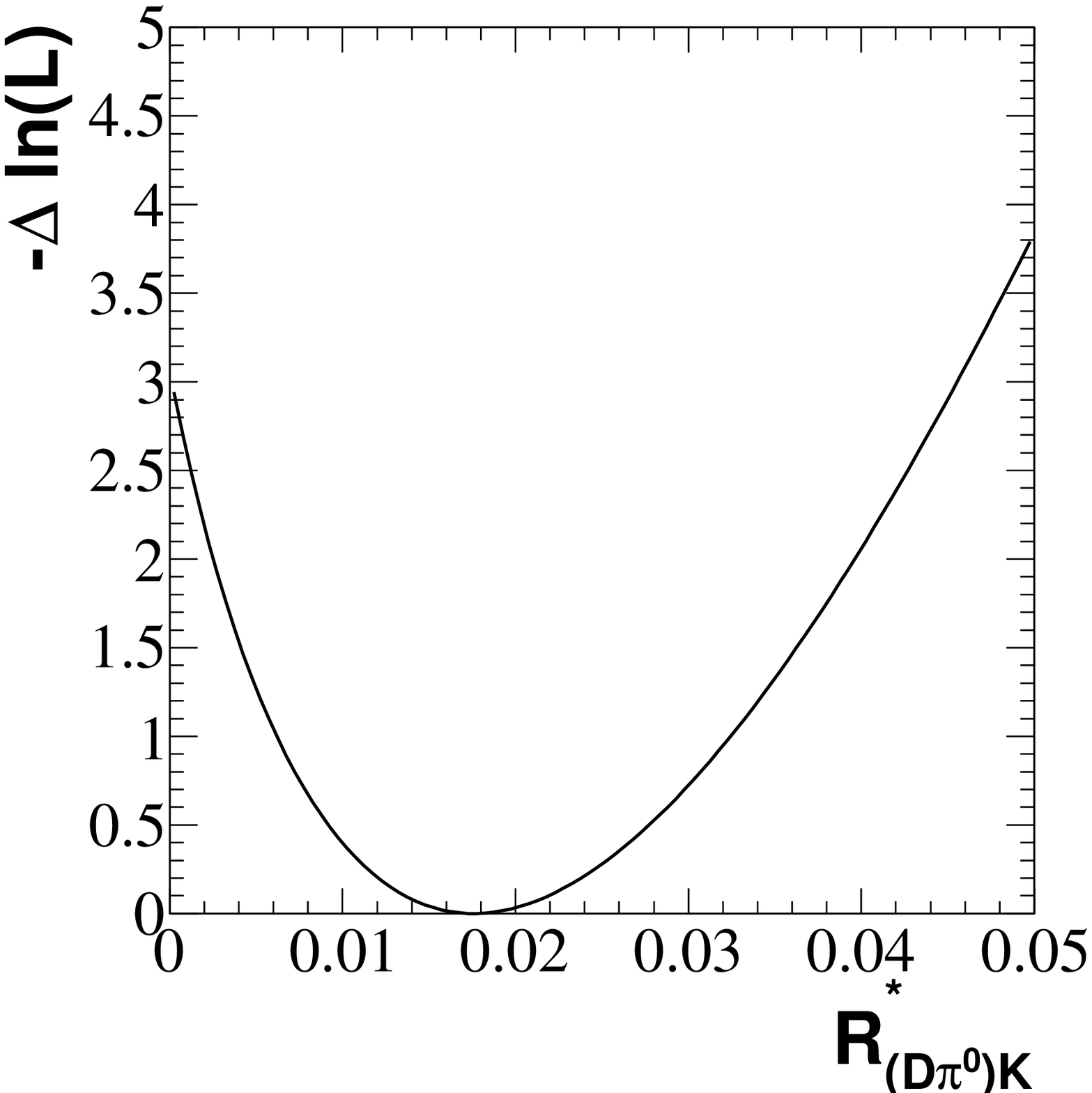,width=0.32\linewidth}
\epsfig{file=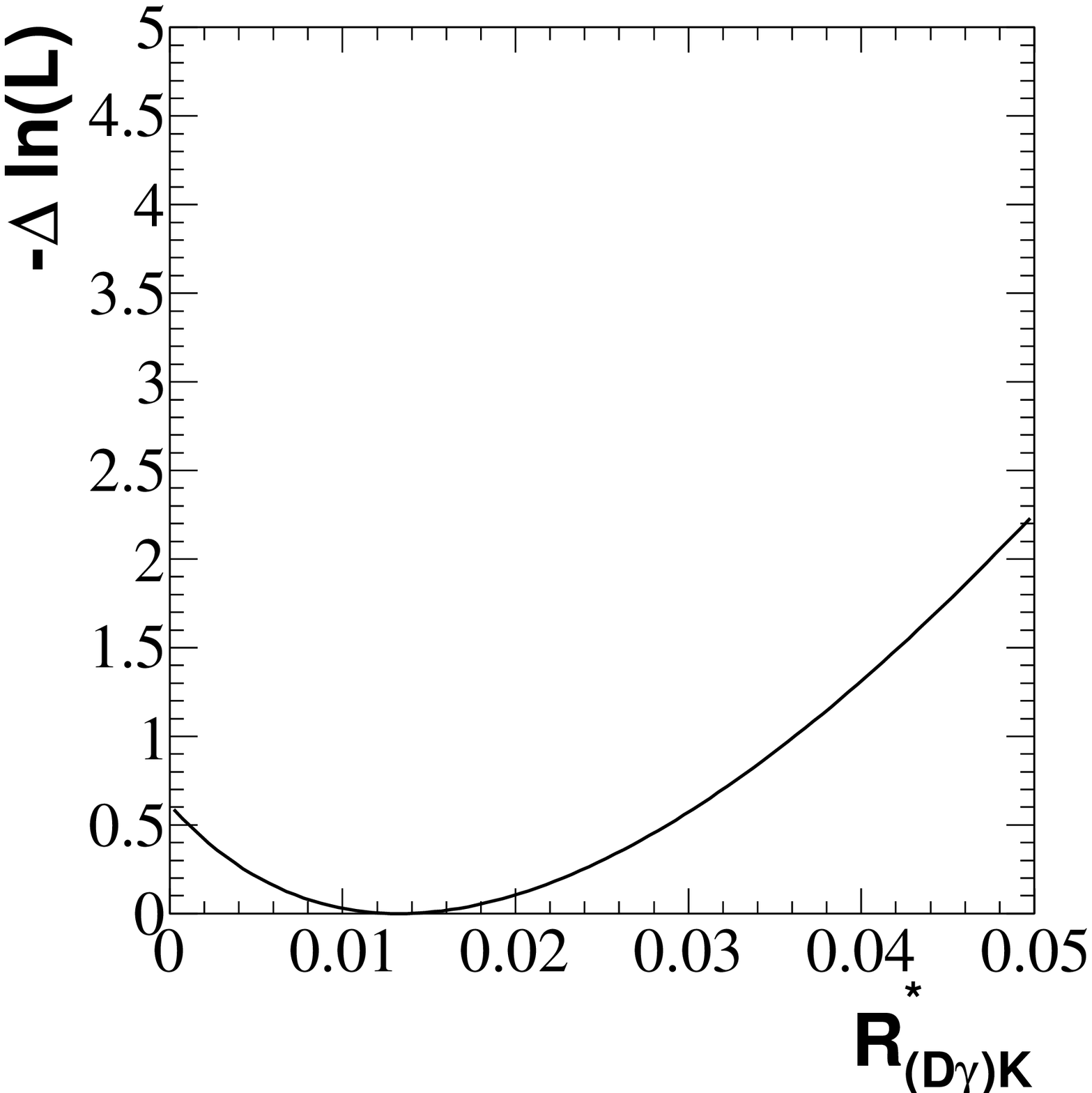,width=0.32\linewidth}
\caption{Negative log-likelihood variation vs \RDDstarK for
$B^\pm\to DK^\pm$ (left), $B^\pm\to D^{*}_{D\piz}K^\pm$ (center)
and $B^\pm\to D^{*}_{D\gamma}K^\pm$ (right). Systematic
uncertainties are not included.} \label{fig:likdatadk}
\end{center}
\end{figure*}

The results for $B\to D^{(*)}K$ are displayed in
Fig.~\ref{fig:dKresultscab} (RS modes) and
Fig.~\ref{fig:dKresultsads} (WS modes). They are summarized in
Table~\ref{tab:fit-results-dk}. Indications of
 signals are observed in the $B\to DK$ and in the
$B\to D^{*}_{D\piz}K$ WS modes, with statistical significances of
$2.2\sigma$ and $2.4\sigma$, respectively
(Fig.~\ref{fig:likdatadk}).  Accounting for the systematic
uncertainties, the significances become $2.1\sigma$ and
$2.2\sigma$, respectively. For $B\to D^{*}_{D\gamma}K$ WS, no
significant signal is observed.

\begin{figure*}
\begin{center}
\epsfig{file=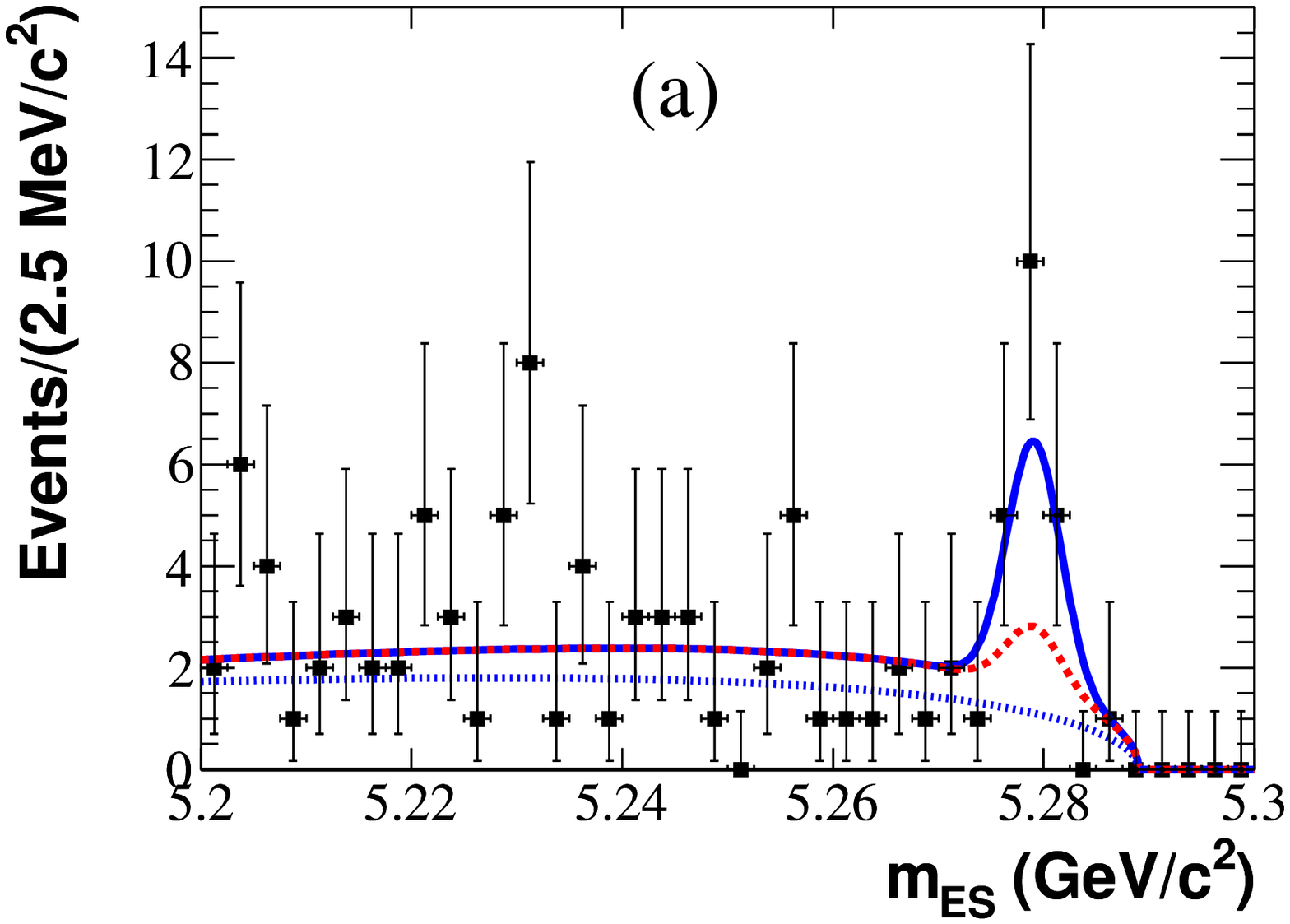,width=0.319\linewidth}
\epsfig{file=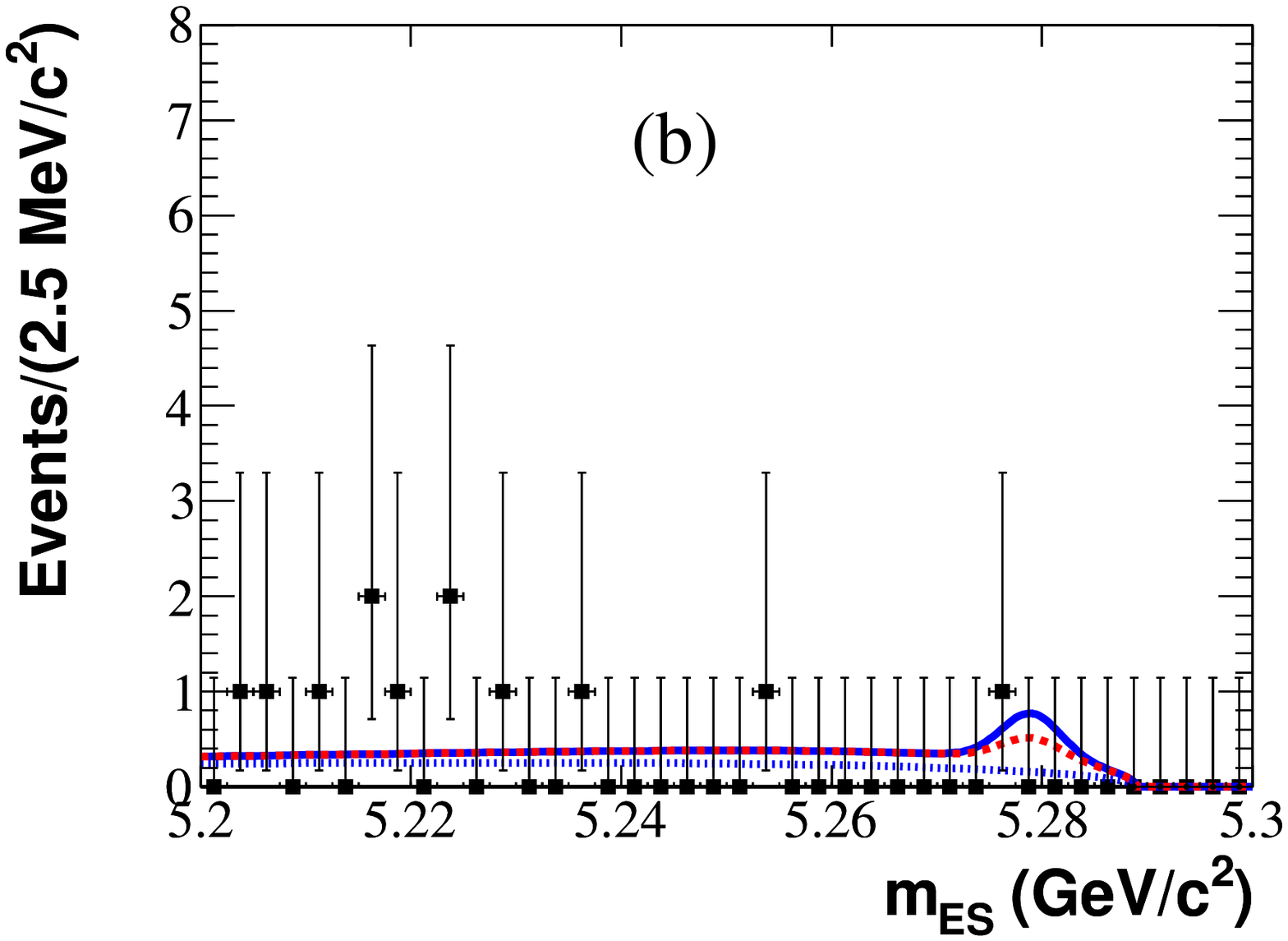,width=0.319\linewidth}
\epsfig{file=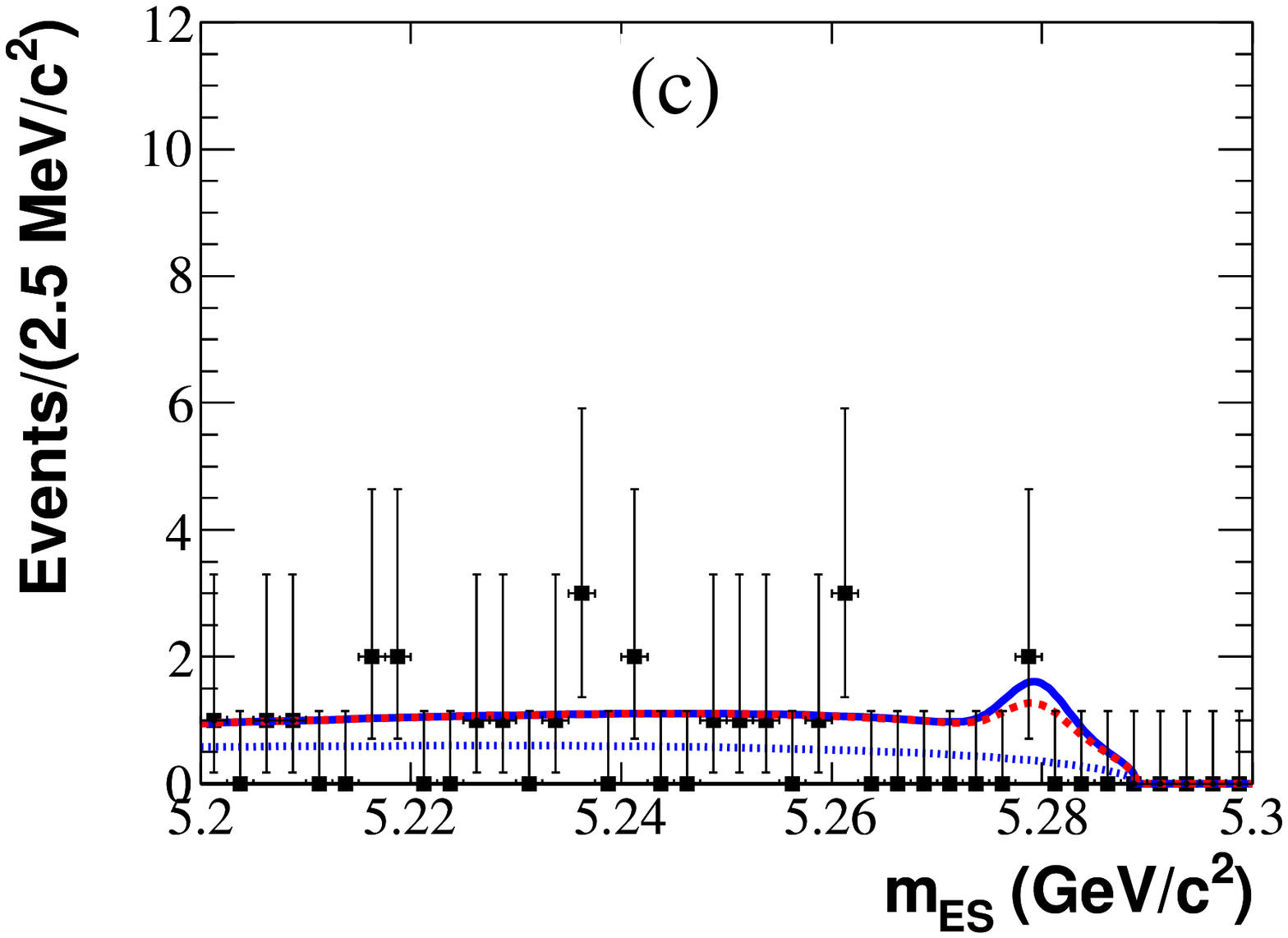,width=0.319\linewidth}
\epsfig{file=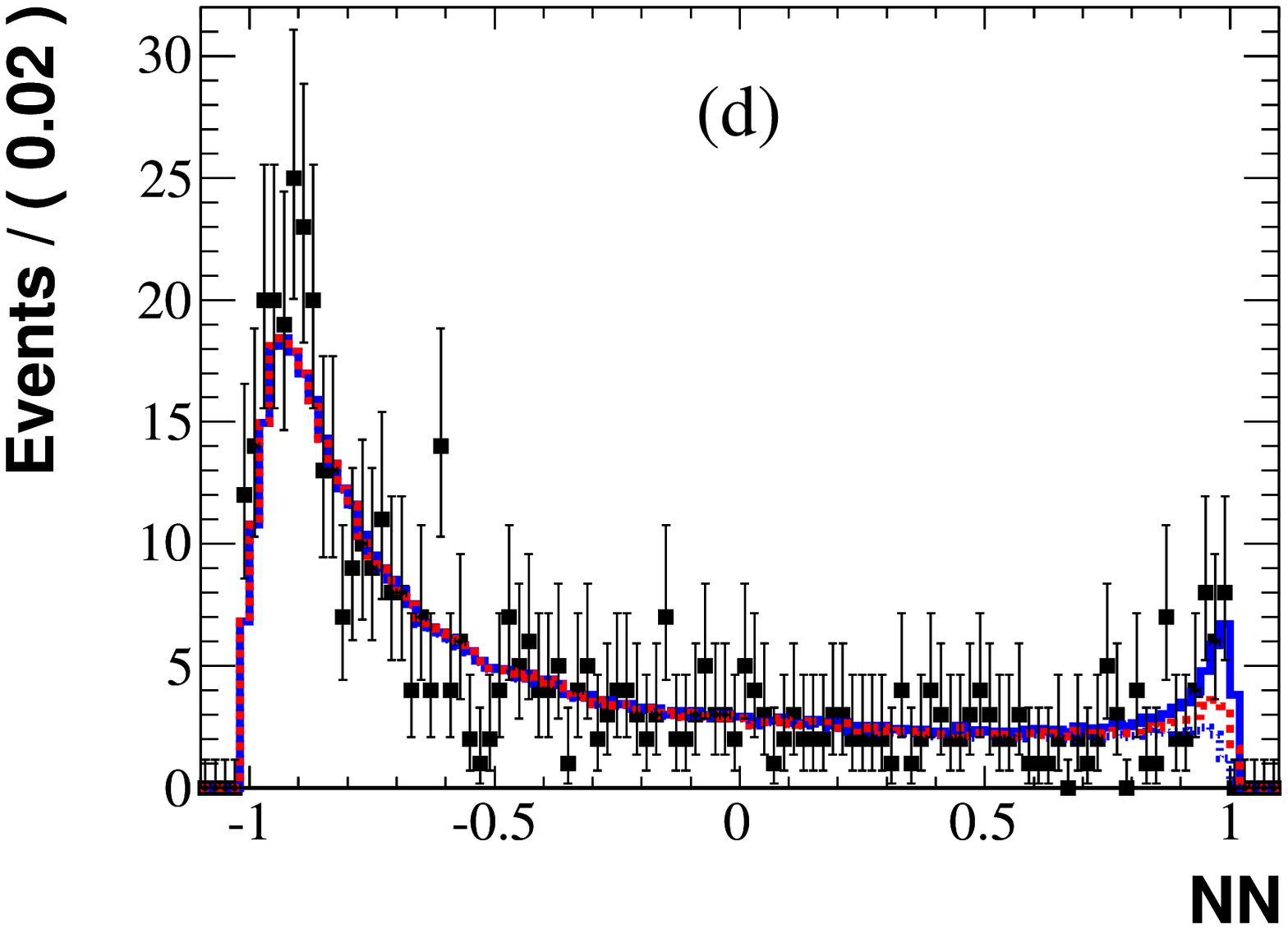,width=0.319\linewidth}
\epsfig{file=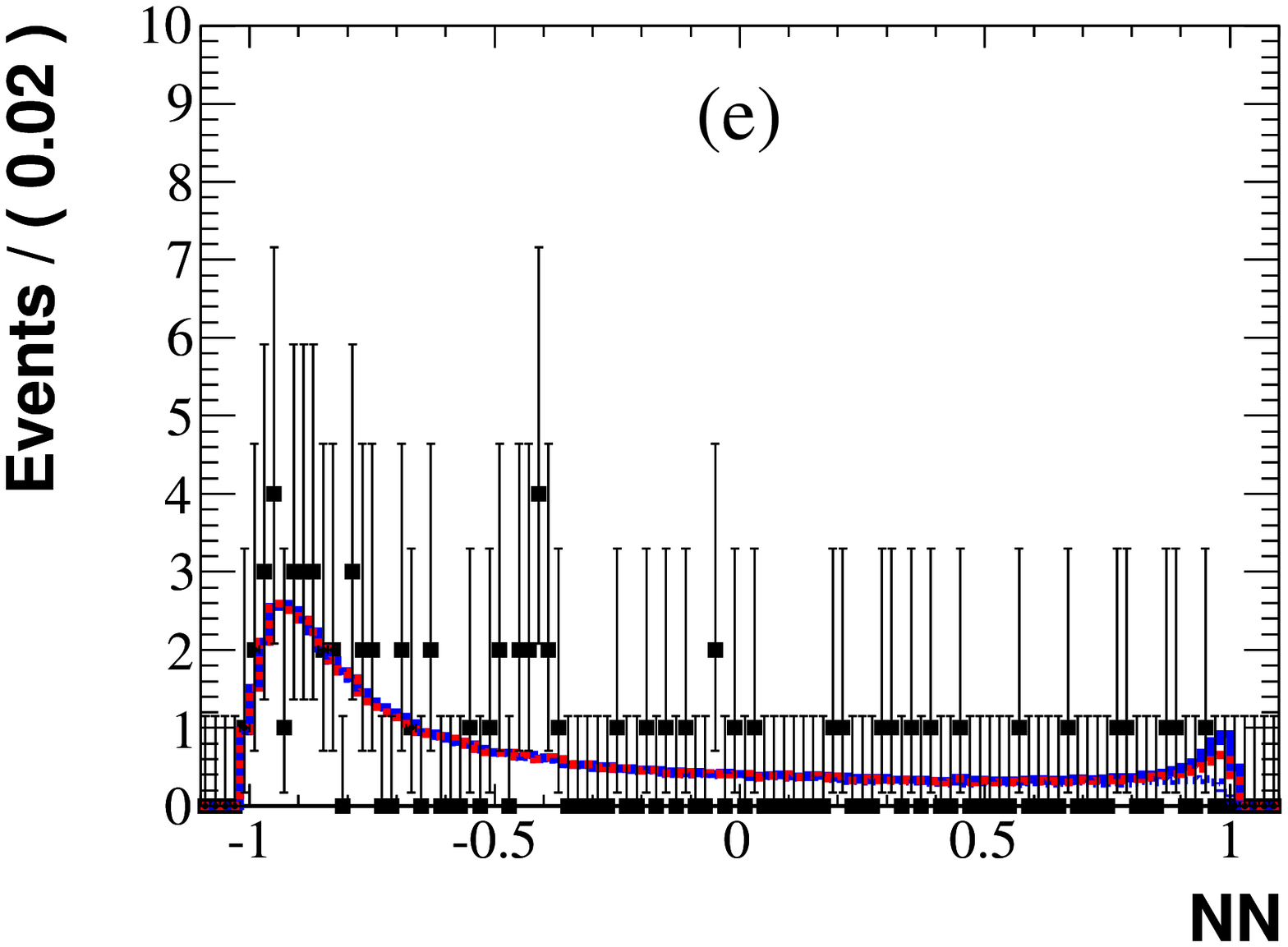,width=0.319\linewidth}
\epsfig{file=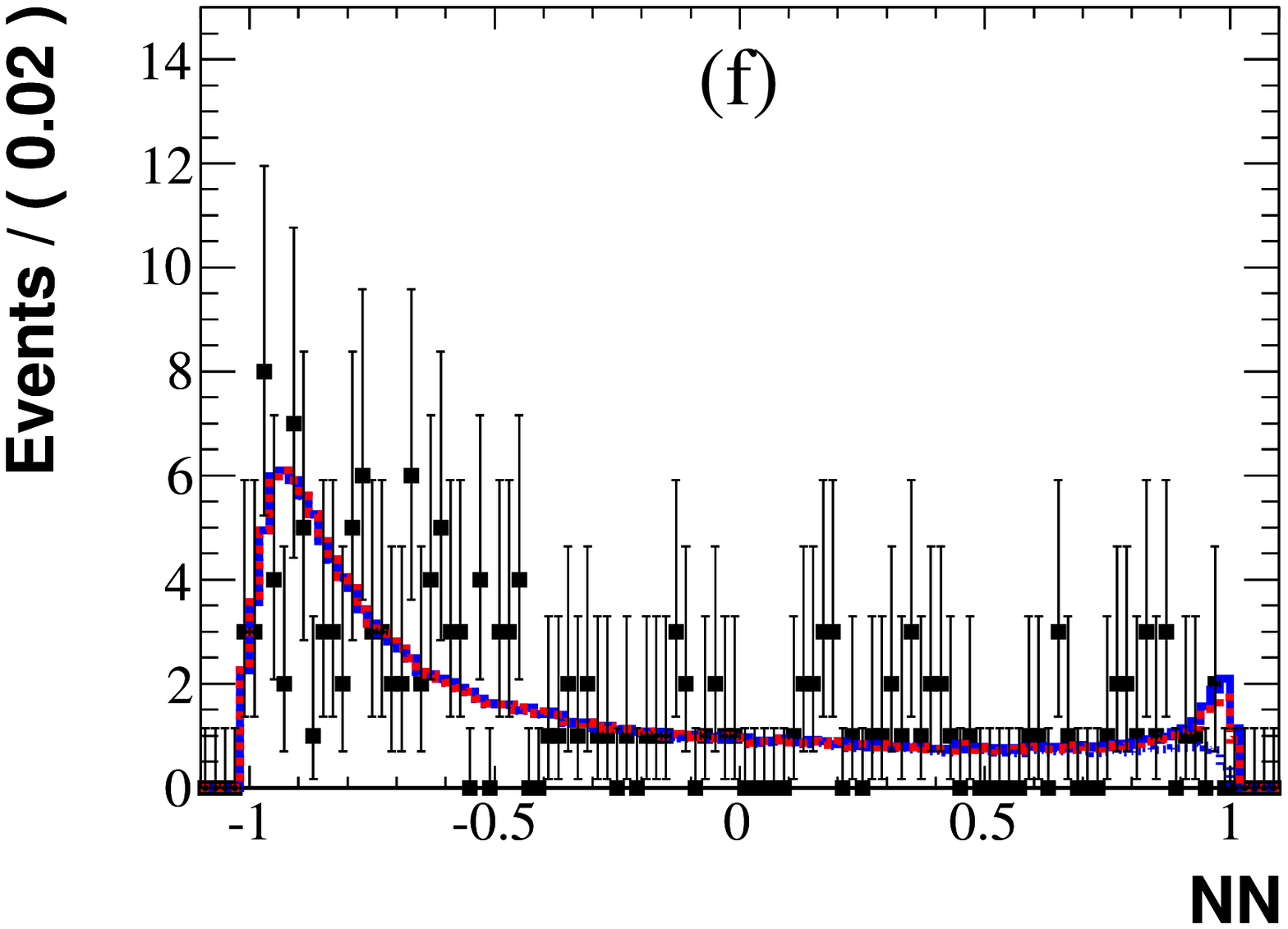,width=0.319\linewidth} \caption{(color
online). Projections on \mes (a, b, c) and $NN$ (d, e, f) of the
fit results for $D\Kp$ (a, d), $D^{*}_{D \piz}\Kp$ (b, e) and
$D^{*}_{D \gamma}\Kp$ (c, f) WS decays, for samples enriched in
signal with the requirements $NN>0.94$ (\mes projections) or
$5.2725<\mes<5.2875\gevcc$ ($NN$ projections). The points with
error bars are data. The curves represent the fit projections for
signal plus background (solid), the sum of all background
components (dashed), and $q\bar q$ background only (dotted).}
\label{fig:dKpresultsads}
\end{center}
\end{figure*}

\begin{figure*}
\begin{center}
\epsfig{file=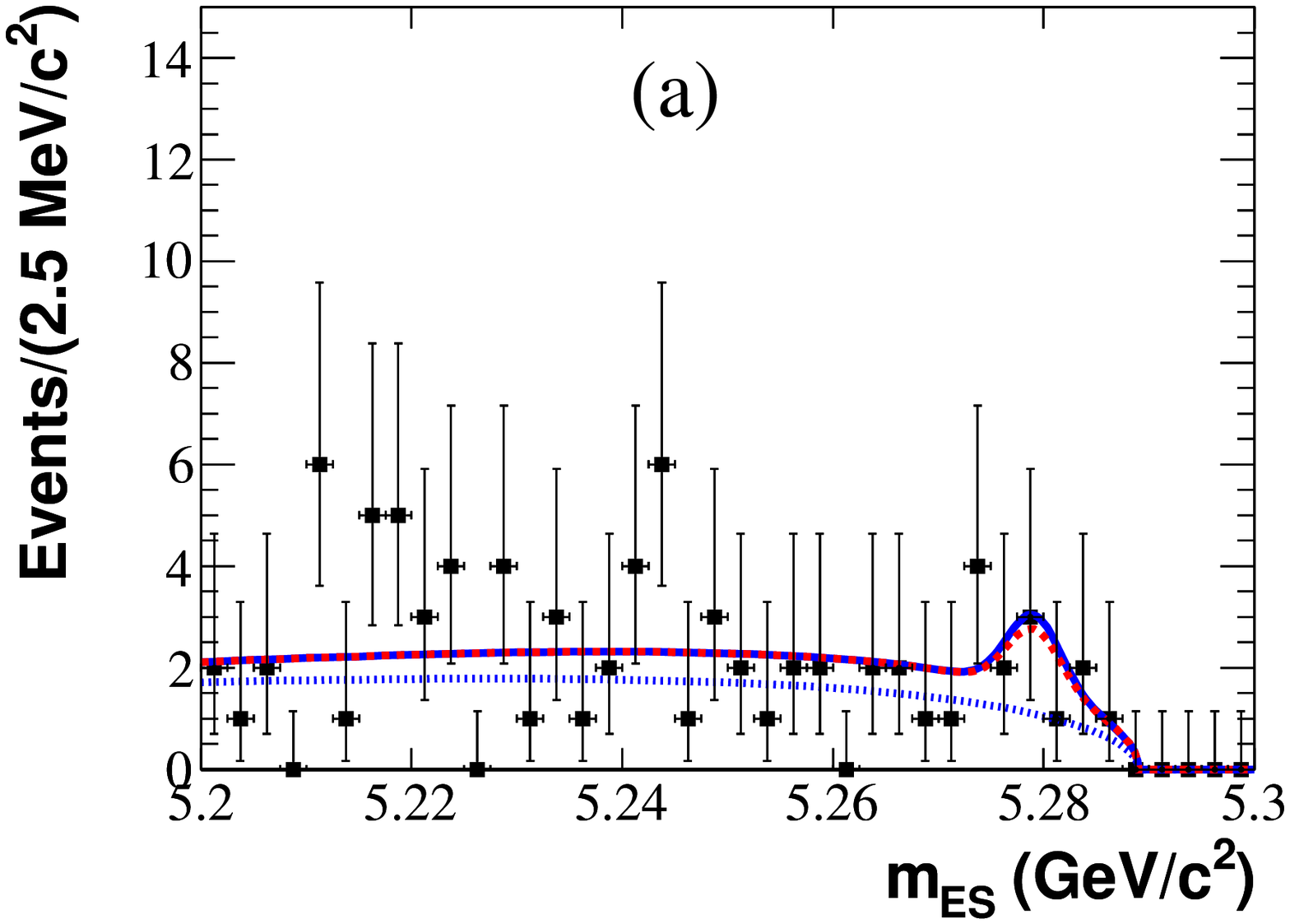,width=0.319\linewidth}
\epsfig{file=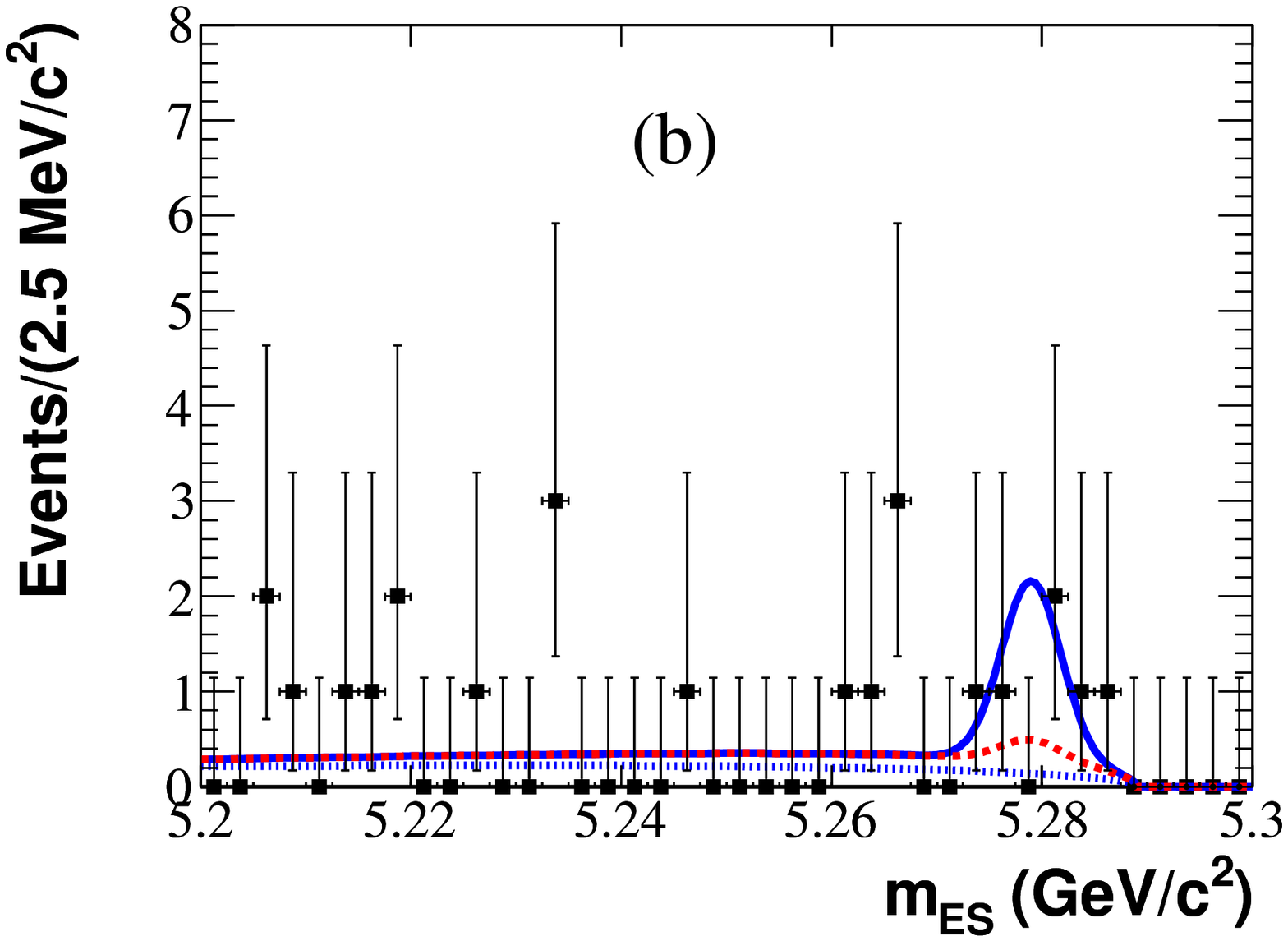,width=0.319\linewidth}
\epsfig{file=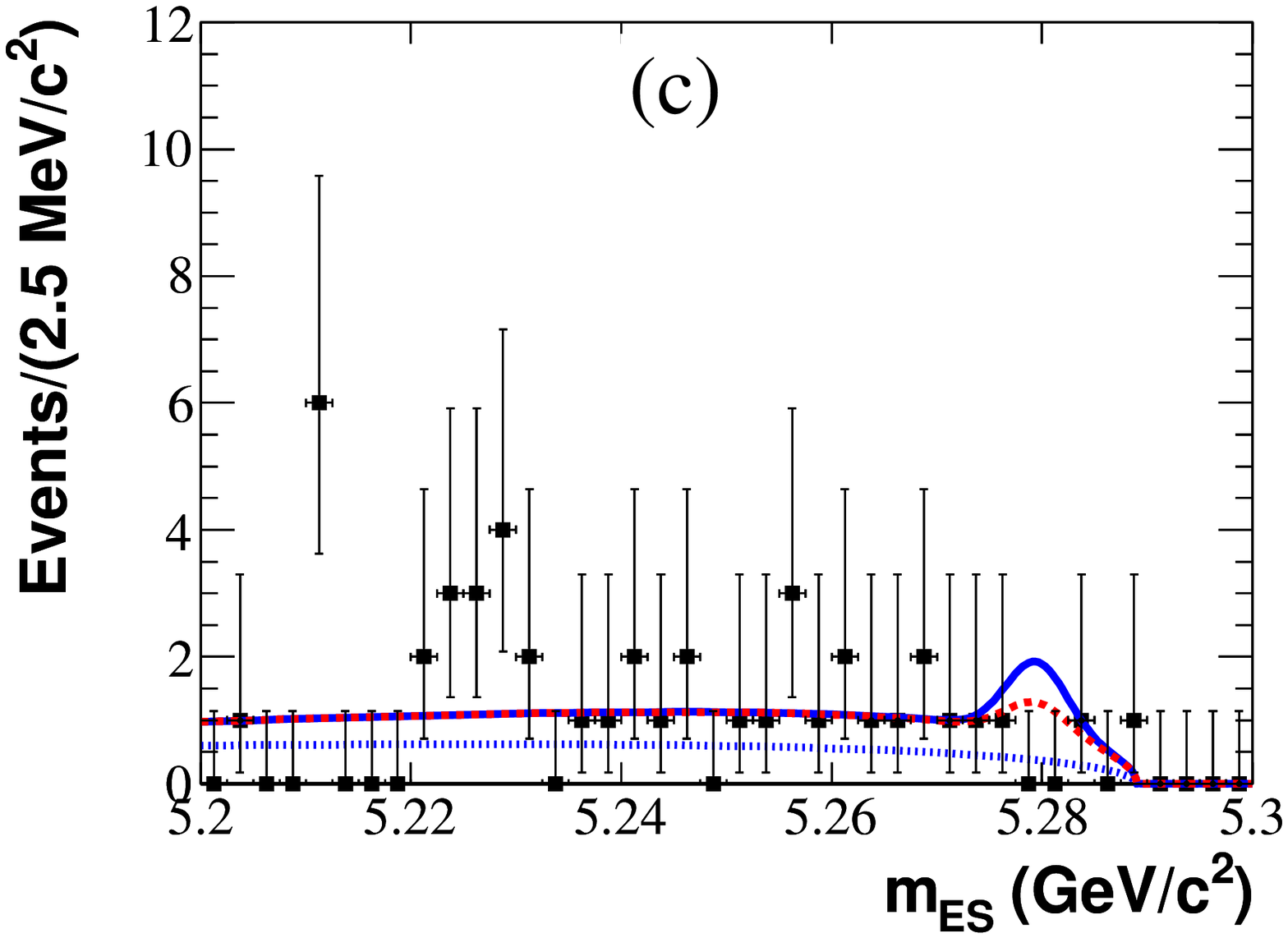,width=0.319\linewidth}
\epsfig{file=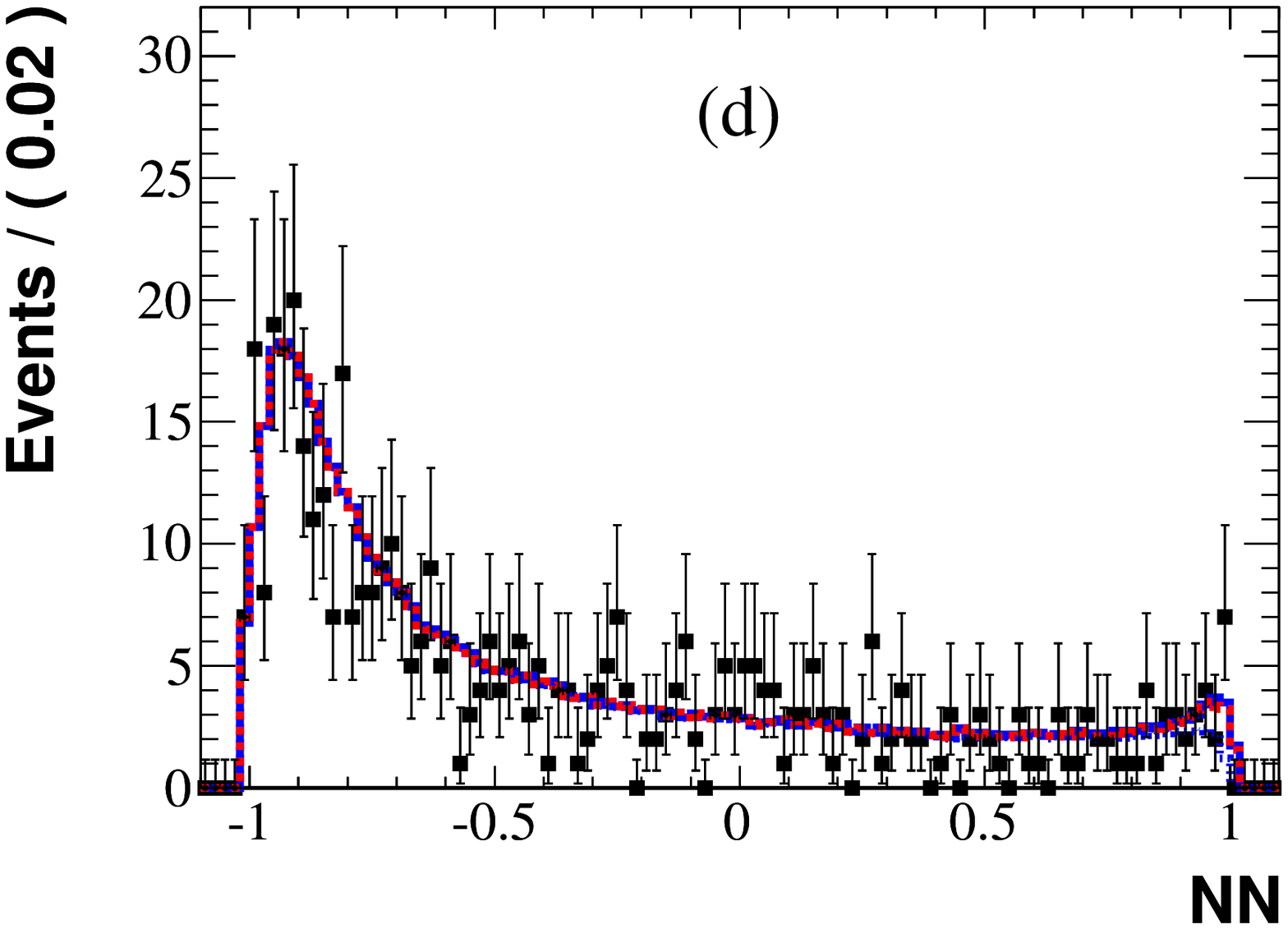,width=0.319\linewidth}
\epsfig{file=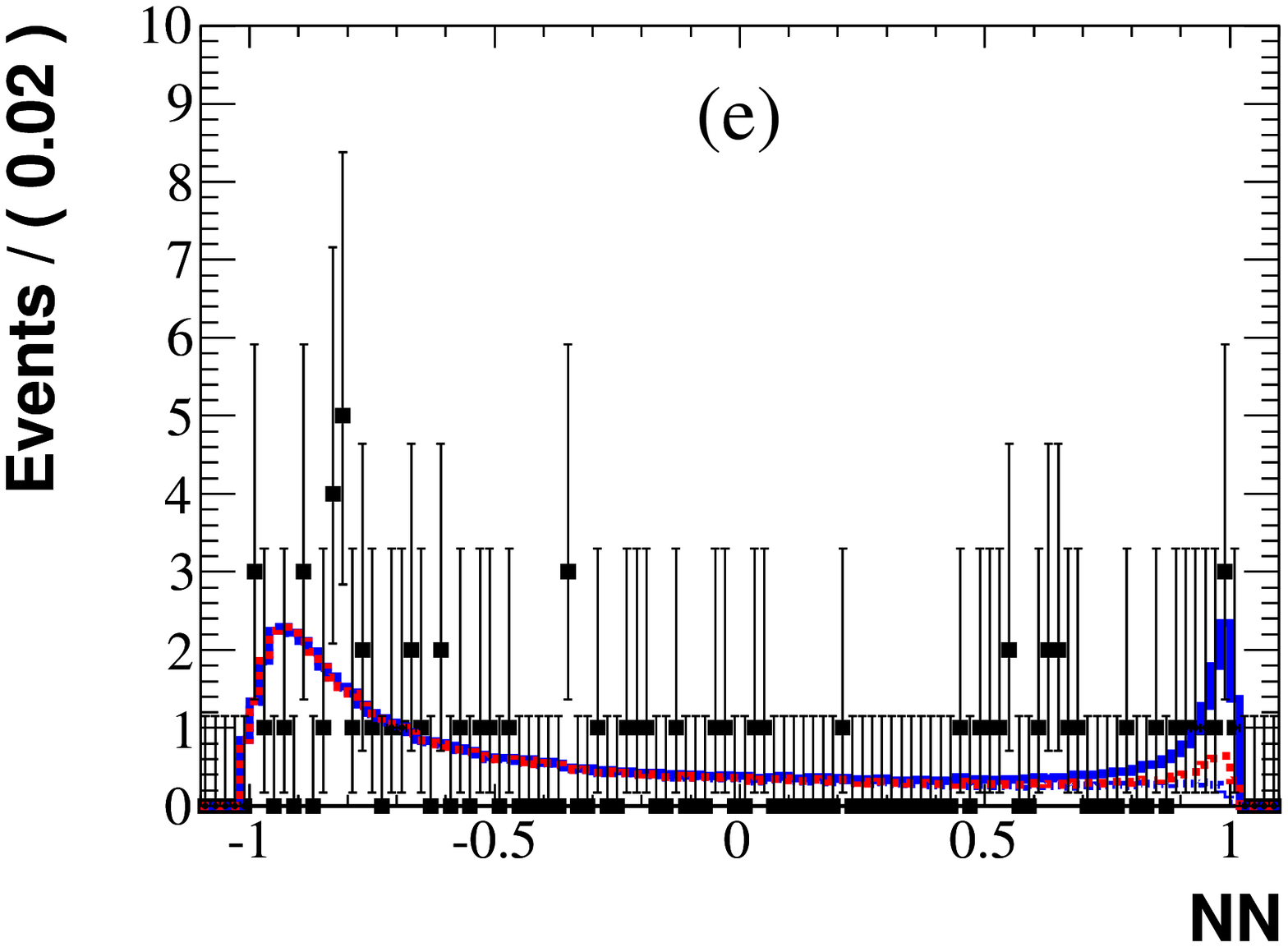,width=0.319\linewidth}
\epsfig{file=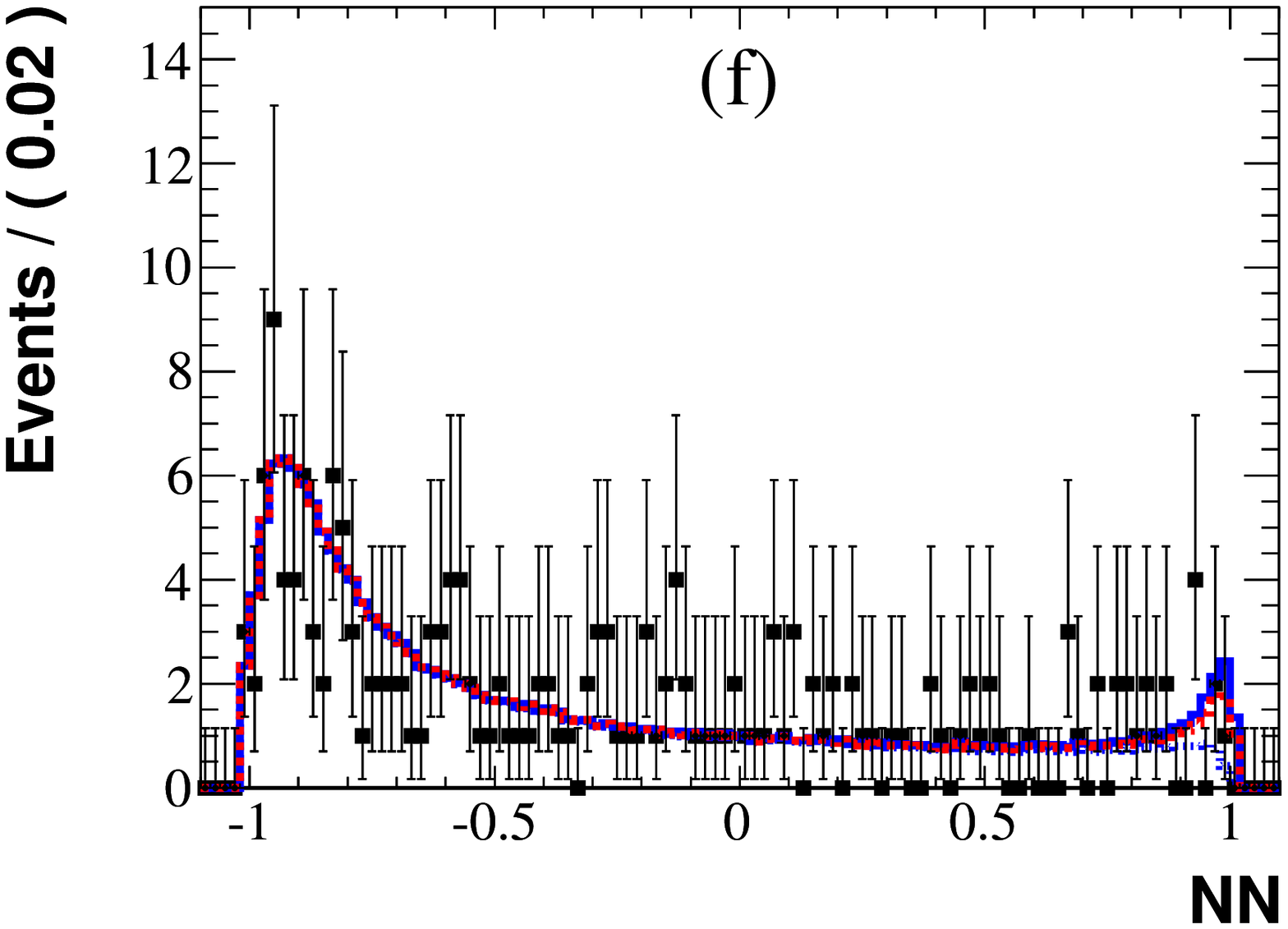,width=0.319\linewidth} \caption{(color
online). Projections on \mes (a, b, c) and $NN$ (d, e, f) of the
fit results for $D\Km$ (a, d), $D^{*}_{D \piz}\Km$ (b, e) and
$D^{*}_{D \gamma}\Km$ (c, f) WS decays, for samples enriched in
signal with the requirements $NN>0.94$ (\mes projections) or
$5.2725<\mes<5.2875\gevcc$ ($NN$ projections). The points with
error bars are data. The curves represent the fit projections for
signal plus background (solid), the sum of all background
components (dashed), and $q\bar q$ background only (dotted). }
\label{fig:dKmresultsads}
\end{center}
\end{figure*}

The systematic uncertainties have been estimated by testing
different fit models and recomputing \RDDstarK, as explained in
Section~\ref{sec:resultsdpi}. A summary of the different
systematic uncertainties is given in
Table~\ref{tab:systematics-dk}. The uncertainties on the $NN$
describing the $B \bar B$ combinatorial background and the
uncertainties on the $B \bar B$ peaking background are the two
main contributions. For $B^{\pm}\to D K^\pm$, we find for the
ratio of the WS to RS decay rates
$$ \RDK = (1.1\pm 0.5 \pm 0.2)\times 10^{-2}.$$
 Expressed in terms of event yields,
the fit result is $19.4 \pm 9.6 \pm 3.5$ WS events. The results of
fits to separate $B^+ \to DK^+$ and $B^- \to DK^-$ data samples
are given in Table~\ref{tab:fit-results-dk}. Projections of the
fits to $B^+$ and $B^-$ data are shown in
Figs.~\ref{fig:dKpresultsads} and ~\ref{fig:dKmresultsads},
respectively. We fit $\RDKp = (2.2\pm 0.9\pm 0.3)\times 10^{-2}$
for the $B^+$ sample, corresponding to $19.2\pm 7.9\pm 2.6$
events. On the contrary, no significant WS signal is observed for
the $B^-$ sample, and we fit $\RDKm = (0.2\pm 0.6\pm 0.2)\times
10^{-2}$. The statistical correlation between $\RDKp$ and $\RDKm$
(or \RDK and \ADK) is insignificant.

The systematic errors on the asymmetries are estimated using the
method discussed previously. The main systematic error on  $\ADK$
is from the uncertainty on the number of peaking \B background
events for the WS channel. This source contributes
$^{+0.11}_{-0.14}$ to $\ADK$, and $\pm 0.08\times 10^{-2}$ to
\RDK, where the changes in the two quantities are 100\% negatively
correlated (increasing the peaking background increases $\ADK$ but
decreases \RDK). The other sources of systematic uncertainty
considered in Table~\ref{tab:systematics-dk} are 100\% correlated
between ${\cal R}^+$ and ${\cal R}^-$, and mostly cancel in the
asymmetry calculation. By comparing the number of \Bp and \Bm
events reconstructed in the $[K^\pm \pi^\mp]_D \pi^\pm$ analysis,
where no significant asymmetry is expected, the uncertainty due to
the detector charge asymmetry is estimated to be below the 1\%
level. Finally, we also account for a possible asymmetry of the
charmless $B^\pm \to K^\pm K^\mp \pi^\pm$ peaking background. The
asymmetry of this background has been measured to be $0\pm
10\%$~\cite{KpKmpipbabar} and we estimate the corresponding
systematic uncertainty by assuming a $\pm$10\% asymmetry of this
background.
 The final result for the asymmetry is:

$$ \ADK = -0.86 \pm 0.47 \
^{+0.12}_{-0.16}\ . $$

\begin{table}[htb]
   \caption{Summary of systematic uncertainties on ${\cal R}$ for $D^{(*)}K$, in units of $10^{-2}$.}
   \begin{center}
\begin{tabular}{lccc}
     \hline\hline
     Error source & $\Delta {\cal R}(10^{-2})$ & $\Delta {\cal R}(10^{-2})$ & $\Delta {\cal R}(10^{-2})$ \\
      & $D K$ & $D^{*}_{D\piz}K$ & $D^{*}_{D\gamma}K$ \\
    \hline
     Signal $NN$                     & $\pm 0.1$ & $\pm 0.1$ & $\pm 0.3$ \\
     \BB  background $NN$            & $\pm 0.1$ & $\pm 0.3$ & $\pm 0.4$ \\
     $q\bar q$ background $NN$       & $\pm 0.1$ & $\pm 0.1$ & $\pm 0.1$ \\
     \BB  comb. bkg shape (\mes)     & $\pm 0.1$ & $\pm 0.1$ & $\pm 0.1$ \\
     Peaking background WS           & $\pm 0.2$ & $\pm 0.3$ & $\pm 0.6$ \\
     Peaking background RS           & $\pm 0.0$ & $\pm 0.1$ & $\pm 0.1$ \\
     Floating \BB  comb. bkg         & -         & $\pm 0.1$ & $\pm 0.2$ \\ \hline
     Combined                        & $\pm 0.2$ & $\pm 0.4$ & $\pm 0.8$ \\
     \hline\hline
   \end{tabular}
   \end{center}
   \label{tab:systematics-dk}
   \end{table}

For $B^{\pm}\to D^{*}_{D\piz} K^\pm$, we find for the ratio of the
WS to RS decay rates
$$ \RDstarKpiz = (1.8\pm 0.9 \pm 0.4)\times 10^{-2}.$$
Expressed in terms of event yields, the fit result is $10.3 \pm
5.5 \pm 2.4$ WS events. The results of fits to separate $B^+ \to
D^{*}K^+$ and $B^- \to D^{*}K^-$ data samples are given in
Table~\ref{tab:fit-results-dk}. Projections of the fits to $B^+$
and $B^-$ data are shown in Figs.~\ref{fig:dKpresultsads} and
~\ref{fig:dKmresultsads}, respectively. We find  $\RDstarKpizm =
(3.7\pm 1.8 \pm 0.9)\times 10^{-2}$ for the $B^-$ sample,
corresponding to $10.2\pm 4.8 \pm 2.4$ events. On the contrary, no
significant WS signal is observed for the $B^+$ sample, and we
find $\RDstarKpizp = (0.5\pm 0.8\pm 0.3)\times 10^{-2}$. The
systematic errors are estimated using the same method as for
$B^{\pm}\to D K^\pm$, separately for $B^+$ and $B^-$ events.
 The main systematic error on the asymmetry \ADstarKpiz
 is from the uncertainty on the number
of peaking \B background events for the WS channel. This source
contributes $\pm 0.09$ to \ADstarKpiz, and $\mp 0.3\times 10^{-2}$
to \RDstarKpiz, where the two quantities are anti-correlated. The
other sources of systematic uncertainties mostly cancel in the
asymmetry calculation, because they induce relative changes on
${\cal R^{*+}}$ and ${\cal R^{*-}}$ which are 100\% correlated.
The final result for the asymmetry is:

$$ \ADstarKpiz = +0.77 \pm 0.35\pm 0.12.$$
The asymmetry for $D^*_{D\piz}K$ has the opposite sign to the
asymmetry for $DK$, in agreement with the shift of approximately
$~180^\circ$ between $\delta_B$ and $\delta^*_B$ suggested by the
measurements of Refs.~\cite{BaBarDalitz, BelleDalitz}.

 For $B\to D^{*}_{D\gamma}K$, we have no significant
signal and fit
$$\RDstarKgam =(1.3\pm 1.4\pm 0.8)\times 10^{-2}.$$
Expressed in terms of event yields, this result corresponds to
$5.9 \pm 6.4 \pm 3.2$ events $D^{*}_{D\gamma}K$ WS. We fit $211\pm
19$ RS \Bm events and $244\pm 20$ RS \Bp events, and find for the
WS to RS ratios $\RDstarKgamm =(1.9\pm 2.3 \pm 1.2)\times 10^{-2}$
and $\RDstarKgamp =(0.9\pm 1.6 \pm 0.7)\times 10^{-2}$. The
corresponding asymmetry is

$$\ADstarKgam = +0.36\pm 0.94^{+0.25}_{-0.41}.$$

\section{DISCUSSION}

We use the $\Bm \to D^{(*)}\Km$ analysis results and a frequentist
statistical approach~\cite{ckmfitter} to extract information on
$r_B$ and $r^{(*)}_B$. In this technique a $\chi^2$ is calculated
using the differences between the measured and theoretical values
(including systematic errors) of the various ADS quantities from
Eqs.~(\ref{eqn:rpm}),~(\ref{eqn:rpmdst1}) and (\ref{eqn:rpmdst2}).
We assume Gaussian measurement uncertainties. This assumption was
checked to be valid and conservative at low $r_B$ values with a
full frequentist approach \cite{BaBarDalitz}. For $\Bm \to D\Km$,
we have for instance
     \begin{eqnarray}
   \label{eqn:chi2}
\chi^2 & = & (\RDKp - \RDKpth(r_B, \gamma, \delta_B, r_D,
\delta_D))^2/\sigma_{\cal R^+}^2 \nonumber \\
       & + &(\RDKm - \RDKmth(r_B, \gamma, \delta_B, r_D,
\delta_D))^2/\sigma_{\cal R^-}^2 \nonumber \\
   & + & (r_D^{(m)}-r_D)^2/\sigma_r^2 \nonumber \\
   & + & (\delta_D^{(m)}-\delta_D)^2/\sigma_{\delta}^2,
    \end{eqnarray}
where $\RDKpmth(r_B, \gamma, \delta_B, r_D, \delta_D)$ is given by
Eq.~(\ref{eqn:rpm}), and where the two last terms constrain $r_D$
and $\delta_D$ to the values $r_D^{(m)}$ and $\delta_D^{(m)}$ of
Ref.~\cite{HFAG} within their errors $\sigma_r$ and
$\sigma_{\delta}$. The choice of (\RDKp, \RDKm) rather than (\RDK,
\ADK) is motivated by the fact that the set of variables (\RDK,
\ADK) is not well-behaved (the uncertainty on \ADK depends on the
central value of \RDK), while (\RDKp, \RDKm) are two statistically
independent observables.
In the same way, the two pairs of ADS observables (\RDstarKpizp,
\RDstarKpizm) and (\RDstarKgamp, \RDstarKgamm) are used to extract
$r^*_B$, while accounting for the relative phase difference in the
two $D^{*}$ decays~\cite{Bondar}. We allow $0\leq r^{(*)}_B\leq
1$, $-180^{\circ}\leq \gamma \leq 180^{\circ}$, and
$-180^{\circ}\leq \delta^{(*)}_B \leq 180^{\circ}$. The minimum of
the $\chi^2$ for the $r^{(*)}_B$, $\gamma$, $\delta^{(*)}_B$,
$r_D$, and $\delta_D$ parameter space is calculated first
($\chi^2_{{\rm min}}$). We then scan the range of $r^{(*)}_B$
minimizing the $\chi^2$ ($\chi^2_{{\rm m}}$) by varying
$\delta^{(*)}_B$, $\gamma$, $r_D$, and $\delta_D$. A confidence
level (C.L.) for $r_B$ is calculated using $\Delta
\chi^2=\chi^2_{\rm m}-\chi^2_{\rm min}$ and one degree of freedom.

\begin{figure}
\begin{center}
\epsfig{file=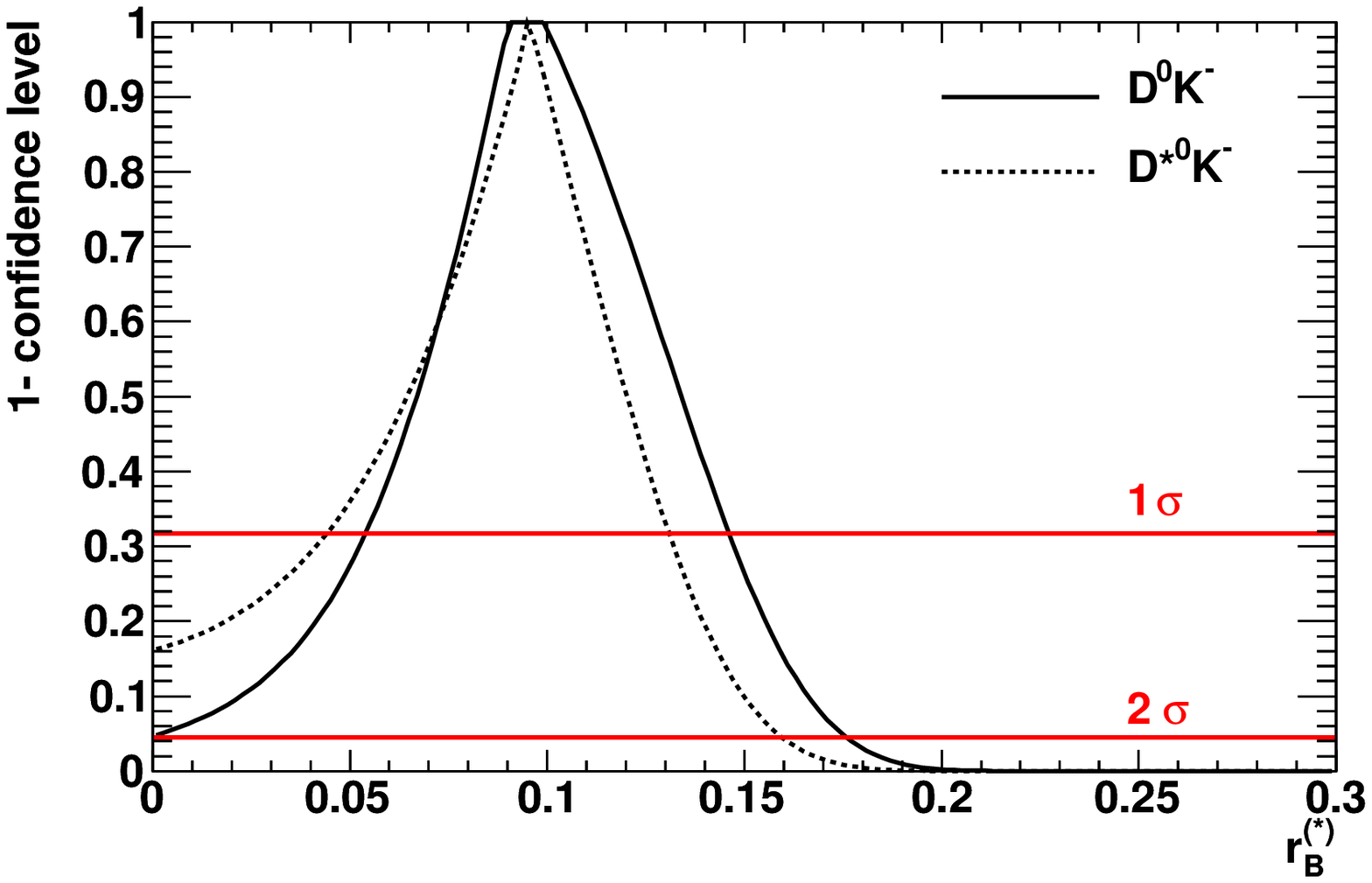,width=0.95\linewidth}
\caption{(color online). Constraints on $r^{(*)}_B$ from the
combined $\Bm\to [K\pi]_{D^{(*)}}K^{-}$ ADS measurements. The
solid (dotted) curve shows the 1 minus the confidence level to
exclude the abscissa value as a function of $r^{(*)}_B$. The
horizontal lines show the exclusion limits at the 1 and 2 standard
deviation levels.}\label{fig:rBDanddstarK}
\end{center}
\end{figure}

The results of this procedure are shown in
Fig.~\ref{fig:rBDanddstarK} for the C.L. curve as a function of
$r^{(*)}_B$. The results are summarized in Tab.
\ref{tab:randrstarB}. For $\Bm\to [K\pi]_{D}K^{-}$, we find the
minimum $\chi^2$ at $r_B=(9.5^{+5.1}_{-4.1})\%$. This leads to the
upper limit: $r_B<16.7\%$ at $90\%$ C.L., to be compared to
$r_B<23\%$
 at $90\%$ C.L. for the previous ADS analysis as
performed by \babar~\cite{ADS-BABAR} with $232 \times10^6$ \BB
pairs, and to $r_B<19\%$ at $90\%$ C.L. for the corresponding ADS
analysis as performed by Belle~\cite{BelleADS} with
$657\times10^6$ \BB pairs. We exclude $r_B=0$ with a C.L. of
95.3\%. Similarly, for $\Bm\to [K\pi]_{D^{*}}K^{-}$
 we find $r^*_B=(9.6^{+3.5}_{-5.1})\%$. This
leads to the upper limit: $r^*_B<15.0\%$ at $90\%$ C.L., to be
compared to $r^*_B<16\%$ at $90\%$ C.L. for the previous \babar\
ADS analysis~\cite{ADS-BABAR}. We exclude $r^*_B=0$ with a C.L. of
83.9\%.

\begin{table}
\caption{Constraints on $r^{(*)}_B$ from the combined $\Bm\to
[K\pi]_{D^{(*)}}K^{-}$ ADS
measurements.}\label{tab:randrstarB}\begin{center}
\begin{tabular}{lcc}
\hline\hline
Parameter & 1 $\sigma$ meas. & $90\%$ C.L. upper limit \\
\hline
$r_B$      & $(9.5^{+5.1}_{-4.1})\%$  & $<16.7\%$  \\
\hline
$r^*_B$ from & & \\
 $D^{*0}\rightarrow D^0\piz$   & $(13.1^{+4.2}_{-6.1})\%$  & $<19.5\%$  \\

 $D^{*0}\rightarrow D^0\gamma$  & $(12.0^{+10.0}_{-12.0})\%$  & $<24.5\%$  \\

 all  $D^{*0}$ decays & $(9.6^{+3.5}_{-5.1})\%$  & $<15.0\%$  \\
\hline \hline
\end{tabular}

\end{center}
\end{table}

\begin{figure}
\begin{center}
\epsfig{file=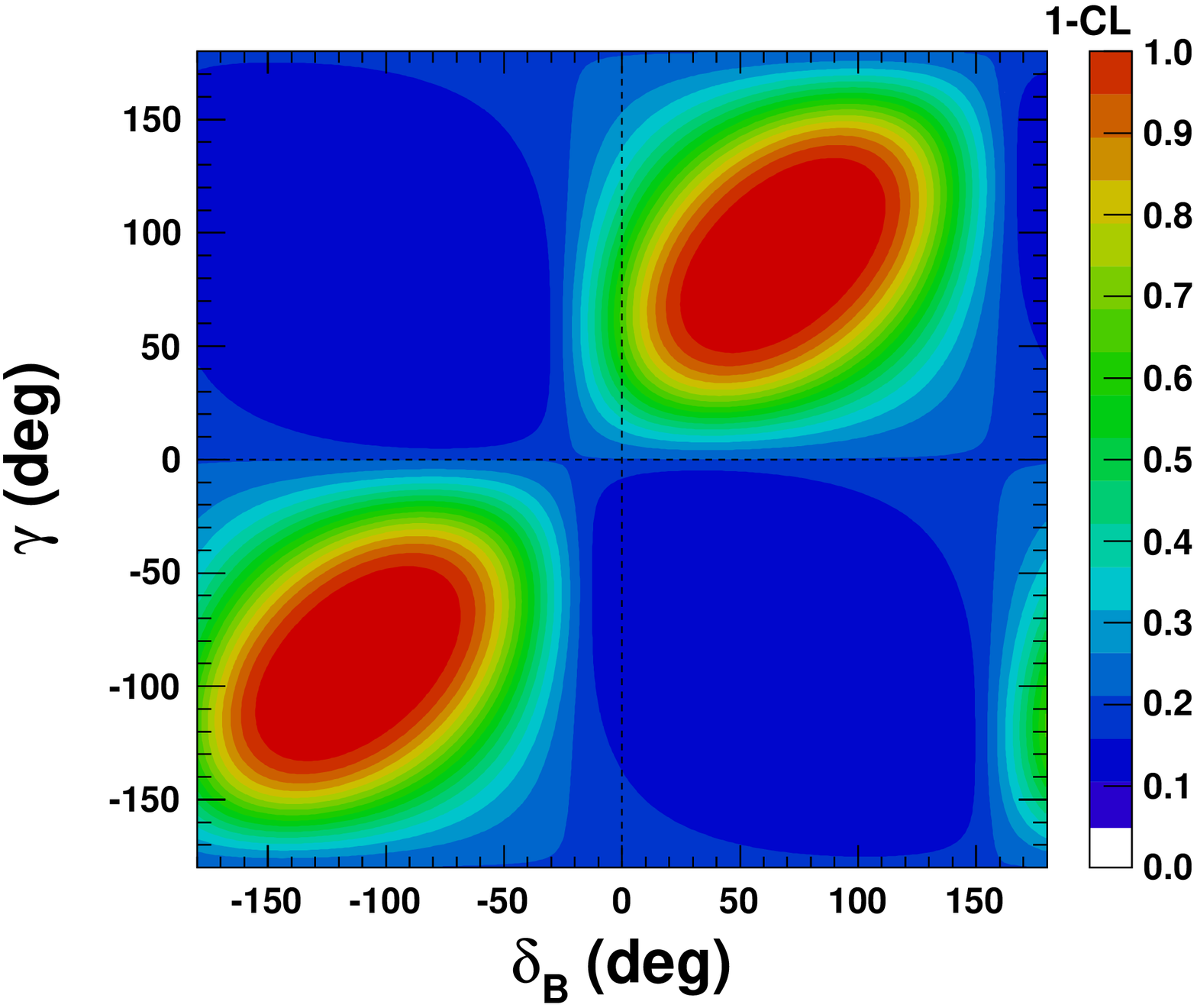,width=0.9\linewidth}
\caption{(color online). One minus confidence level isocontours
%$1\sigma$ (striped areas) and $2\sigma$ (shaded) contours
on $\gamma$ vs $\delta_B$ from the $\Bm\to [K\pi]_{D}K^{-}$ ADS
measurement.} \label{fig:deltaB}
\end{center}
\end{figure}

\begin{figure}
\begin{center}
\epsfig{file=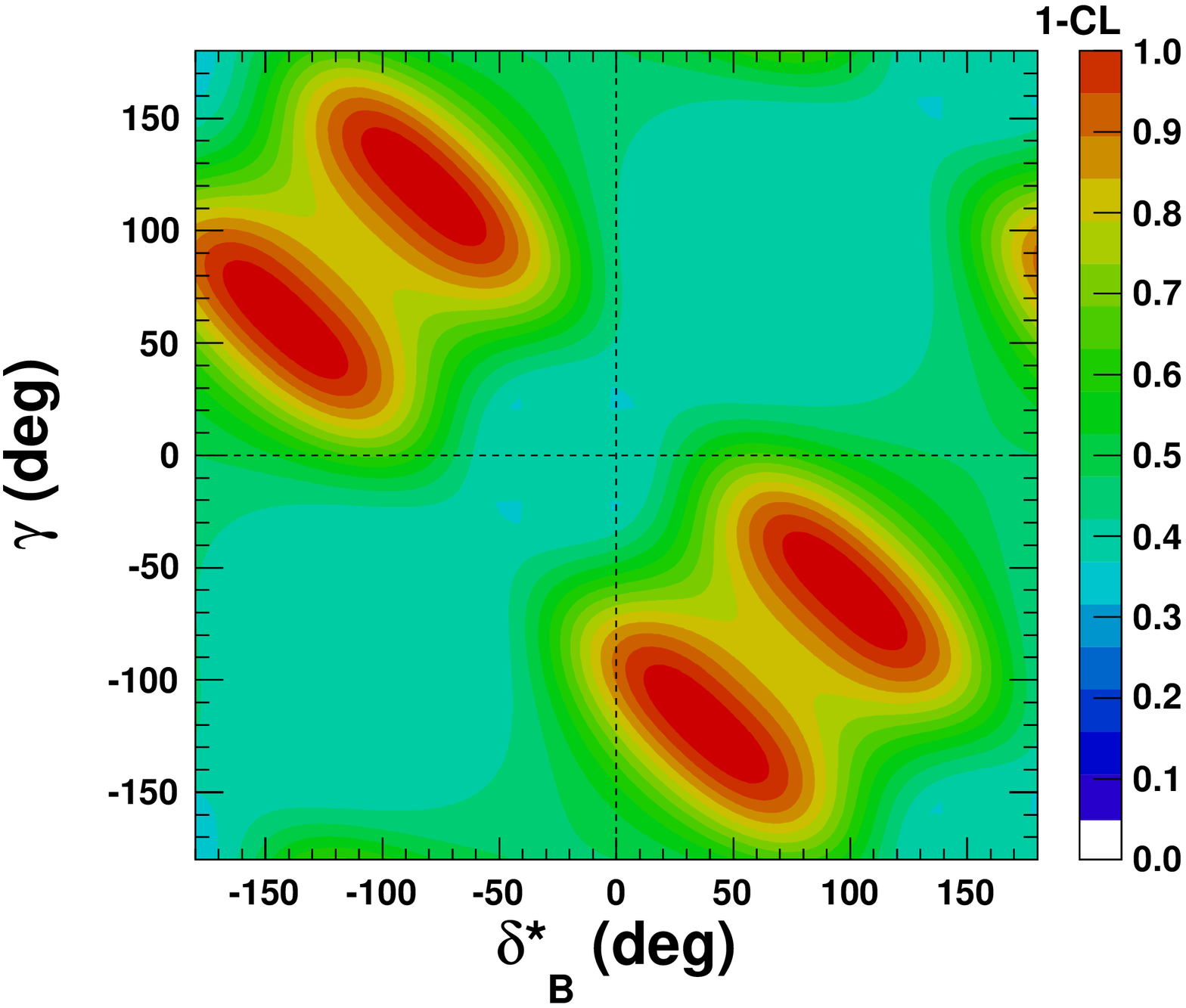,width=0.9\linewidth}
\caption{(color online). One minus confidence level isocontours
%$1\sigma$ (striped areas) and $2\sigma$(shaded) contours
on $\gamma$ vs $\delta^*_B$ from the combined
$\Bm\to [K\pi]_{D^{*}}K^{-}$ ADS measurements.}
\label{fig:deltastarB}
\end{center}
\end{figure}

\begin{figure}
\begin{center}
\epsfig{file=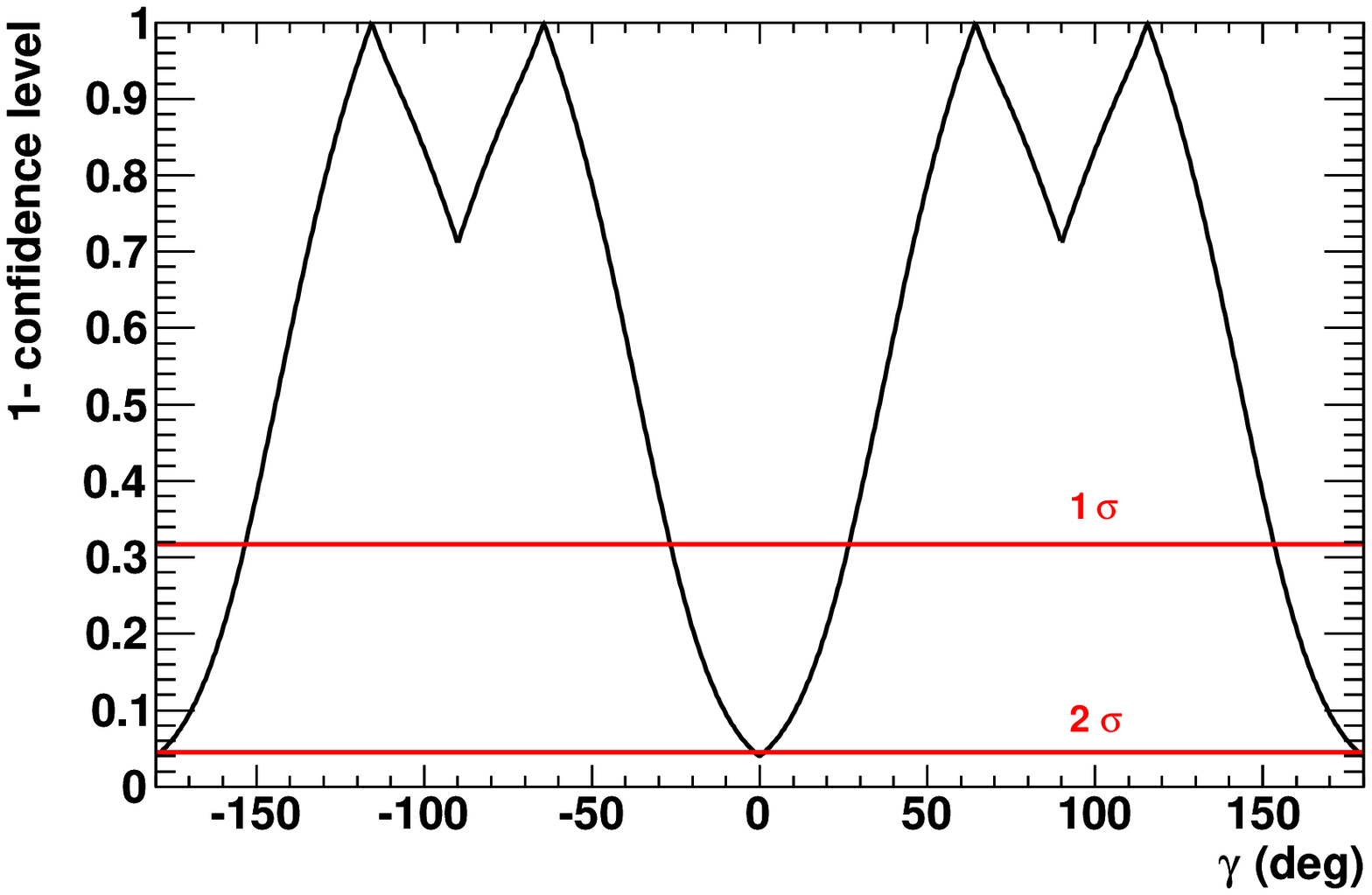,width=0.95\linewidth}
\caption{(color online). Constraints on $\gamma$ from the combined
$\Bm\to D^{(*)}[K^+\pi^-]K^{-}$ ADS measurements. The solid curve
shows the (1-C.L.) to exclude the abscissa value. The horizontal
lines show the exclusion limits at the 1 and 2 standard deviation
levels. } \label{fig:gamma}
\end{center}
\end{figure}

Using the above procedure we also determine the 2D confidence
intervals for $\gamma$ vs $\delta^{(*)}_B$ shown in Figs.
~\ref{fig:deltaB} and \ref{fig:deltastarB}. Choosing the solution
with $0<\gamma<180^\circ$ favors a positive sign for the strong
phase $\delta_B$ ($\ADK<0$), and a negative sign for the strong
phase $\delta^*_B$ ($\ADstarKpiz>0$). This result is in good
agreement with the values of the strong phases determined in
Refs.~\cite{BaBarDalitz, BelleDalitz}. Finally,
Fig.\ref{fig:gamma} shows the C.L. curve as a function of $\gamma$
when combining the $DK$ and $D^*K$ results.

\section{SUMMARY} \label{sec:Summary}

   In summary, using a data sample of 467 million \BB  pairs, we present an updated search of the decays $B^-\to D^{(*)}K^-$
   where the neutral $D$ meson decays into the $K^+\pim$ final state (WS). The analysis method is first applied to $B^-\to D^{(*)}\pi^-$, where the
   $D$ decays into the Cabibbo favored ($K^-\pi^+$) and doubly suppressed modes ($K^+\pim$).
   We measure  $\RDpi = (3.3 \pm 0.6 \pm
   0.4)\times 10^{-3}$, $\RDstarpipiz = (3.2 \pm 0.9 \pm
   0.8)\times 10^{-3}$ and $\RDstarpigam = (2.7 \pm 1.4 \pm
   2.2)\times 10^{-3}$, in good agreement with the ratio $R_D$ of the suppressed to favored
   $\Dz\to K\pi$ decay rates, $R_D=(3.36\pm 0.08)\times 10^{-3}$~\cite{HFAG}.
   Both the branching fraction ratios and the \CP asymmetries measured for those modes, $\ADpi = (3
\pm 17 \pm 4)\times 10^{-2}$, $\ADstarpipiz = (9 \pm 27 \pm
5)\times 10^{-2}$ and $\ADstarpigam = (65 \pm 55\ ^{+20}_{-24}
)\times 10^{-2}$, are consistent with the expectations discussed
in Section~\ref{sec:Introduction}.

We see indications of signals for the $B\to DK$ and $B\to
D^{*}_{D\piz}K$ wrong-sign modes, with significances of
$2.1\sigma$ and $2.2\sigma$, respectively. The ratios of the WS to
RS branching fractions are measured to be $ \RDK = (1.1\pm 0.5 \pm
0.2)\times 10^{-2}$ and $ \RDstarKpiz = (1.8\pm 0.9 \pm 0.4)\times
10^{-2}$ for $B\to DK$ and $B\to D^{*}_{D\piz}K$, respectively.
The separate measurements of \RDDstarKpm for \Bp and \Bm events
indicates large \CP asymmetries, with $ \ADK = -0.86 \pm 0.47 \
^{+0,12}_{-0.16} $ for $B\to DK$ and $ \ADstarKpiz = +0.77 \pm
0.35\pm 0.12 $ for $B\to D^{*}K$, $D^{*}\to D\piz$. For the $B\to
D^{*}_{D\gamma}K$ WS mode, we see no statistically significant
evidence of a signal. We measure $ \RDstarKgam = (1.3\pm 1.4\pm
0.8 )\times 10^{-2}$ and $ \ADstarKgam = +0.36 \pm 0.94\
^{+0.25}_{-0.41}$. These results are used to extract the following
constraints on $r^{(*)}_B$:

   \begin{eqnarray}
r_B &=& (9.5^{+5.1}_{-4.1})\%,\nonumber \\
r^*_B &=& (9.6^{+3.5}_{-5.1})\%. \nonumber
   \end{eqnarray}

Assuming $0<\gamma<180^\circ$, we also extract constraints on the
strong phases $\delta_B^{(*)}$, in good agreement with other
measurements~Ref.~\cite{BaBarDalitz, BelleDalitz}.

\section{ACKNOWLEDGMENTS}
\label{sec:Acknowledgments}

% Standard acknowledgments paragraph; must always be included.
We are grateful for the 
extraordinary contributions of our \pep2\ colleagues in
achieving the excellent luminosity and machine conditions
that have made this work possible.
The success of this project also relies critically on the 
expertise and dedication of the computing organizations that 
support \babar.
The collaborating institutions wish to thank 
SLAC for its support and the kind hospitality extended to them. 
This work is supported by the
US Department of Energy
and National Science Foundation, the
Natural Sciences and Engineering Research Council (Canada),
the Commissariat \`a l'Energie Atomique and
Institut National de Physique Nucl\'eaire et de Physique des Particules
(France), the
Bundesministerium f\"ur Bildung und Forschung and
Deutsche Forschungsgemeinschaft
(Germany), the
Istituto Nazionale di Fisica Nucleare (Italy),
the Foundation for Fundamental Research on Matter (The Netherlands),
the Research Council of Norway, the
Ministry of Education and Science of the Russian Federation, 
Ministerio de Ciencia e Innovaci\'on (Spain), and the
Science and Technology Facilities Council (United Kingdom).
Individuals have received support from 
the Marie-Curie IEF program (European Union), the A. P. Sloan Foundation (USA) 
and the Binational Science Foundation (USA-Israel).

\end{document}